\documentclass{article}
\usepackage{times}
\usepackage{amsmath,amssymb,amsfonts,amsthm}
\usepackage{hyperref}
\def\l{\label}
\def\ct{\cite}
\def\r{\ref}
\def\lgth{[\,\mbox{length}\,]}
\def\gam{\gamma}
\def\del{\delta}
\def\eps{\epsilon}
\def\Th{\Theta}
\def\sig{\sigma}
\def\om{\omega}
\def\udot{\dot{u}}
\def\3nab{\tilde{\nabla}}
\def\vec#1{\mbox{\boldmath$#1$}}
\def\fd{{\bf e}}
\def\p{\partial}
\def\lgl{\langle}
\def\rgl{\rangle}
\def\la {\langle}
\def\ra {\rangle}
\def\c{\mbox{curl}}
\def\div{\mbox{div}}

\def\hsp5{\hspace{5mm}}
\newcommand{\sfrac}[2]{{\textstyle{#1\over#2}}}
\def\case#1/#2{\textstyle\frac{#1}{#2}}

\def\ts {\textstyle}
\def\be {\begin{equation}}
\def\ee {\end{equation}}
\def\bea {\begin{eqnarray}}
\def\eea {\end{eqnarray}}
\def\cs{c_{s}^{2}}

\def\vb {v_{\!_B}}

\def\st {\sigma_{\!_T}}
\def\ne {n_{\!_E}}
\def\ts {\textstyle}
\def\bi {\bibitem}
\def\case#1/#2{\textstyle\frac{#1}{#2} }

\def\cqg{{\em Class. Quantum Grav.\/} }
\def\grg{{\em Gen. Rel. Grav.\/} }
\def\prd{{\em Phys. Rev.\/} D }
\def\prl{{\em Phys. Rev. Lett.\/} }
\def\apj{{\em Astrophys. J.\/} }
\def\jmp{{\em J. Math. Phys.\/} }
\def\mn{{\em Mon. Not. Roy. Astr. Soc.\/} }
\title{{\Huge\sc COSMOLOGICAL MODELS}\\
{\Large\sc Carg\`{e}se Lectures 1998}}
\author{{\LARGE\sc George F R Ellis\thanks{Electronic address:
George.Ellis@uct.ac.za}}\\ \\
{and}\\ \\
{\LARGE\sc Henk van Elst\thanks{Electronic address:
hvanelst@merkur-fh.org}}\\             \\
{\em Cosmology Group\/} \\
{\em Department of Mathematics and Applied Mathematics\/} \\
{\em University of Cape Town, Rondebosch 7701, Cape Town,
South Africa\/} 
}
\date{{September 2, 2008}}
\begin{document}
\maketitle
\vspace{20mm}
\begin{abstract}
The aim of this set of lectures is a systematic presentation of a
$1+3$ covariant approach to studying the geometry, dynamics, and
observational properties of relativistic cosmological models.  In
giving (i) the basic $1+3$ covariant relations for a cosmological
fluid, the present lectures cover some of the same ground as a
previous set of Carg\`{e}se lectures \cite{ell73}, but they then go
on to give (ii) the full set of corresponding tetrad equations,
(iii) a classification of cosmological models with exact
symmetries, (iv) a brief discussion of some of the most useful
exact models and their observational properties, and (v) an
introduction to the gauge-invariant and $1+3$ covariant
perturbation theory of
almost-Friedmann--Lema\^{\i}tre--Robertson--Walker universes, with
a fluid description for the matter and a kinetic theory description
of the radiation.
\end{abstract}
\vspace{20mm}
\centerline{\bigskip\noindent e-print
\href{http://arxiv.org/abs/gr-qc/9812046}{arXiv:gr-qc/9812046v5}}
\newpage
\tableofcontents
\newpage

\section{Basic relations}
A {\bf cosmological model} represents the {\bf Universe} at a
particular scale. We will assume that on large scales,
space-time geometry is described by Einstein's {\bf general theory
of relativity} (see, e.g., d'Inverno \ct{din}, Wald \ct{wald},
Hawking and Ellis \ct{he73}, or Stephani \ct{ste90}). Then a
cosmological model is defined by specifying \ct{ehl61}--\ct{ell73}:
\\

* the {\bf space-time geometry} represented on some specific
 averaging scale and determined by the {\bf metric}
 $g_{ab}(x^\mu)$, which
 --- because of the requirement of compatibility with observations
 --- must either have some expanding Robertson--Walker (`RW')
 geometries as a regular limit (see \ct{kra93}), or else be
 demonstrated to have observational properties compatible with the
 major features of current astronomical observations of the
 Universe;

* the {\bf matter} present, represented on the same averaging
  scale, and its {\bf physical behaviour} (the energy-momentum
  tensor of each matter component, the equations governing the
  behaviour of each such component, and the interaction terms
  between them), which must represent physically plausible matter
  (ranging from early enough times to the present day, this will
  include most of the interactions described by present-day
  physics); and

* the {\bf interaction of the geometry and the matter}
  --- how matter
  determines the geometry, which in turn determines the motion of
  the matter (see e.g. \ct{whe68}). We assume this is through
  {\bf Einstein's relativistic gravitational field equations}
  (`EFE') given by\footnote{Throughout this review we employ
  geometrised units characterised by $c = 1 = 8\pi
  G/c^{2}$. Consequently, all 
  geometrical variables occurring have physical dimensions that are
  integer powers of the dimension $\lgth$. The index convention is
  such that space-time and spatial indices with respect to a
  general basis are denoted by $a, b, \dots = 0, 1, 2, 3$ and
  $\alpha, \beta, \dots = 1, 2, 3$, respectively, while space-time
  and spatial indices in a coordinate basis are $\mu, \nu, \dots
  = 0, 1, 2, 3$ and $i, j, \dots = 1, 2, 3,$, respectively.}
\be
\l{eq:efe}
G_{ab} \equiv R_{ab} - {\sfrac12}\,R\,g_{ab}
= T_{ab} - \Lambda\,g_{ab} \ ,
\ee
which, because of the {\bf twice-contracted Bianchi identities},
guarantee the conservation of total energy-momentum
\be
\l{eq:cons}
\nabla_{b}G^{ab} = 0 \hsp5 \Rightarrow \hsp5
\nabla_{b}T^{ab} = 0 \ ,
\ee
provided the {\bf cosmological constant} $\Lambda$ satisfies the
relation $\nabla_{a}\Lambda = 0$, i.e., it is constant in time and
space. \\

Together, these determine the combined dynamical evolution of the
model and the matter in it. The description must be sufficiently
complete to determine

* the {\bf observational relations} predicted by the model for both
  discrete sources and background radiation, implying a
  well-developed theory of {\bf structure growth} for very small
  and for very large physical scales (i.e., for light atomic nuclei
  and for galaxies and clusters of galaxies), and of {\bf radiation
  absorbtion and emission}.\\

To be useful in an explanatory role, a cosmological model must be
easy to describe --- that means they have symmetries or special
properties of some kind or other. The usual choices for the matter
description will be some combination of

* a fluid with a physically well-motivated equation of state, for
 example a perfect fluid with specified equation of state (beware
 of imperfect fluids, unless they have well-defined and motivated
 physical properties);

* a mixture of fluids, usually with different 4-velocities;

* a set of particles represented by a kinetic theory description;

* a scalar field $\phi$, with a given potential $V(\phi)$ (at early
 times);

* an electromagnetic field described by Maxwell's field
  equations. \\

As intimated above, the observational relations implied by
cosmological models must be compared with astronomical
observations. This determines those solutions that can usefully be
considered as viable cosmological models of the real Universe. A
major aim of the present lectures is to point out that this class
is wider than just the standard
{\bf Friedmann--Lema\^{\i}tre--Robertson--Walker} (`FLRW')
{\bf cosmologies}; indeed those models cannot be realistic on
all scales of description, but they may also be inaccurate on large
scales, or at very early and very late times. To examine this, we
need to consider the subspace of the {\bf space of all cosmological
solutions} that contains models with observational properties like
those of the real Universe at some stage of their histories. Thus
we are interested in the {\it full} state space of solutions,
allowing us to see how realistic models are related to each other
and to higher symmetry models, including particularly the FLRW
models. \\

These lectures develop general techniques for examining this, and
describe some specific models of interest.  The first part looks at
exact general relations valid in all cosmological models, the
second part at exact cosmological solutions of the EFE, and the
third part at approximate equations and solutions: specifically,
`almost-FLRW' models, linearised about a FLRW geometry.

\section{$1+3$ covariant description}

{\bf Space-times} can be described via 

(a) the {\bf metric} $g_{ij}(x^k)$ described in a particular set of
{\bf local coordinates}, with its differential properties, as
embodied by the connection, given through the Christoffel
symbols;

(b) the {\bf metric} described by means of particular {\bf tetrads},
with its connection given through the Ricci rotation
coefficients;

(c) {\bf $1+3$ covariantly} defined variables. In anisotropic
cases, tetrad vectors can be uniquely defined in a $1+3$ covariant
way and this approach merges into (b).

Here we will concentrate on the {\bf $1+3$ covariant approach},
based on \ct{ehl61,kt62,ell71,ell73,maa97}, but dealing also with
the tetrad approach which serves as a completion to the $1+3$
covariant approach. The basic point here is that because we have
complete coordinate freedom in General Relativity, it is preferable
where possible to describe physics and geometry by tensor relations
and quantities; these then remain valid whatever coordinate system
is chosen.

\subsection{Variables}
\subsubsection{Average 4-velocity of matter}
In a cosmological space-time $({\cal M},{\bf g})$, at late times
there will be a family of preferred worldlines representing the
average motion of {\bf matter} at each point\footnote{We are here
assuming a fluid description can be used on a large enough scale
\ct{ehl61,ell71}. The alternative is that the matter distribution
is hierarchically structured at all levels or fractal (see
e.g. \ct{rib} and refrences there), so that a fluid description
does not apply. The success of the FLRW models encourages us to use
the approach taken here.} (notionally, these represent the
histories of clusters of galaxies, with associated `fundamental
observers'); at early times there will be uniquely defined notions
of the average velocity of matter (at that time, interacting gas
and radiation), and corresponding preferred worldlines. In each
case their {\bf 4-velocity} is\footnote{Merging from the one
concept to the other as structure formation takes place.}
\be
u^a = {dx^a \over d\tau} \ , \hsp5 u_a u^a = -\,1 \ ,
\ee
where $\tau$ is {\bf proper time} measured along the fundamental
worldlines.  We assume this 4-velocity is unique: that is, there is
a well-defined preferred motion of matter at each space-time
event. At recent times this is taken to be the 4-velocity defined
by the vanishing of the dipole of the cosmic microwave background
radiation (`CBR'): for there is precisely one 4-velocity which will
set this dipole to zero. It is usually assumed that this is the
same as the average 4-velocity of matter in a suitably sized volume
\ct{ell71}; indeed this assumption is what underlies studies of
large scale motions and the `Great Attractor'.

Given $u^a$, there are defined unique {\bf projection tensors}
\bea
U^a{}_b = -\,u^a\,u_b & \hsp5 \Rightarrow & \hsp5
U^a{}_c\,U^c{}_b = U^a{}_b \ ,
~U^a{}_a = 1 \ , ~U_{ab}\,u^b = u_a \ ,\\
h_{ab} = g_{ab} + u_a\,u_b & \hsp5 \Rightarrow & \hsp5
h^a{}_c\,h^c{}_b = h^a{}_b \ ,
~ h^a{}_a = 3 \ , ~h_{ab}\,u^b = 0 \ .
\eea
The first projects parallel to the 4-velocity vector $u^a$, and the
second determines the (orthogonal) metric properties of the
instantaneous rest-spaces of observers moving with 4-velocity
$u^a$. There is also defined a {\bf volume element} for the
rest-spaces:
\be
\eta_{abc} = u^{d}\,\eta_{dabc} \hsp5 \Rightarrow \hsp5
\eta_{abc} = \eta_{[abc]} \ ,
~\eta_{abc}\,u^c = 0 \ ,
\ee
where $\eta_{abcd}$ is the 4-dimensional volume element
($\eta_{abcd} = \eta_{[abcd]}$, $\eta_{0123} =
\sqrt{|\det\,g_{ab}\,|}$).

Moreover, two derivatives are defined: the {\bf covariant time
derivative} $\,\dot{}\,$ {\bf along the fundamental worldlines}, where
for any tensor $T^{ab}{}_{cd}$
\be
\dot{T}^{ab}{}_{cd} = u^{e}\nabla_{e}T^{ab}{}_{cd} \ ,
\ee
and the {\bf fully orthogonally projected covariant derivative} $\3nab$,
where for any tensor $T^{ab}{}_{cd}$
\be
\3nab_{e}T^{ab}{}_{cd} = h^a{}_f\,h^b{}_g\,h^p{}_c\,h^q{}_d\,
h^r{}_e\nabla_r\,T^{fg}{}_{pq} \ ,
\ee
with total projection on all free indices. The tilde serves as a
reminder that if $u^{a}$ has {\em non-zero\/} vorticity, $\3nab$ is
{\it not} a proper 3-dimensional covariant derivative (see
Eq. (\r{eq:tilt}) below). Finally, following \ct{maa97} (and see
also \ct{maaellsik97}), we use angle brackets to denote orthogonal
projections of vectors and the orthogonally projected symmetric
trace-free part of tensors:
\be
v^{\la a\ra} = h^{a}{}_{b}\,v^{b} \ , \hsp5
T^{\la ab\ra} = [\ h^{(a}{}_{c}\,h^{b)}{}_{d}
- \sfrac{1}{3}\,h^{ab}\,h_{cd}\ ]\,T^{cd} \ ;
\ee
for convenience the angle brackets are also used to denote
othogonal projections of covariant time derivatives along $u^{a}$
({\bf Fermi derivatives}):
\be
\dot{v}{}^{\la a\ra} = h^{a}{}_{b}\,\dot{v}{}^{b} \ , \hsp5
\dot{T}{}^{\la ab\ra} = [\ h^{(a}{}_{c}\,h^{b)}{}_{d}
- \sfrac{1}{3}\,h^{ab}\,h_{cd}\ ]\,\dot{T}{}^{cd} \ .
\ee
%

{\it Exercise}: Show that the projected time and space derivatives
of $U_{ab}$, $h_{ab}$ and $\eta_{abc}$ all vanish.

\subsubsection{Kinematical quantities}
We split the first covariant derivative of $u_a$ into its
irreducible parts, defined by their symmetry properties:
\be
\l{eq:kin}
\nabla_{a}u_{b} = -\,u_a\,\udot_b + \3nab_{a}u_{b} =
-\,u_a\,\udot_b + {\sfrac13}\,\Th\,h_{ab} + \sigma_{ab}
+ \om_{ab} \ ,
\ee
where the trace $\Theta = \3nab_au^a{}$ is the {\bf rate
of volume expansion} scalar of the fluid (with $H = \Theta/3$ the
{\bf Hubble scalar}); $\sig_{ab} = \3nab_{\la a}u_{b\ra}$
is the trace-free symmetric {\bf rate of shear} tensor
($\sig_{ab} = \sig_{(ab)}$, $\sig_{ab}\,u^b = 0$,
$\sig^a{}_{a}= 0$), describing the rate of distortion of the
matter flow; and $\om_{ab} = \3nab_{[a}u_{b]}$ is the
skew-symmetric {\bf vorticity} tensor ($\om_{ab} =
\om_{[ab]}$, $\om_{ab}\,u^b = 0$),\footnote{The vorticity here is
defined with respect to a right-handedly oriented spatial basis.}
describing the rotation of the matter relative to a non-rotating
(Fermi-propagated) frame. The stated meaning for these quantities
follows from the evolution equation for a {\bf relative position
vector} $\eta^a_\bot = h^a{}_b\eta^b$, where $\eta^a$ is a
deviation vector for the family of fundamental worldlines,
i.e., $u^b\nabla_b\eta^a =
\eta^b \nabla_bu^a\,$. Writing $\eta^a_\bot = \delta\ell\,e^a$,
$e_a e^a = 1$, we find the relative distance $\delta\ell$ obeys the
propagation equation
\be
\frac{(\delta \ell){}\dot{}}{\delta\ell} = {\sfrac13}\,\Th
+ (\sigma_{ab}e^ae^b) \ ,
\ee
(the generalised Hubble law), and the relative direction vector
$e^a$ the propagation equation
\be
\dot{e}{}^{\la a\ra} = (\sigma^a{}_b
- (\sigma_{cd}e^ce^d)\,h^a{}_b - \omega^a{}_b)\,e^b \ ,
\ee
giving the observed rate of change of position in the sky of
distant galaxies. Finally $\dot{u}^a = u^b\nabla_bu^a$ is the {\bf
relativistic acceleration} vector, representing the degree to
which the matter moves under forces other than gravity plus inertia
(which cannot be covariantly separated from each other in General
Relativity: they are different aspects of the same effect). The
acceleration vanishes for matter in free fall (i.e., moving under
gravity plus inertia alone).

\subsubsection{Matter tensor}
The matter {\bf energy-momentum tensor} $T_{ab}$ can be
decomposed relative to $u^a$ in the form\footnote{We should really
write $\mu = \mu(u^a)$, etc; but usually assume this dependence is
understood.}
\bea
T_{ab} = \mu\,u_a\,u_b + q_a\,u_b + u_a\,q_b + p\,h_{ab} + \pi_{ab}
\l{eq:stress} \ ,\\
\hsp5 q_a\,u^a = 0 \ , ~\pi^a{}_a = 0 \ , ~\pi_{ab} = \pi_{(ab)} \ ,
~\pi_{ab}\,u^b = 0 \ , \nonumber
\eea
where $\mu = (T_{ab}u^{a}u^{b})$ is the {\bf relativistic energy
density} relative to $u^a$, $q^{a} = -\,T_{bc}\,u^{b}\,h^{ca}$ is
the {\bf relativistic momentum density}, which is also the energy
flux relative to $u^a$, $p = {\sfrac13}\,(T_{ab}h^{ab})$ is the
{\bf isotropic pressure}, and $\pi_{ab} = T_{cd}\,h^{c}{}_{\la
a}\,h^{d}{}_{b\ra}$ is the trace-free {\bf anisotropic pressure}
(stress).

The physics of the situation is in the {\bf equations of
state} relating these quantities; for example, the commonly
imposed restrictions
\be
\l{eq:pf}
q^{a} = \pi_{ab} = 0 \hsp5 \Leftrightarrow \hsp5
T_{ab} = \mu\,u_a\,u_b + p\,h_{ab}
\ee
characterise a {\bf perfect fluid} with, in general, equation of state
$p = p(\mu,s)$. If in addition we assume that $p = 0$, we have the
simplest case: pressure-free matter (`dust' or `Cold Dark
Matter'). Otherwise, we must specify an equation of state
determining $p$ from $\mu$ and possibly other thermodynamical
variables. Whatever these relations may be, we usually require that
various {\bf energy conditions} hold: one or all of
\be
\mu > 0 \ , \hsp5 (\mu+p) > 0 \ , \hsp5 (\mu+3p) > 0 \ ,
\ee
(the latter, however, being violated by scalar fields in
inflationary universe models), and additionally demand the {\bf
isentropic speed of sound} $\cs = (\p p/\p\mu)_{s = {\rm
const}}$ obeys
\be
0 \leq \cs \leq 1 \hsp5 \Leftrightarrow \hsp5
0 \leq \left(\frac{\p p}{\p\mu}\right)_{s = {\rm const}} \leq 1 \ , 
\ee
as required for local stability of matter (lower bound) and
causality (upper bound), respectively.

\subsubsection{Maxwell field strength tensor}
The {\bf Maxwell field strength tensor} $F_{ab}$ of an
electromagnetic field is split relative to $u^{a}$ into {\bf
electric} and {\bf magnetic field} parts by the relations (see
\ct{ell73})
\bea
E_a = F_{ab}\,u^b & \Rightarrow & E_{a}u^{a} = 0 \ , \\
H_a = {\sfrac12}\,\eta_{abc}\,F^{bc}
& \Rightarrow & H_{a}u^{a} = 0 \ .
\eea
%

\subsubsection{Weyl curvature tensor}
In analogy to $F_{ab}$, the {\bf Weyl conformal curvature tensor}
$C_{abcd}$ is split relative to $u^a$ into {\bf `electric'} and
{\bf `magnetic' Weyl curvature} parts according to
\bea
E_{ab} = C_{acbd}\,u^c\,u^d
& \Rightarrow &
E^a{}_a = 0 \ , ~E_{ab} = E_{(ab)} \ , ~E_{ab}\,u^b = 0 \ , \\
H_{ab} = {\sfrac12}\,\eta_{ade}\,C^{de}{}_{bc}\,u^{c}
& \Rightarrow &
H^a{}_a = 0 \ , ~H_{ab} = H_{(ab)} \ , ~H_{ab}\,u^b = 0 \ .
\eea
These represent the `free gravitational field', enabling
gravitational action at a distance (tidal forces, gravitational
waves), and influence the motion of matter and radiation through
the {\bf geodesic deviation equation} for timelike and null
congruences, respectively \ct{lev26}--\ct{sze66}. Together with the
{\bf Ricci curvature tensor} $R_{ab}$ (determined locally at
each point by the matter tensor through Einstein's field
equations (\r{eq:efe})), these quantities
completely represent the space-time {\bf Riemann curvature
tensor} $R_{abcd}$, which in fully $1+3$-decomposed form
becomes\footnote{Here $P$ is the perfect fluid part, $I$ the
imperfect fluid part, $E$ that due to the electric Weyl curvature, and
$H$ that due to the magnetic Weyl curvature. This obscures the
similarities in these equations between $E$ and $\pi$, and between
$H$ and $q$; however, this partial symmetry is broken by the field
equations, so the splitting given here (due to M Shedden) is
conceptually useful.}
\bea
\l{riem}
R^{ab}{}_{cd} & = & R_P^{ab}{}_{cd} + R_I^{ab}{}_{cd} 
+ R_E^{ab}{}_{cd} + R_H^{ab}{}_{cd}\, ,\nonumber \\
R_P^{ab}{}_{cd} & = & \sfrac{2}{3}\,(\mu+3p-2\Lambda)\,u^{[a}\,
u_{[c}\, h^{b]}{}_{d]} + \sfrac{2}{3}\,(\mu+\Lambda)\,h^{a}{}_{[c}
\,h^{b}{}_{d]} \, ,\nonumber \\ 
R_I^{ab}{}_{cd} & = & -\,2\,u^{[a}\,h^{b]}{}_{[c}\,q_{d]}
- 2\,u_{[c}\,h^{[a}{}_{d]}\,q^{b]}
- 2\,u^{[a}\,u_{[c}\,\pi^{b]}{}_{d]} 
+ 2\,h^{[a}{}_{[c}\,\pi^{b]}{}_{d]} \,,\\
R_E^{ab}{}_{cd} & = & 4\,u^{[a}\,u_{[c}\,E^{b]}{}_{d]} 
+ 4\,h^{[a}{}_{[c}\,E^{b]}{}_{d]} \,,\nonumber\\
R_H^{ab}{}_{cd} & = & 2\,\eta^{abe}\,u_{[c}\,H_{d]e} 
+ 2\,\eta_{cde}\,u^{[a}\,H^{b]e} \ . \nonumber
\eea
%

\subsubsection{Auxiliary quantities}
It is useful to define some associated kinematical quantities: the
{\bf vorticity vector}
\be
\omega^a = {\sfrac12}\,\eta^{abc}\,\omega_{bc}
\hsp5 \Rightarrow \hsp5 \omega_a\,u^a = 0 \ ,
~\omega_{ab}\,\om^b = 0 \ ,
\ee
the magnitudes
\be
\omega^2 = {\sfrac12}\,(\omega_{ab}\omega^{ab}) \geq 0 \ ,
~~\sigma^2 = {\sfrac12}\,(\sigma_{ab}\sigma^{ab}) \geq 0 \ ,
\ee
and the {\bf average length scale} $S$ determined by

\be
\l{eq:ell}
\frac{\dot{S}}{S} = {\sfrac13}\,\Theta \ ,
\ee
so the volume of a fluid element varies as $S^3$. Further it is
helpful to define particular {\bf spatial gradients} orthogonal
to $u^a$, characterising the inhomogeneity of space-time:
\be
X_a \equiv \3nab_a\mu \ , \hsp5 Z_a \equiv \3nab_a\Theta \ .
\ee
The energy density $\mu$ (and also $\Th$) satisfies the important
{\bf commutation relation} for the $\3nab$-derivative \ct{ebh90}
\be
\l{eq:tilt}
\3nab_{[a}\3nab_{b]}\mu = \eta_{abc}\,\om^{c}\,\dot{\mu} \ .
\ee
This shows that if $\omega^{a}\,\dot{\mu} \neq 0$ in an open set,
then $X_a \neq 0$ there, so non-zero vorticity implies anisotropic
number counts in an expanding universe \ct{god52} (this is because
there are then no 3-surfaces orthogonal to the fluid flow; see
\ct{ell71}).

\subsection{$1+3$ covariant propagation and constraint equations}
\l{subsec:13ceqs}
There are three sets of equations to be considered, resulting from
Einstein's field equations (\r{eq:efe}) and their associated
integrability conditions.

\subsubsection{Ricci identities}
The first set arise from the {\bf Ricci identities} for the vector
field $u^a$, i.e.,
\be
2\,\nabla_{[a}\nabla_{b]}u^{c} = R_{ab}{}^{c}{}_{d}\,u^{d} \ .
\ee
On substituting in from (\r{eq:kin}), using (\r{eq:efe}), and
separating out the orthogonally projected part into trace,
symmetric trace-free, and skew symmetric parts, and the parallel
part similarly, we obtain three propagation equations and three
constraint equations. The {\bf propagation equations} are,

{\bf 1}. The {\bf Raychaudhuri equation} \ct{ray55}
\be
\l{eq:ray}
\dot{\Th} - \3nab_{a}\udot^{a}
= - \,\sfrac{1}{3}\,\Th^{2} + (\udot_{a}\udot^{a})
- 2\,\sig^{2} + 2\,\om^{2} - \sfrac{1}{2}\,(\mu+3p)
+ \Lambda \ ,
\ee
which is the {\it basic equation of gravitational attraction\/}
\ct{ehl61}--\ct{ell73}, showing the repulsive nature of a positive
cosmological constant, leading to identification of $(\mu+3p)$ as
the active gravitational mass density, and underlying the basic
singularity theorem (see below).

{\bf 2}. The {\bf vorticity propagation equation}
\be
\l{eq:omdot}
\dot{\om}^{\lgl a\rgl} - \sfrac{1}{2}\,\eta^{abc}\,
\3nab_{b}\udot_{c}
= - \,\sfrac{2}{3}\,\Th\,\om^{a} + \sig^{a}\!_{b}\,\om^{b} \ ;
\ee
together with (\r{eq:en1}) below, showing how vorticity
conservation follows if there is a perfect fluid with acceleration
potential $\Phi$ \ct{ehl61,ell73}, since then, on using
(\r{eq:tilt}), $\eta^{abc}\,\3nab_{b}\udot_{c} = \eta^{abc}\,
\3nab_{b}\3nab_{c}\Phi = 2\,\om^{a}\,\dot{\Phi}$,

{\bf 3}. The {\bf shear propagation equation}
\bea
\l{eq:sigdot}
\dot{\sig}^{\lgl ab\rgl} - \3nab{}^{\lgl a}\udot^{b\rgl}
= - \,\sfrac{2}{3}\,\Th\,\sig^{ab} + \udot^{\lgl a}\,
\udot^{b\rgl} - \sig^{\lgl a}\!_{c}\,\sig^{b\rgl c}
- \om^{\lgl a}\,\om^{b\rgl} - (E^{ab}-\sfrac{1}{2}\,\pi^{ab}) \ ,
\eea
the anisotropic pressure source term $\pi_{ab}$ vanishing for a
perfect fluid; this shows how the tidal gravitational field, the
electric Weyl curvature $E_{ab}$, directly induces shear (which
then feeds into the Raychaudhuri and vorticity propagation
equations, thereby changing the nature of the fluid flow).\\

The {\bf constraint equations} are,

{\bf 1}. The {\bf $(0\alpha)$-equation}
\be
\l{eq:onu}
0 = (C_{1})^{a} = \3nab_{b}\sig^{ab} - \sfrac{2}{3}\,\3nab^{a}\Th
+ \eta^{abc}\,[\ \3nab_{b}\om_{c} + 2\,\udot_{b}\,\om_{c}\ ]
+ q^{a} \ ,
\ee
showing how the momentum flux (zero for a perfect fluid) relates to
the spatial inhomogeneity of the expansion;

{\bf 2}. The {\bf vorticity divergence identity}
\be
0 = (C_{2}) = \3nab_{a}\om^{a} - (\udot_{a}\om^{a}) \ ;
\ee

{\bf 3}. The {\bf $H_{ab}$-equation}
\be
\l{hconstr}
0 = (C_{3})^{ab} = H^{ab} + 2\,\udot^{\lgl a}\,
\om^{b\rgl} + \3nab^{\lgl a}\om^{b\rgl}
- (\c\,\sig)^{ab} \ ,
\ee
characterising the magnetic Weyl curvature as being constructed from
the `distortion' of the vorticity and the `curl' of the shear,
$(\c\,\sig)^{ab} = \eta^{cd\lgl a}\,\3nab_{c}\sig^{b\rgl}\!_{d}$.

\subsubsection{Twice-contracted Bianchi identities}
The second set of equations arise from the {\bf twice-contracted
Bianchi identities} which, by Einstein's field equations (\r{eq:efe}),
imply the conservation equations (\r{eq:cons}). Projecting parallel
and orthogonal to $u^a$, we obtain the propagation equations
\be
\l{eq:cons1}
\dot{\mu} + \3nab_{a}q^{a} = - \,\Th\,(\mu+p) - 2\,(\udot_{a}q^{a})
- (\sig_{ab}\pi^{ab})
\ee
and
\be
\l{eq:cons2}
\dot{q}^{\lgl a\rgl} + \3nab^{a}p + \3nab_{b}\pi^{ab}
= - \,\sfrac{4}{3}\,\Th\,q^{a} - \sig^{a}\!_{b}\,q^{b}
- (\mu+p)\,\udot^{a} - \udot_{b}\,\pi^{ab}
- \eta^{abc}\,\om_{b}\,q_{c} \ ,
\ee
which constitute the {\bf energy conservation equation}
and the {\bf momentum conservation equation}, respectively.
For perfect fluids, characterised by Eq. (\r{eq:pf}),
these reduce to
\be
\l{eq:en}
\dot{\mu} = -\,\Th\,(\mu+p) \ ,
\ee
and the constraint equation
\be 
\l{eq:en1}
0 = \3nab_{a}p + (\mu+p)\,\udot_{a} \ .
\ee
This shows that $(\mu+p)$ is the inertial mass density, and also governs the
conservation of energy. It is clear that if this quantity is zero
(an effective cosmological constant) or negative, the behaviour of
matter will be anomalous.\\

{\it Exercise}: Examine what happens in the two cases (i) $(\mu+p)
= 0$, (ii) $(\mu+p) < 0$.

\subsubsection{Other Bianchi identities}
The third set of equations arise from the {\bf Bianchi identities}
\be
\nabla_{[a}R_{bc]de} = 0 \ .
\ee
Double contraction gives Eq. (\r{eq:cons}), already considered. On
using the splitting of $R_{abcd}$ into $R_{ab}$ and $C_{abcd}$, the
above $1+3$ splitting of those quantities, and Einstein's field equations,
the {\bf once-contracted Bianchi identities} give two further propagation
equations and two further constraint equations, which are similar
in form to Maxwell's field equations in an expanding universe
(see \ct{haw66,ell73}).

The {\bf propagation equations} are,
\bea
(\dot{E}^{\lgl ab\rgl}+\sfrac{1}{2}\,\dot{\pi}^{\lgl ab\rgl})
- (\c\,H)^{ab} + \sfrac{1}{2}\,\3nab^{\lgl a}q^{b\rgl}
& = & - \,\sfrac{1}{2}\,(\mu+p)\,\sig^{ab}
- \Th\,(E^{ab}+\sfrac{1}{6}\,\pi^{ab}) \\
& & \hsp5 + \ 3\,\sig^{\lgl a}\!_{c}\,(E^{b\rgl c}
-\sfrac{1}{6}\,\pi^{b\rgl c}) - \udot^{\lgl a}\,q^{b\rgl}
\nonumber \\
& & \hsp5 + \ \eta^{cd\lgl a}\,[\ 2\,\udot_{c}\,H_{d}{}^{b\rgl}
+ \om_{c}\,(E_{d}{}^{b\rgl}+\sfrac{1}{2}\,\pi_{d}{}^{b\rgl})\ ] \ ,
\nonumber
\eea
the {\bf $\dot{E}$-equation}, and
\bea
\dot{H}^{\lgl ab\rgl} + (\c\,E)^{ab} -\sfrac12(\c\,\pi)^{ab} 
& = & - \,\Th\,H^{ab} + 3\,\sig^{\lgl a}\!_{c}\,H^{b\rgl c}
+ \sfrac{3}{2}\, \om^{\lgl a}\,q^{b\rgl} \\
& & \hsp5 - \ \eta^{cd\lgl a}\,[\ 2\,\udot_{c}\,E_{d}{}^{b\rgl}
- \sfrac{1}{2}\,\sig^{b\rgl}\!_{c}\,q_{d}
- \om_{c}\,H_{d}{}^{b\rgl}\ ] \ , \nonumber
\eea
the {\bf $\dot{H}$-equation}, where we have defined the `curls'
\bea
(\c\,H)^{ab} & = & \eta^{cd\lgl a}\,\3nab_{c}H_{d}{}^{b\rgl} \ , \\
(\c\,E)^{ab} & = & \eta^{cd\lgl a}\,\3nab_{c}E_{d}{}^{b\rgl} \ , \\
(\c\,\pi)^{ab} & = & \eta^{cd\lgl a}\,\3nab_{c}\pi_{d}{}^{b\rgl} \ .
\eea
These equations show how gravitational radiation arises: taking the
time derivative of the $\dot{E}$-equation gives a term of the form
$(\c\,H){}\dot{}\,$; commuting the derivatives and substituting
from the $\dot{H}$-equation eliminates $H$, and results in a term
in $\ddot{E}$ and a term of the form $(\c\,\c\,E)$, which together
give the wave operator acting on $E$ \ct{haw66,dunbasell};
similarly the time derivative of the $\dot{H}$-equation gives a
wave equation for $H$.

The {\bf constraint equations} are
\bea
\l{eq:divE}
0 & = & (C_{4})^{a} \ = \ \3nab_{b}(E^{ab}+\sfrac{1}{2}\,\pi^{ab})
- \sfrac{1}{3}\,\3nab^{a}\mu + \sfrac{1}{3}\,\Th\,q^{a}
- \sfrac{1}{2}\,\sig^{a}\!_{b}\,q^{b} - 3\,\om_{b}\,H^{ab}
\nonumber \\
& & \hspace{25mm} - \ \eta^{abc}\,[\ \sig_{bd}\,H_{c}{}^{d}
- \sfrac{3}{2}\,\om_{b}\,q_{c}\ ] \ ,
\eea
the {\bf $(\div\,E)$-equation} with source the spatial gradient
of the energy density, which can be regarded as a vector analogue
of the Newtonian Poisson equation \ct{vanell98a}, enabling tidal
action at a distance, and
\bea
\l{eq:divH}
0 & = & (C_{5})^{a} \ = \ \3nab_{b}H^{ab} + (\mu+p)\,\om^{a}
+ 3\,\om_{b}\,(E^{ab}-\sfrac{1}{6}\,\pi^{ab}) \nonumber \\
& & \hspace{25mm} + \ \eta^{abc}\,[\ \sfrac{1}{2}\,\3nab_{b}q_{c}
+ \sig_{bd}\,(E_{c}{}^{d} +\sfrac{1}{2}\,\pi_{c}{}^{d})\ ] \ ,
\eea
the {\bf $(\div\,H)$-equation}, with source the fluid
vorticity. These equations show respectively that scalar modes will
result in a non-zero divergence of $E_{ab}$ (and hence a non-zero
$E$-field), and vector modes in a non-zero divergence of $H_{ab}$
(and hence a non-zero $H$-field).

\subsubsection{Maxwell's field equations}
Finally, we turn for completeness to the $1+3$ decomposition of
{\bf Maxwell's field equations}
\be
\l{eq:max}
\nabla_{b}F^{ab} = j_{\rm e}^{a} \ , \hspace{20mm}
\nabla_{[a}F_{bc]} = 0 \ .
\ee
As shown in \ct{ell73}, the {\bf propagation equations} can be
written as
\bea
\dot{E}{}^{\la a\ra} - \eta^{abc}\,\3nab_{b}H_{c}
& = & -\,j_{\rm e}^{\la a\ra} - \sfrac{2}{3}\,\Th\,E^{a}
+ \sig^{a}{}_{b}\,E^{b} + \eta^{abc}\,[\ \udot_{b}\,H_{c}
+ \om_{b}\,E_{c}\ ] \ , \\
\dot{H}{}^{\la a\ra} + \eta^{abc}\,\3nab_{b}E_{c}
& = & -\,\sfrac{2}{3}\,\Th\,H^{a}
+ \sig^{a}{}_{b}\,H^{b} - \eta^{abc}\,[\ \udot_{b}\,E_{c}
- \om_{b}\,H_{c}\ ] \ ,
\eea
while the {\bf constraint equations} assume the form
\bea
0 & = & (C_{E}) \ = \ \3nab_{a}E^{a} - 2\,(\om_{a}H^{a})
- \rho_{\rm e} \ , \\
0 & = & (C_{H}) \ = \ \3nab_{a}H^{a} + 2\,(\om_{a}E^{a}) \ ,
\eea
where $\rho_{\rm e} = (-j_{{\rm e}\,a}u^{a})$.

\subsection{Pressure-free matter (`dust')}
A particularly useful dynamical restriction is
\be
0 = p = q^a = \pi_{ab} \ ~ \Rightarrow  ~\dot{u}_a = 0\,,
\ee
so the matter (often described as `baryonic') is represented only
by its 4-velocity $u^a$ and its energy density $\mu > 0$.  The
implication follows from momentum conservation: (\r{eq:en1}) shows
that the matter moves geodesically (as expected from the
equivalence principle).  This is the case of {\it pure
gravitation}: it separates out the (non-linear) gravitational
effects from all the fluid dynamical effects. The vanishing of the
acceleration greatly simplifies the above set of equations.

\subsection{Irrotational flow}
If we have a barotropic perfect fluid:
\be
0 = q^{a} = \pi_{ab} \ , \hsp5p = p(\mu) \ ,
\hsp5 \Rightarrow \hsp5 \eta^{abc}\,\3nab_{b}\udot_{c} = 0 \ ,
\ee
then $\om^{a} = 0$ is involutive: i.e.,
$$
\om^a = 0 ~{\rm initially} \hsp5 \Rightarrow \hsp5
\dot{\om}^{\la a\ra} = 0 \hsp5 \Rightarrow \hsp5 \omega^a = 0
~{\rm \, at \,all \,later \,times}
$$
follows from the vorticity conservation equation (\r{eq:omdot})
(and is true also in the special case $p = 0$). When the vorticity
vanishes:

{\bf 1}. The fluid flow is hypersurface-orthogonal, and there
exists a cosmic time function $t$ such that $u_a =
-\,g(x^b)\,\nabla_{a}t$; if additionally the acceleration vanishes,
we can set $g$ to unity;

{\bf 2}. The metric of the orthogonal 3-spaces is $h_{ab}$,

{\bf 3}. From the Gauss embedding equation and the Ricci
identities for $u^a$, the Ricci tensor of these 3-spaces is given
by \ct{ehl61,ell71} 
\bea
\l{eq:3rab}
{}^3\!R_{ab} & = & -\,\dot{\sig}_{\la ab\ra} - \Th\,\sig_{ab}
+ \3nab_{\la a}\udot_{b\ra} + \udot_{\la a}\,\udot_{b\ra}
+ \pi_{ab} 
+ \ \sfrac{1}{3}\,h_{ab}\ [\ 2\,\mu
- \sfrac{2}{3}\,\Th^{2} + 2\,\sig^{2} + 2\,\Lambda\ ] \ ,
\eea
which relates ${}^3\!R_{ab}$ to $E_{ab}$ via (\r{eq:sigdot}), 
and their Ricci scalar is given by
\be
\l{eq:3R}
{}^3\!R = 2\,\mu - \sfrac{2}{3}\,\Th^{2} + 2\,\sig^{2} + 2\,\Lambda
\ ,
\ee
which is a generalised Friedmann equation, showing how the matter
tensor determines the 3-space average curvature. These equations
fully determine the curvature tensor ${}^3\!R_{abcd}$ of the
orthogonal 3-spaces \ct{ell71}. 

\subsection{Implications}
Altogether, in general we have six propagation equations and six
constraint equations; considered as a set of evolution equations
for the $1+3$ covariant variables, they are a first-order system of
equations. This set is determinate once the fluid equations of
state are given; together they then form a complete set of
equations (the system closes up, but is essentially infinite
dimensional because of the spatial derivatives that occur). The
total set is normal hyperbolic at least in the case of a perfect
fluid, although this is not obvious from the above form; it is
shown by completing the equations to tetrad form (see the next
section) and then taking combinations of the equations to give a
symmetric hyperbolic normal form (see \ct{fri98,vanell98c}). We can
determine many of the properties of specific solutions directly
from these equations, once the nature of these solutions has been
prescribed in $1+3$ covariant form (see for example the FLRW and
Bianchi Type I cases considered below).\\

The {\em key issue\/} that arises is consistency of the constraints
with the evolution equations. It is believed that they are {\it
generally consistent} for physically reasonable and well-defined
equations of state, i.e., they are consistent if no restrictions are
placed on their evolution other than implied by the constraint
equations and the equations of state (this has not been proved in
general, but is very plausible; however, it has been shown for
irrotational dust \ct{maa97,vel97}). It is this that makes
consistent the overall hyperbolic nature of the equations with the
`instantaneous' action at a distance implicit in the Gauss-like
equations (specifically, the $(\div\,E)$-equation), the point being
that the `action at a distance' nature of the solutions to these
equations is built into the initial data, which ensures that the
constraints are satisfied initially, and are conserved thereafter
because the time evolution preserves these constraints
(cf. \ct{ellsch72}).  A particular aspect of this is that when
$\omega^a = 0$, the generalised Friedmann equation (\r{eq:3R}) is
an integral of the Raychaudhuri equation (\r{eq:ray}) and energy
equation (\r{eq:en}).\\

One must be very cautious with imposing simplifying assumptions
(such as, e.g., vanishing shear) in order to obtain solutions:
this can lead to major restrictions on the possible flows, and one
can be badly misled if their consistency is not investigated
carefully \ct{ell96,vanell98a}. Cases of particular interest are
shear-free fluid motion (see \ct{ell67}--\ct{sze97}) and various
restrictions on the Weyl curvature tensor, including the `silent
universes', characterised by $H_{ab} = 0$ (and $p = 0$)
\ct{matetal94,hveetal97}, or models with
$\3nab_{b}H^{ab} = 0$ \ct{sop98}.

\subsubsection{Energy equation}
It is worth commenting here that, because of the equivalence
principle, there is no agreed energy conservation equation for the
gravitational field itself, nor is there a definition of its
entropy (indeed some people --- Freeman Dyson, for example
\ct{dys71} --- claim it has no entropy). Thus the above set of
equations does not contain expressions for gravitational
energy\footnote{ There are some proposed `super-energy' tensors,
e.g., the Bel--Robinson tensor \ct{bel}, but they do not play
a significant role in the theory.} or entropy, and the concept of
energy conservation does not play the major role for gravitation
that it does in the rest of physics, neither is there any agreed
view on the growth of entropy of the gravitational
field.\footnote{Entropy is well understood in the case of black
holes, but not for gravitational fields in the expanding Universe.}
However, energy conservation of the matter content of space-time,
expressed by the divergence equation $\nabla_{b}T^{ab} = 0$, is of
course of major importance.\\

If we assume a perfect fluid with a (linear) $\gamma$-law equation
of state, then (\r{eq:en}) shows that
\be
\l{eq:en2}
p = (\gamma-1)\,\mu \ , ~\dot{\gamma} = 0
\hsp5 \Rightarrow \hsp5 \mu = M/S^{3\gamma} \ , 
~\dot{M} = 0 \ .
\ee
One can approximate ordinary fluids in this way with $1\leq
\gamma\leq 2$ in order that the causality and energy conditions
are valid, with `dust' and Cold Dark Matter (`CDM') corresponding
to $\gamma = 1 \Rightarrow \mu = M/S^{3}$, and radiation to $\gamma
= \sfrac{4}{3} \Rightarrow \mu = M/S^{4}$.\\

{\it Exercise}: Show how to generalise this to more realistic
equations of state, taking account of entropy and of matter
pressure (see e.g. \ct{ehl61}--\ct{ell73}).\\

In the case of a mixture of non-interacting matter, radiation and
CDM having the {\em same\/} 4-velocity, represented as a single
perfect fluid, the total energy density is simply the sum of these
components: $\mu = \mu_{\rm dust} + \mu_{\rm CDM} + \mu_{\rm
radn}$. (NB: This is only possible in universes with spatially
homogeneous radiation energy density, because the matter will move
on geodesics which by the momentum conservation equation implies
$\3nab_{a}p_{\rm radn} = 0 \Leftrightarrow \3nab_{a}\mu_{\rm radn}
= 0$.  This will not be true for a general inhomogeneous or
perturbed FLRW model, but will be true in exact FLRW and orthogonal
Bianchi models.)\\

{\it Exercise}: The pressure can still be related to the energy
density by a $\gamma$-law as in (\r{eq:en2}) in this case of
non-interacting matter and radiation, but $\gamma$ will no longer
be constant. What is the equation giving the variation of (i)
$\gamma$, (ii) the speed of sound, with respect to the scale factor
in this case? (See \ct{me88}.) \\

A scalar field has a perfect fluid energy-momentum tensor if the
surfaces $\{\phi = \mbox{const}\}$ are spacelike and we choose
$u^a$ normal to these surfaces. Then it approximates the equation
of state (\r{eq:en2}) in the `slow-rolling' regime, with $\gamma
\approx 0$, and in the velocity-dominated regime, with $\gamma
\approx 2$. In the former case the energy conditions are no longer
valid, so `inflationary' behaviour is possible, which changes the
nature of the attractors in the space of space-times in an
important way.\\

{\it Exercise}: Derive expressions for $\mu$, $p$, $(\mu+p)$,
$(\mu+3p)$ in this case. Under what conditions can a scalar field
have (a) $(\mu+p) = 0$, (b) $(\mu+3p) = 0$, (c) $(\mu+3p) < 0$?

\subsubsection{Basic singularity theorem}
Using the definition (\r{eq:ell}) of $S$, the Raychaudhuri equation
can be rewritten in the form (cf. \ct{ray55})
\be
\l{eq:ray1}
3\,\frac{\ddot{S}}{S} = - \,2\,(\sigma^2-\omega^2)
+ \3nab_{a}\udot^{a} + (\udot_{a}\udot^{a})
- \sfrac{1}{2}\,(\mu+3p) + \Lambda \ ,
\ee
showing how the curvature of the curve $S(\tau)$ along each
worldline (in terms of proper time $\tau$ along that worldline) is
determined by the kinematical quantities, the total energy density
and pressure\footnote{This form of the equation is valid for
imperfect fluids also: the quantities $q^a$ and $\pi_{ab}$ do not
directly enter this equation.} in the combination $(\mu+3p)$, and
the cosmological constant $\Lambda$. This gives the basic
\begin{quotation}
{\bf Singularity Theorem} \ct{ray55,ehl61,ell71,ell73}: In a
universe where $(\mu+3p) >0$, $\Lambda \leq 0$, and $\dot{u}^a =
\om^{a} = 0$ at all times, at any instant when $H_0 =
\sfrac{1}{3}\,\Theta_0 > 0$, there must have been a time $t_0 <
1/H_0$ ago such that $S \rightarrow 0$ as $t
\rightarrow t_0$; a space-time singularity occurs there, where $\mu
\rightarrow \infty$ and $p \rightarrow \infty$ for ordinary matter
(with $(\mu+p) > 0$).
\end{quotation}
The further {\bf singularity theorems} of Hawking and Penrose (see
\ct{hawpen70,he73,tce89}) utilize this result (and its null
version) as an essential part of their proofs.\\

Closely related to this are two other results: the statements that
(a) a static universe model containing ordinary matter requires
$\Lambda > 0$ (Einstein's discovery of 1917), and (b) the Einstein
static universe is unstable (Eddington's discovery of 1930). Proofs
are left to the reader; they follow directly from (\r{eq:ray1}).

\subsubsection{Relations between important parameters}
Given the definitions
\be
H_0 = {\dot{S}_0 \over S_0} \ , 
~~~q_0 = -\,{1 \over H_0^2}{\ddot{S}_0 \over S_0} \ ,
~~~ \Omega_0 = {\mu_0 \over 3 H_0^2} \ , 
~~~w_0 = {p_0\over \mu_0} \ ,
~~~\Omega_\Lambda = {\Lambda \over 3H_0^2} \ ,
\l{eq:def_par}
\ee
for the present-day values of the Hubble scalar (`constant'),
deceleration parameter, density parameter, pressure to density
ratio, and cosmological constant parameter, respectively, then from
(\r{eq:ray1}) we obtain
\be
q_0 = 2\,{(\sigma^2 -\omega^2)_0 \over H_0^2}
- {(\3nab_{a}\udot^{a})_0 + (\udot_{a}\udot^{a})_0 \over 3 H_0^2}
+ \sfrac12\,\Omega_0\,(1+3w_0) - \Omega_\Lambda \ .
\l{eq:ray2}
\ee
Now CBR anisotropies let us deduce that the first two terms on the
right are very small today, certainly less than $10^{-3}$, and we
can reasonably estimate from the nature of the matter that $p_0 \ll
\mu_0$ and the third term on the right is also very small, so we
estimate that in realistic Universe models, at the present time
\be
\l{eq:ray3}
q_0 \approx \sfrac12\,\Omega_0 - \Omega_\Lambda \ .
\ee
(Note we can estimate the magnitudes of the terms which have been
neglected in this approximation.)  This shows that a cosmological
constant can cause an acceleration (negative $q_0$); if it vanishes,
as commonly assumed, the expression simplifies: 
\be
\Lambda = 0 ~~\Rightarrow ~~q_0 \approx \sfrac12\,\Omega_0\,,
\ee
expressing how the matter density present causes a deceleration of
the Universe. If we assume no vorticity ($\omega^a = 0$), then from
(\r{eq:3R}) we can estimate
\be
\l{eq:fried1}
{}^3\!R_0 \approx 6\,H_0^2\,(\,\Omega_0 - 1 + \Omega_\Lambda\,) \ ,
\ee
where we have dropped a term $(\sigma_0/H_0)^2$. If $\Lambda = 0$,
then ${}^3\!R_0 \approx 6\,H_0^2\,(\,\Omega_0 - 1\,)$, showing that
$\Omega_0 = 1$ is the critical value separating irrotational
universes with positive spatial curvature ($\Omega_0 > 1
\Rightarrow {}^3\!R_0>0$) from those with negative spatial
curvature ($\Omega_0 < 1 \Rightarrow {}^3\!R_0<0$).\\

Present day values of these parameters are almost certainly in the
ranges \cite{colell97}: baryon density: $0.01 \leq \Omega_0^{\rm
baryons} \leq 0.03$, total matter density: $0.1 \leq \Omega_0 \leq
0.3 ~\mbox{to}~1$ (implying that much matter may not be baryonic),
Hubble constant: $45\,{\rm km}/{\rm sec}/{\rm Mpc} \leq H_0 \leq
80\,{\rm km}/{\rm sec}/{\rm Mpc}$, deceleration parameter: $-\,0.5
\leq q_0 \leq 0.5$, cosmological constant: $0 \leq \Omega_\Lambda
\leq 1$.

\subsection{Newtonian case}
Newtonian equations can be developed completely in parallel
\ct{ray57,hec61,ell71} and are very similar, but simpler; e.g.,
the Newtonian version of the Raychaudhuri equation is
\be
\dot{\Theta} + \sfrac13\,\Theta^2 + 2\,(\sigma^2 -\omega^2)
- \mbox{D}_{\alpha}a^\alpha + \sfrac12\,\rho - \Lambda = 0 \ ,
\ee
where $\rho$ is the matter density and $a_\alpha = \dot{v}_\alpha +
\mbox{D}_{\alpha}\Phi$ is the Newtonian analogue of the relativistic
`acceleration vector', with $\dot{}$ the convective derivative and
$\Phi$ the Newtonian potential (with suitably generalised boundary
conditions \ct{hs55,hs56}). The Newtonian analogue of $E_{ab}$ is
\be
E_{\alpha\beta} = \mbox{D}_{\alpha}\mbox{D}_{\beta}\Phi
- {\sfrac13}\,(\mbox{D}_{\gam}\mbox{D}^{\gam}\Phi)\,h_{\alpha\beta}
\ , \l{pot}
\ee
where $h_{\alpha\beta}$ denotes the metric, and $\mbox{D}_{\alpha}$
the covariant derivative, of Euclidean space. For the latter
$\mbox{D}_{\alpha}h_{\beta\gam} = 0$ and $[\,\mbox{D}_{\alpha},
\mbox{D}_{\beta}\,] = 0$. There is no analogue of $H_{ab}$ in
Newtonian theory \ct{ell71}, as shown by a strict limit process
leading from relativistic to Newtonian solutions \ct{be96}.\\

{\it Exercise}: Under what conditions will a relativistic
cosmological solution allow a representation (\r{pot}) for the
electric part of the Weyl curvature tensor? Will the potential
$\Phi$ occurring here necessarily also relate to the acceleration
of the timelike reference worldlines?

\subsection{Solutions}
Useful solutions are defined by considering appropriate
restrictions on the kinematical quantities, Weyl curvature tensor,
or space-time geometry, for a specified plausible matter
content. Given such restrictions,

(a) we need to understand the {\bf dynamical evolution} that
results, particularly fixed points, attractors, etc., in terms of
suitable variables,

(b) we particularly seek to determine and characterise {\bf
involutive subsets} of the space of space-times: these are
subspaces mapped into themselves by the dynamical evolution of the
system, and so are left invariant by that evolution. The constraint
and evolution equations must be consistent with each other on such
subsets. A characterisation of these subspaces goes a long way to
characterising the nature of self-consistent solutions of the full
non-linear EFE.

As far as possible we aim to do this for the exact equations. We
are also concerned with

(c) {\bf linearisation} of the equations about known simple
solutions, and determination of properties of the resulting
linearised solutions, in particular considering whether they
accurately represent the behaviour of the full non-linear theory in
a neighbourhood of the background solution (the issue of {\bf
linearisation stability}),

(d) derivation of the {\bf Newtonian limit} and its properties from
the General Relativity equations, and understanding how accurately
this represents the properties of the full relativistic equations
(and of its linearised solutions); see \ct{vanell98a} for a
discussion.

\section{Tetrad description}
\l{sec:onf}
The $1+3$ covariant equations are immediately transparent in terms
of representing relations between $1+3$ covariantly defined
quantities with clear geometrical and/or physical
significance. However, they do {\it not} form a complete set of
equations guaranteeing the existence of a corresponding metric and
connection. For that we need to use a {\bf tetrad description}.
The equations determined will then form a complete set, which will
contain as a subset all the $1+3$ covariant equations just derived
(albeit presented in a slightly different form). For completeness
we will give these equations for a general dissipative relativistic
fluid (recent presentations, giving the following form of the
equations, are \ct{hveugg97,vanell98c}). First we briefly summarize
a generic tetrad formalism, and then its application to
cosmological models (cf. \ct{ell67,mm73}).

\subsection{General tetrad formalism}
A {\bf tetrad} is a set of four mutually orthogonal unit basis
vector fields $\{\,\vec{e}_{a}\,\}_{a = 0,1,2,3}$, which can be
written in terms of a local coordinate basis by means of the
{\bf tetrad components} $e_{a}{}^i(x^j)$:
\be
\vec{e}_a = e_a{}^i(x^j)\,{\p \over \p x^i}
\hsp5 \Leftrightarrow \hsp5
\vec{e}_a(f) = e_a{}^i(x^j)\,{\p f\over \p x^i}\,,\hsp5 
e_{a}{}^{i} = \vec{e}_{a}(x^{i}) \ ,
\ee
(the latter stating that the $i$-th component of the $a$-th tetrad
vector is just the directional derivative of the $i$-th coordinate
in the direction $\vec{e}_{a}$). This can be thought of as just a
general change of vector basis, leading to a change of tensor
components of the standard tensorial form: $T^{ab}{}_{cd} =
e^a{}_i\,e^b{}_j\,e_c{}^k\,e_d{}^l\,T^{ij}{}_{kl}$ with obvious
inverse, where the inverse components $e^a{}_i(x^j)$ (note the
placing of the indices!) are defined by
\be
e_{a}{}^{i}\,e^{a}{}_{j} = \delta^{i}{}_{j} \hsp5 \Leftrightarrow \hsp5
e_{a}{}^{i}\,e^{b}{}_{i} = \delta_{a}{}^{b} \ .
\ee
However, it is a change from an integrable basis to a
non-integrable one, so non-tensorial relations (specifically: the
form of the metric and connection components) are a bit different
than when coordinate bases are used. A change of one tetrad basis
to another will also lead to transformations of the standard tensor
form for all tensorial quantities: if $\vec{e}_{a} =
\Lambda_a{}^{a'}(x^i)\,\vec{e}_{a'}$ is a change of tetrad basis
with inverse $\vec{e}_{a'} = \Lambda_{a'}{}^{a}(x^i)\,\vec{e}_{a}$
(each of these matrices representing a Lorentz transformation),
then $T^{ab}{}_{cd} = \Lambda_{a'}{}^{a}\,\Lambda_{b'}{}^{b}\,
\Lambda_{c}{}^{c'}\,\Lambda_{d}{}^{d'}\,T^{a'b'}{}_{c'd'}$.
Again, the inverse is obvious.\footnote{The tetrad components of any
quantity are invariant when the coordinate basis is changed (for a
fixed tetrad), and coordinate components are invariant when a
change of tetrad basis is made (for a fixed set of coordinates);
however, either change will alter the tetrad components relative to
the given coordinates.}\\
 
The components of the {\bf metric} in the tetrad form are given by
\be
g_{ab} = g_{ij}\,e_a{}^i\,e_b{}^j = \vec{e}_{a}\cdot\,\vec{e}_{b}
= \eta_{ab} \ ,
\ee
where $\eta_{ab} = \mbox{diag}\,(\,-\,1, \,+\,1, \,+\,1,
\,+\,1\,)$, showing that the basis vectors  are unit vectors
mutually orthogonal to each other (because the components 
$g_{ab}$ are just the scalar products of these vectors with each
other). The inverse equation
\be
g_{ij}(x^k) = \eta_{ab}\,e^a{}_i(x^k)\,e^b{}_j(x^k)
\ee
explicitly constructs the coordinate components of the metric from
the (inverse) tetrad components $e^a{}_i(x^j)$. We can raise and
lower tetrad indices by use of the metric $g_{ab} = \eta_{ab}$ and
its inverse $g^{ab} = \eta^{ab}$.\\

The {\bf commutation functions} related to the tetrad are the
quantities $\gamma^a{}_{bc}(x^i)$ defined by the {\bf
commutators} of the basis vectors:\footnote{Remember that the
commutator of any two vectors $X$, $Y$ is $[X,Y] = XY - YX$.}
\be
\l{eq:55}
[\,\vec{e}_{a},\,\vec{e}_{b}\,] = \gamma^c{}_{ab}(x^i)\,\vec{e}_{c}
\hsp5 \Rightarrow \hsp5
\gamma^a{}_{bc}(x^i) = - \,\gamma^a{}_{cb}(x^i) \ .
\ee
It follows (apply this relation to the local coordinate $x^i$)
that in terms of the tetrad components,
\be
\gamma^a{}_{bc}(x^i) = e^a{}_i\,(\,e_b{}^j\,\p_je_c{}^i
- e_c{}^j\,\p_je_b{}^i\,)
= -\,2\,e_b{}^i\,e_c{}^j\,\nabla_{[i}e^a{}_{j]} \ .
\ee
These quantities vanish iff the basis $\{\,\vec{e}_{a}\,\}$ is a
coordinate basis: that is, there exist local coordinates $x^{i}$
such that $\vec{e}_{a} = \delta_{a}{}^{i}\,\p/\p x^{i}$, iff
$[\,\vec{e}_{a},\,\vec{e}_{b}\,] = 0 \Leftrightarrow
\gamma^a{}_{bc} = 0$.\\

The {\bf connection components} $\Gamma^a{}_{bc}$ for the tetrad
(`Ricci rotation coefficients') are defined by the relations
\be
\nabla_{\vec{e}_{b}}\vec{e}_{a} = \Gamma^c{}_{ab}\,\vec{e}_{c}
\hsp5 \Leftrightarrow \hsp5
\Gamma^c{}_{ab} = e^c{}_i\,e_b{}^j\,\nabla_{j}e_a{}^i{} \ ,
\ee
i.e., it is the $c$-component of the covariant derivative in the
$b$-direction of the $a$-vector. It follows that all covariant
derivatives can be written out in tetrad components in a way
completely analogous to the usual tensor form, for example
$\nabla_{a}T_{bc} = \vec{e}_{a}(T_{bc}) - \Gamma^d{}_{ba}\,T_{dc}
- \Gamma^d{}_{ca}\,T_{bd}$, where for any function $f$,
$\vec{e}_{a}(f) = e_{a}{}^{i}\,\p f/\p x^i$ is the derivative
of $f$ in the direction $\vec{e}_{a}$. In particular, because
$\vec{e}_{a}(g_{bc}) = 0$ for $g_{ab} = \eta_{ab}$, applying
this to the metric gives
\be
\nabla_{a}g_{bc} = 0 \hsp5 \Leftrightarrow \hsp5
-\,\Gamma^{d}{}_{ba}\,g_{dc} - \Gamma^{d}{}_{ca}\,g_{bd} = 0
\hsp5 \Leftrightarrow \hsp5
\Gamma_{(ab)c} = 0 \ ,
\ee
--- the rotation coefficients are skew in their first two indices,
when we raise and lower the first indices only. We obtain from this
and the assumption of vanishing torsion the tetrad relations that
are the analogue of the usual Christoffel relations:
\be
\gam^{a}{}_{bc} = -\,(\Gamma^a{}_{bc} - \Gamma^a{}_{cb}) \ , \hsp5
\Gamma_{abc} = \sfrac12\,(\,g_{ad}\,\gamma^{d}{}_{cb}
- g_{bd}\,\gamma^{d}{}_{ca} + g_{cd}\,\gamma^{d}{}_{ab}\,) \ .
\ee
This shows that the rotation coefficients and the commutation
functions are each just linear combinations of the other.\\

Any set of vectors whatever must satisfy the {\bf Jacobi
identities}:
$$[\,X,\,[\,Y,\,Z\,]\,] + [\,Y,\,[\,Z,\,X]\,] +
[\,Z,\,[\,X,\,Y\,]\,] = 0 \ ,
$$
which follow from the definition of a commutator. Applying this to
the basis vectors $\vec{e}_{a}$, $\vec{e}_{b}$ and $\vec{e}_{c}$ gives
the identities
\be
\l{eq:jac}
\vec{e}_{[a}(\gamma^d{}_{bc]}) + \gamma^e{}_{[ab}\,\gamma^d{}_{c]e} = 0
\ ,
\ee
which are the integrability conditions that the
$\gamma^a{}_{bc}(x^i)$ are the commutation functions for the set of
vectors $\vec{e}_{a}$.\\

If we apply the Ricci identities to the tetrad basis vectors
$\vec{e}_{a}$, we obtain the Riemann curvature tensor components in the
form
\be
R^a{}_{bcd} = \vec{e}_{c}(\Gamma^a{}_{bd}) - \vec{e}_{d}(\Gamma^a{}_{bc})
+ \Gamma^a{}_{ec}\,\Gamma^e{}_{bd}
- \Gamma^a{}_{ed}\,\Gamma^e{}_{bc}
- \Gamma^a{}_{be}\,\gamma^e{}_{cd} \ . 
\ee
Contracting this on $a$ and $c$, one obtains Einstein's field equations
in the form
\be
R_{bd} = \vec{e}_{a}(\Gamma^a{}_{bd}) - \vec{e}_{d}(\Gamma^a{}_{ba})
+ \Gamma^a{}_{ea}\,\Gamma^e{}_{bd}
- \Gamma^a{}_{de}\,\Gamma^e{}_{ba}  
= T_{bd} - \sfrac12\,T\,g_{bd} + \Lambda\,g_{bd} \ . 
\ee
It is not immediately obvious that this is symmetric, but this
follows because (\r{eq:jac}) implies $R_{a[bcd]} = 0 \Rightarrow
R_{ab} = R_{(ab)}$.

\subsection{Tetrad formalism in cosmology}
\l{subsec:13teqs}
For a cosmological model we choose $\vec{e}_{0}$ to be the
future-directed unit tangent of the matter flow, $u^{a}$. This
fixing implies that the initial six-parameter freedom of using
Lorentz transformations has been reduced to a three-parameter
freedom of rotations of the spatial frame
$\{\,\vec{e}_{\alpha}\,\}$. The 24 algebraically
independent frame components of the space-time connection
$\Gamma^{a}{}_{bc}$ can then be split into the set (see
\ct{ell67,ellmac69,vanell98c})
\bea
\l{onfgam1}
\Gamma_{\alpha 00} & = & \udot_{\alpha} \\
\Gamma_{\alpha 0\beta} & = & \sfrac{1}{3}\,\Th\,\delta_{\alpha\beta}
+ \sig_{\alpha\beta} - \eps_{\alpha\beta\gam}\,\om^{\gam} \\ 
\Gamma_{\alpha\beta 0} & = & \eps_{\alpha\beta\gam}\,\Omega^{\gam} \\
\l{onfgam4}
\Gamma_{\alpha\beta\gam} & = & 2\,a_{[\alpha}\,\delta_{\beta]\gam}
+ \eps_{\gam\delta[\alpha}\,n_{\beta]}{}^{\delta}
+ \sfrac{1}{2}\,\eps_{\alpha\beta\delta}\,n_{\gam}{}^{\delta} \ .
\eea
The first two sets contain the kinematical variables. In the third
is the rate of rotation $\Omega^{\alpha}$ of the spatial frame
$\{\,\vec{e}_{\alpha}\,\}$ with respect to a {\bf
Fermi-propagated basis}. Finally, the quantities $a^{\alpha}$ and
$n_{\alpha\beta} = n_{(\alpha\beta)}$ determine the 9 spatial
rotation coefficients. The commutator equations (\r{eq:55}) applied
to any space-time scalar $f$ take the form
\bea
\l{onfcomts}
[\,\vec{e}_{0}, \vec{e}_{\alpha}\,]\,(f) & = &
\udot_{\alpha}\ \vec{e}_{0}(f) - [\ \sfrac{1}{3}\,\Th\,
\del_{\alpha}{}^{\beta} + \sig_{\alpha}{}^{\beta}
+ \eps_{\alpha}{}^{\beta}{}_{\gam}\,(\om^{\gam}
+\Omega^{\gam})\ ]\ \vec{e}_{\beta}(f)
\\ \nonumber \\
\l{onfcomss}
[\,\vec{e}_{\alpha}, \vec{e}_{\beta}\,]\,(f) & = &
2\,\eps_{\alpha\beta\gam}\,\om^{\gam}\ \vec{e}_{0}(f)
+ [\ 2\,a_{[\alpha}\,\delta_{\beta]}{}^{\gam}
+ \eps_{\alpha\beta\delta}\,n^{\delta\gam}\ ]\ \vec{e}_{\gam}(f) \ ;
\eea
%

The full set of equations for a gravitating fluid can be written as
a set of constraints and a set of evolution equations, which
include the tetrad form of the $1+3$ covariant equations given
above, but complete them by giving all Ricci and Jacobi identities
for the basis vectors. We now give these equations.

\subsubsection{Constraints}
The following set of relations does not contain any frame
derivatives with respect to $\vec{e}_{0}$. Hence, we refer to these
relations as `constraints'. From the Ricci identities for $u^{a}$
and the Jacobi identities we have the $(0\alpha)$-equation
$(C_{1})^{\alpha}$, which, in Hamiltonian treatments of the EFE, is
also referred to as the `momentum constraint', the vorticity
divergence identity $(C_{2})$ and the $H_{ab}$-equation
$(C_{3})^{\alpha\beta}$, respectively; the once-contracted Bianchi
identities yield the $(\div\,E)$- and $(\div\,H)$-equations
$(C_{4})^{\alpha}$ and $(C_{5})^{\alpha}$ \ct{ell71,hveugg97}; the
constraint $(C_{J})^{\alpha}$ again arises from the Jacobi
identities while, finally, $(C_{G})^{\alpha\beta}$ and $(C_{G})$
stem from the EFE. In detail,
\pagebreak
\bea
\l{onfdivsig}
0 & = & (C_{1})^{\alpha} \ = \ (\vec{e}_{\beta} - 3\,a_{\beta})\,
(\sig^{\alpha\beta}) - \sfrac{2}{3}\,\delta^{\alpha\beta}\,
\vec{e}_{\beta}(\Th) - n^{\alpha}\!_{\beta}\,\om^{\beta} + q^{\alpha}
\nonumber \\
& & \hspace{25mm} + \ \epsilon^{\alpha\beta\gam}\,
[\ (\vec{e}_{\beta} + 2\,\udot_{\beta} - a_{\beta})\,
(\om_{\gam}) - n_{\beta\delta}\,\sig_{\gam}{}^{\delta}\ ]
\\ \nonumber \\
\l{onfdivom}
0 & = & (C_{2}) \ = \ (\vec{e}_{\alpha} - \udot_{\alpha}
- 2\,a_{\alpha})\,(\om^{\alpha})
\\ \nonumber \\
\l{onfhconstr}
0 & = & (C_{3})^{\alpha\beta} \ = \ H^{\alpha\beta}
+ (\delta^{\gam\lgl\alpha}\,\vec{e}_{\gam} + 2\,\udot^{\lgl\alpha}
+ a^{\lgl\alpha})\,(\om^{\beta\rgl}) - \sfrac{1}{2}\,
n_{\gam}{}^{\gam}\,\sig^{\alpha\beta}
+ 3\,n^{\lgl\alpha}\!_{\gam}\,\sig^{\beta\rgl\gam} \nonumber \\
& & \hspace{25mm} - \ \eps^{\gam\delta\lgl\alpha}\,[\ (\vec{e}_{\gam}
- a_{\gam})\,(\sig_{\delta}{}^{\beta\rgl})
+ n_{\gam}{}^{\beta\rgl}\,\om_{\delta}\ ]
\\ \nonumber \\
\l{onfdive}
0 & = & (C_{4})^{\alpha} \ = \ (\vec{e}_{\beta} - 3\,a_{\beta})
\,(E^{\alpha\beta}+\sfrac{1}{2}\,\pi^{\alpha\beta})
- \sfrac{1}{3}\,\delta^{\alpha\beta}\,\vec{e}_{\beta}(\mu)
+ \sfrac{1}{3}\,\Th\,q^{\alpha}
- \sfrac{1}{2}\,\sig^{\alpha}\!_{\beta}\,q^{\beta}
- 3\,\om_{\beta}\,H^{\alpha\beta} \nonumber \\
& & \hspace{25mm} - \ \eps^{\alpha\beta\gam}\,[\ \sig_{\beta\delta}
\,H_{\gam}{}^{\delta} - \sfrac{3}{2}\,\om_{\beta}
\,q_{\gam} + n_{\beta\delta}\,(E_{\gam}{}^{\delta}
+\sfrac{1}{2}\,\pi_{\gam}{}^{\delta})\ ]
\\ \nonumber \\
\l{onfdivh}
0 & = & (C_{5})^{\alpha} \ = \ (\vec{e}_{\beta} - 3\,a_{\beta})\,
(H^{\alpha\beta}) + (\mu+p)\,\om^{\alpha}
+ 3\,\om_{\beta}\,(E^{\alpha\beta}
-\sfrac{1}{6}\,\pi^{\alpha\beta}) - \sfrac{1}{2}\,
n^{\alpha}\!_{\beta}\,q^{\beta} \nonumber \\
& & \hspace{25mm} + \ \eps^{\alpha\beta\gam}\,[\ \sfrac{1}{2}
\,(\vec{e}_{\beta} - a_{\beta})\,(q_{\gam})
+ \sig_{\beta\delta}\,(E_{\gam}{}^{\delta}+\sfrac{1}{2}\,
\pi_{\gam}{}^{\delta}) - n_{\beta\delta}\,H_{\gam}{}^{\delta}\ ]
\\ \nonumber \\
\l{onfjac}
0 & = & (C_{J})^{\alpha} \ = \ (\vec{e}_{\beta} - 2\,a_{\beta})\,
(n^{\alpha\beta}) + \sfrac{2}{3}\,\Th\,\om^{\alpha}
+ 2\,\sig^{\alpha}\!_{\beta}\,\om^{\beta}
+ \eps^{\alpha\beta\gam}\,[\ \vec{e}_{\beta}(a_{\gam})
- 2\,\om_{\beta}\,\Omega_{\gam}\ ]
\\ \nonumber \\
\l{onfgauss}
0 & = & (C_{G})^{\alpha\beta} \ = \ {}^{*}\!S^{\alpha\beta}
+ \sfrac{1}{3}\,\Th\,\sig^{\alpha\beta}
- \sig^{\lgl\alpha}\!_{\gam}\,\sigma^{\beta\rgl\gam}
- \om^{\lgl\alpha}\,\om^{\beta\rgl} + 2\,\om^{\lgl\alpha}\,
\Omega^{\beta\rgl} - (E^{\alpha\beta}+\sfrac{1}{2}\,
\pi^{\alpha\beta})
\\ \nonumber \\
\l{onffried}
0 & = & (C_{G}) \ = \ {}^{*}\!R + \sfrac{2}{3}\,\Th^{2}
- (\sig_{\alpha\beta}\sig^{\alpha\beta})
+ 2\,(\om_{\alpha}\om^{\alpha}) - 4\,(\om_{\alpha}\Omega^{\alpha})
- 2\,\mu - 2\,\Lambda \ ,
\eea
where
\bea
\l{onftf3ric}
{}^{*}\!S_{\alpha\beta} & = & \fd_{\lgl\alpha}(a_{\beta\rgl})
+ b_{\lgl\alpha\beta\rgl} - \eps^{\gam\delta}{}_{\lgl\alpha}\,
(\fd_{|\gam|} - 2\,a_{|\gam|})\,(n_{\beta\rgl\delta})
\\ \nonumber \\
\l{onf3rscl}
{}^{*}\!R & = &  2\,(2\,\fd_{\alpha} - 3\,a_{\alpha})\,
(a^{\alpha}) - \sfrac{1}{2}\,b_{\alpha}{}^{\alpha}
\\ \nonumber \\
b_{\alpha\beta} & = & 2\,n_{\alpha\gam}\,n_{\beta}{}^{\gam}
- n_{\gam}{}^{\gam}\,n_{\alpha\beta} \ .
\eea
If $\om^{\alpha} = 0$, so that $u^{a}$ become the normals to a
family of 3-spaces of constant time, the last two constraints in
the set correspond to the symmetric trace-free and trace parts of
the Gauss embedding equation (\r{eq:3rab}). In this case, one
also speaks of $(C_{G})$ as the generalised Friedmann equation,
alias the `Hamiltonian constraint' or the `energy constraint'.

\subsubsection{Evolution of spatial commutation functions}
The 9 spatial commutation functions $a^{\alpha}$ and
$n_{\alpha\beta}$ are generally evolved by equations (40) and (41)
given in \ct{hveugg97}; these originate from the Jacobi
identities. Employing each of the constraints $(C_{1})^{\alpha}$ to
$(C_{3})^{\alpha\beta}$ listed in the previous paragraph, we can
eliminate $\vec{e}_{\alpha}$ frame derivatives of the kinematical
variables $\Th$, $\sig_{\alpha\beta}$ and $\om^{\alpha}$ from their
right-hand sides. Thus, we obtain the following equations for the
evolution of the spatial connection components:
\bea
\l{onfadot}
\vec{e}_{0}(a^{\alpha}) & = & - \,\sfrac{1}{3}\,(\Th\,
\delta^{\alpha}\!_{\beta} - \sfrac{3}{2}\,
\sig^{\alpha}\!_{\beta})\,(\udot^{\beta} + a^{\beta})
+ \sfrac{1}{2}\,n^{\alpha}\!_{\beta}\,\om^{\beta}
- \sfrac{1}{2}\,q^{\alpha} \nonumber \\
& & \hsp5 - \ \sfrac{1}{2}\,\eps^{\alpha\beta\gam}\,
[\ (\udot_{\beta} + a_{\beta})\,\om_{\gam} - n_{\beta\delta}\,
\sig_{\gam}{}^{\delta} - (\vec{e}_{\beta}+\udot_{\beta}
-2\,a_{\beta})\,(\Omega_{\gam})\ ]
+ \sfrac{1}{2}\,(C_{1})^{\alpha}
\\ \nonumber \\
\l{onfndot}
\vec{e}_{0}(n^{\alpha\beta}) & = & - \,\sfrac{1}{3}\,\Th\,
n^{\alpha\beta} - \sig^{\lgl\alpha}\!_{\gam}\,
n^{\beta\rgl\gam} + \sfrac{1}{2}\,\sig^{\alpha\beta}\,
n_{\gam}{}^{\gam} - (\udot^{\lgl\alpha} + a^{\lgl\alpha})\,
\om^{\beta\rgl} - H^{\alpha\beta} + (\delta^{\gam\lgl\alpha}\,
\vec{e}_{\gam} + \udot^{\lgl\alpha})\,(\Omega^{\beta\rgl})
\nonumber \\
& & \hsp5 - \ \sfrac{2}{3}\,\delta^{\alpha\beta}\,
[\ 2\,(\udot_{\gam} + a_{\gam})\,\om^{\gam}
- (\sig_{\gam\delta}n^{\gam\delta})
+ (\vec{e}_{\gam} + \udot_{\gam})\,(\Omega^{\gam})\ ]
\nonumber \\
& & \hsp5 - \ \eps^{\gam\delta\lgl\alpha}\,
[\ (\udot_{\gam} + a_{\gam})\,\sig_{\delta}{}^{\beta\rgl}
- (\om_{\gam}+ 2\,\Omega_{\gam})\,n_{\delta}{}^{\beta\rgl}\ ]
- \sfrac{2}{3}\,\delta^{\alpha\beta}\,(C_{2}) + (C_{3})^{\alpha\beta}
\ .
\eea
%

\subsubsection{Evolution of kinematical variables}
The evolution equations for the 9 kinematical variables $\Th$,
$\om^{\alpha}$ and $\sig_{\alpha\beta}$ are provided by the Ricci
identities for $u^{a}$, i.e.,
\bea
\l{onfthdot}
\vec{e}_{0}(\Th) - \vec{e}_{\alpha}(\udot^{\alpha})
& = & -\,\sfrac{1}{3}\,\Th^{2}
+ (\udot_{\alpha} - 2\,a_{\alpha})\,\udot^{\alpha}
- (\sig_{\alpha\beta}\sig^{\alpha\beta})
+ 2\,(\om_{\alpha}\om^{\alpha}) \nonumber \\
& & \hsp5 - \ \sfrac{1}{2}\,(\mu+3p) + \Lambda
\\ \nonumber \\
\l{onfomdot}
\vec{e}_{0}(\om^{\alpha}) - \sfrac{1}{2}\,\eps^{\alpha\beta\gam}\,
\vec{e}_{\beta}(\udot_{\gam})
& = & - \,\sfrac{2}{3}\,\Th\,\om^{\alpha} + \sig^{\alpha}\!_{\beta}
\,\om^{\beta} - \sfrac{1}{2}\,n^{\alpha}\!_{\beta}\,\udot^{\beta}
- \sfrac{1}{2}\,\eps^{\alpha\beta\gam}\,[\ a_{\beta}\,
\udot_{\gam} - 2\,\Omega_{\beta}\,\om_{\gam}\ ]
\\ \nonumber \\
\l{onfsigdot}
\vec{e}_{0}(\sig^{\alpha\beta}) - \delta^{\gam\lgl\alpha}\,
\vec{e}_{\gam}(\udot^{\beta\rgl})
& = & - \,\sfrac{2}{3}\,\Th\,\sig^{\alpha\beta}
+ (\udot^{\lgl\alpha} + a^{\lgl\alpha})\,\udot^{\beta\rgl}
- \sig^{\lgl\alpha}\!_{\gam}\,\sig^{\beta\rgl\gam}
- \om^{\lgl\alpha}\,\om^{\beta\rgl} \nonumber \\
& & \hsp5 - \ (E^{\alpha\beta}-\sfrac{1}{2}\,\pi^{\alpha\beta})
+ \eps^{\gam\delta\lgl\alpha}\,[\ 2\,\Omega_{\gam}\,
\sig_{\delta}{}^{\beta\rgl} - n_{\gam}{}^{\beta\rgl}\,
\udot_{\delta}\ ] \ .
\eea
%

\subsubsection{Evolution of matter and Weyl curvature variables}
Finally, we have the equations for the 4 matter variables $\mu$ and
$q^{\alpha}$ and the 10 Weyl curvature variables $E_{\alpha\beta}$
and $H_{\alpha\beta}$, which are obtained from the twice-contracted
and once-contracted Bianchi identities, respectively:
\bea
\l{onfmudot}
\vec{e}_{0}(\mu) + \vec{e}_{\alpha}(q^{\alpha})
& = & - \,\Th\,(\mu+p)
- 2\,(\udot_{\alpha} - a_{\alpha})\,q^{\alpha}
- (\sig_{\alpha\beta}\pi^{\alpha\beta}) \
\\ \nonumber \\
\l{onfqdot}
\vec{e}_{0}(q^{\alpha}) + \delta^{\alpha\beta}\,\vec{e}_{\beta}(p)
+ \vec{e}_{\beta}(\pi^{\alpha\beta})
& = & - \sfrac{4}{3}\,\Th\,q^{\alpha} - \sig^{\alpha}\!_{\beta}\,
q^{\beta} - (\mu+p)\,\udot^{\alpha} - (\udot_{\beta}
- 3\,a_{\beta})\,\pi^{\alpha\beta} \nonumber \\
& & \hsp5 - \ \eps^{\alpha\beta\gam}\,[\ (\om_{\beta}
-\Omega_{\beta})\,q_{\gam} - n_{\beta\delta}\,\pi_{\gam}{}^{\delta}
\ ]
\\ \nonumber \\
\l{onfedot}
\vec{e}_{0}(E^{\alpha\beta}+\sfrac{1}{2}\,\pi^{\alpha\beta})
- \eps^{\gam\delta\lgl\alpha}\,\vec{e}_{\gam}(H_{\delta}{}^{\beta\rgl})
& + &\ \sfrac{1}{2}\,\delta^{\gam\lgl\alpha}\,\vec{e}_{\gamma}
(q^{\beta\rgl})
= - \,\sfrac{1}{2}\,(\mu+p)\,\sig^{\alpha\beta} \nonumber \\
& & \hsp5 - \,\Th\,(E^{\alpha\beta}+\sfrac{1}{6}\,
\pi^{\alpha\beta})
+ 3\,\sigma^{\lgl\alpha}\!_{\gam}\,(E^{\beta\rgl\gam}
-\sfrac{1}{6}\,\pi^{\beta\rgl\gam}) \nonumber \\
& & \hsp5 + \ \sfrac{1}{2}\,n_{\gam}{}^{\gam}\,H^{\alpha\beta}
- 3\,n^{\lgl\alpha}\!_{\gam}\,H^{\beta\rgl\gam}
- \sfrac{1}{2}\,(2\,\udot^{\lgl\alpha}
+ a^{\lgl\alpha})\,q^{\beta\rgl} \nonumber \\
& & \hsp5 + \ \eps^{\gam\delta\lgl\alpha}\,[\ (2\,\udot_{\gam}
- a_{\gam})\,H_{\delta}{}^{\beta\rgl} \\
& & \hspace{20mm} + \ (\om_{\gam}+2\,\Omega_{\gam})\,
(E_{\delta}{}^{\beta\rgl}+\sfrac{1}{2}\,\pi_{\delta}{}^{\beta\rgl})
+ \sfrac{1}{2}\,n_{\gam}{}^{\beta\rgl}\,q_{\delta}\ ] \nonumber
\\ \nonumber \\
\l{onfhdot}
\vec{e}_{0}(H^{\alpha\beta}) + \eps^{\gam\delta\lgl\alpha}\,
\vec{e}_{\gam}(E_{\delta}{}^{\beta\rgl}-\sfrac{1}{2}\,
\pi_{\delta}{}^{\beta\rgl})
& = & - \,\Th\,H^{\alpha\beta} + 3\,\sig^{\lgl\alpha}\!_{\gam}\,
H^{\beta\rgl\gam} + \sfrac{3}{2}\,\om^{\lgl\alpha}\,q^{\beta\rgl}
\nonumber \\
& & \hsp5 - \ \sfrac{1}{2}\,n_{\gam}{}^{\gam}\,(E^{\alpha\beta}
-\sfrac{1}{2}\,\pi^{\alpha\beta}) + 3\,n^{\lgl\alpha}\!_{\gam}\,
(E^{\beta\rgl\gam}-\sfrac{1}{2}\,\pi^{\beta\rgl\gam}) \nonumber \\
& & \hsp5 + \ \eps^{\gam\delta\lgl\alpha}\,[\ a_{\gam}\,
(E_{\delta}{}^{\beta\rgl}-\sfrac{1}{2}\,\pi_{\delta}{}^{\beta\rgl})
- 2\,\udot_{\gam}\,E_{\delta}{}^{\beta\rgl} \\
& & \hspace{20mm} + \sfrac{1}{2}\,\sig_{\gam}{}^{\beta\rgl}
\,q_{\delta} + (\om_{\gam}+2\,\Omega_{\gam})\,
H_{\delta}{}^{\beta\rgl}\ ] \ . \nonumber
\eea
%

{\it Exercise}: (a) Show how most of these equations are the tetrad
version of corresponding $1+3$ covariant equations. For which of
the tetrad equations is this not true? (b) Explain why there are no
equations for $\vec{e}_{0}(\Omega^{\alpha})$ and
$\vec{e}_{0}(\dot{u}^{\alpha})$. [Hint: What freedom is there in
choosing the tetrad?]

\subsection{Complete set}
For a prescribed set of matter equations of state, this gives the
complete set of tetrad relations, which can be used to characterise
particular families of solutions in detail. It clearly contains all
the $1+3$ covariant equations above, plus others required to form a
complete set. It can be recast into a symmetric hyperbolic
form \ct{vanell98c} (at least for perfect fluids), showing the
hyperbolic nature of the equations and determining their
characteristics. Detailed studies of exact
solutions will need a coordinate system and vector basis, and
usually it will be advantageous to use tetrads for this purpose,
because the tetrad vectors can be chosen in physically preferred
directions (see \ct{ell67,steell68} for the use of tetrads to study
locally rotationally symmetric space-times, and
\ct{ellmac69,waiell97} for Bianchi universes; these cases are both
discussed below).\\

Finally it is important to note that when tetrad vectors are chosen
uniquely in an invariant way (e.g., as eigenvectors of a
non-degenerate shear tensor or of the electric Weyl curvature
tensor), then --- because they are uniquely defined from $1+3$
covariant quantities --- all the rotation coefficients above are
in fact covariantly defined scalars, so all these equations are
invariant equations. The only times when it is not possible to
define unique tetrads in this way is when the space-times are
isotropic or locally rotationally symmetric, as discussed below.

\section{FLRW models and observational relations}
A particularly important involutive subspace of the space of
cosmological space-times is that of the
{\bf Friedmann--Lema\^{\i}tre} (`FL') {\bf models}, based on the
everywhere-isotropic {\bf Robertson--Walker} (`RW') {\bf geometry}.
It is characterised by a {\bf perfect fluid} matter tensor and the
condition that {\bf local isotropy} holds everywhere:
\be
\l{eq:frw}
0 = \udot^{a} = \sigma_{ab} = \om^{a} \hsp5 \Leftrightarrow \hsp5
0 = E_{ab} = H_{ab} \hsp5 \Rightarrow \hsp5
0 = X_{a} = Z_{a} = \3nab_{a}p \ ,
\ee
the first conditions stating the kinematical quantities are locally
isotropic, the second that these models are conformally flat,
and the third that they are spatially homogeneous. \\

{\it Exercise}: Show that the implications in this relation follow
from the $1+3$ covariant equations in the previous section when $p
= p(\mu)$, thus showing that isotropy everywhere implies spatial
homogeneity in this case.

\subsection{Coordinates and metric}
It follows then that (see \ct{ell87}):

{\bf 1}. {\bf Matter-comoving local coordinates} can be
found\footnote{There are many
other coordinate systems in use, for example with different
definitions of the radial distance $r$.} so that the metric takes
the form
\be
\l{eq:frw1}
ds^2 = -\,dt^2 + S^2(t)\,(\,dr^2 + f^2(r)\,d\Omega^2\,) \ ,
\hsp5 u^a = \delta^a{}_0 \ ,
\ee
where $d\Omega^2 = d\theta^2 + \sin^2\theta\,d\phi^2$, $u_a =
-\,\nabla_{a}t$, and $\dot{S}/S = \sfrac13\,\Theta$, characterising
$S(t)$ as the {\bf scale factor} for distances between any pair of
fundamental observers. The expansion of matter depends only on one
scale length, so it is isotropic (there is no distortion or
rotation).

{\bf 2}. The Ricci tensor ${}^3\!R_{ab}$ is isotropic, so the
3-spaces $\{t = \mbox{const}\}$ are 3-spaces of {\bf constant
(scalar) curvature} $6k/S^2$ where $k$ can be normalised to
$\pm\,1$, if it is non-zero. Using the geodesic deviation
equation in these 3-spaces, one finds that
(see \ct{ell87,hveell98b})
\be
\l{eq:fr}
f(r) = \sin r \ , ~r \ , ~\sinh r \hsp5 \mbox{if} \hsp5
k = +\,1 \ , ~0 \ , ~-\,1 \ .
\ee
Thus, when $k = +\,1$, the {\bf surface area}
$4\pi\,S^2(t)\,f^2(r)$ of a
geodesic 2-sphere in these spaces, centred on the (arbitrary) point
$r = 0$, increases to a maximum at $r = \pi/2$ and then decreases
to zero again at the antipodal point $r = \pi$; hence the point at
$r = 2\pi$ has to be the same point as $r = 0$, and these 3-spaces
are necessarily closed, with finite total volume. In the other
cases the 3-spaces are usually unbounded, and the surface areas of
these 2-spaces increase without limit; however, unusual topologies
still allow the spatial sections to be closed
\ct{ell71a}.\\

{\it Exercise}: Find the obvious orthonormal tetrad associated
with these coordinates, and determine their commutators and Ricci
rotation coefficients.

\subsection{Dynamical equations}
The remaining non-trivial equations are the energy equation
(\r{eq:en}), the Raychaudhuri equation (\r{eq:ray}), which
now takes the form
\be
3\,\frac{\ddot{S}}{S} + \sfrac12\,(\mu+3p) = 0 \ ,
\ee
and the Friedmann equation that follows from (\r{eq:3R}):
\be
{}^3\!R = 2\,\mu - \sfrac23\,\Theta^2 = {6k \over S^2 } \ ,
\l{friedmann}
\ee
where $k$ is a constant. Any two of these equations imply the third
if $\dot{S} \neq 0$ (the latter equation being a first integral of
the other two). All one has to do then to determine the dynamics is
to solve the Friedmann equation. The solution depends on what form
is assumed for the matter: Usually it is taken to be a perfect
fluid with equation of state $p = p(\mu)$, or as a sum of such
fluids, or as a scalar field with given potential $V(\phi)$.  For
the $\gamma$-law discussed above, the energy equation integrates to
give (\r{eq:en2}), which can then be used to represent $\mu$ in the
Friedmann equation.\\

{\it Exercise}: Show that on using the tetrad found above, all the
other $1+3$ covariant and tetrad equations are identically true
when these equations are satisfied.

\subsubsection{Basic parameters}
As well as the parameters $H_0$, $\Omega_0$, $\Omega_\Lambda$ and
$q_0$, the FLRW models are characterised by the {\bf spatial
curvature parameter} $K_0 = {k/S^2_0} = {{}^3\!R_0/6}$. These
parameters are
related by the equations (\r{eq:ray3}) and (\r{eq:fried1}), which
are now exact rather than approximate relations.

\subsubsection{Singularity and ages}
The existence of the big bang, and age limits on the Universe,
follow directly from the Raychaudhuri equation, together with the
energy assumption $(\mu+3p) > 0$ (true at least when quantum fields
do not dominate), because the Universe is expanding today
($\Theta_0 > 0$). That is, the singularity theorem above applies in
particular to FLRW models. Furthermore, from the Raychaudhuri
equation, in any FLRW model, the fundamental age relation holds
(see e.g. \ct{ell87}):
\begin{quotation}
{\bf Age Theorem}: In an expanding FLRW model with vanishing
cosmological constant and satisfying the active gravitational mass
density energy condition, ages are strictly constrained by the
Hubble expansion rate: namely, at every instant, the age $t_0$ of
the model (the time since the big bang) is less than the inverse
Hubble constant at that time:
\be
\l{eq:age}
(\mu+3p) > 0 \ , ~\Lambda = 0 \hsp5 \Rightarrow \hsp5
t_0 < 1/H_0 \ .
\ee
\end{quotation}
More precise ages $t_0(H_0,\Omega_0)$ can be determined for any
specific cosmological model from the Friedmann equation
(\r{friedmann}); in particular, in a matter-dominated early
universe the same result will hold with a factor $2/3$ on the
right-hand side, while in a radiation dominated universe the factor
will be $1/2$. Note that this relation applies in the early
universe when the expansion rate was much higher, and, hence, shows
that the hot early epoch ended shortly after the initial
singularity \ct{ell87}.\\
 
The age limits are one of the central issues in modern cosmology
\ct{ell96b,colell97}. Hipparchos satellite measurements suggest a
lowering of the age estimates of globular clusters to about $1.2
\times 10^9\,{\rm years}$, together with a decrease in the estimate
of the Hubble constant to about $H_0 \approx 50\,{\rm km}/{\rm
sec}/{\rm Mpc}$. This corresponds to a Hubble time $1/H_0$ of about
$1.8 \times 10^9\,{\rm years}$, implying there is no problem, but
red giant and cepheid measurements suggest $H_0 \approx 72 - 77\,
{\rm km}/{\rm sec}/{\rm Mpc}$ \ct{h098}, implying the situation is
very tight indeed. However, recent supernovae measurements
\ct{per88} suggest a positive cosmological constant, allowing
violation of the age constraint, and hence easing the situation.
All these figures should still be treated with caution; the issue
is fundamental to the viability of the FLRW models, and still needs
resolution.

\subsection{Exact and approximate solutions}
If $\Lambda = 0$ and the energy conditions are satisfied, FLRW
models expand forever from a big bang if $k = -\,1$ or $k = 0$, and
recollapse in the future if $k = +\,1$. A positive value of
$\Lambda$ gives a much wider choice for behaviours
\ct{rob33,refsta66}.

\subsubsection{Simplest models}
a) {\bf Einstein static model}: $S(t) = \mbox{const}$, $k = +\,1$,
$\Lambda = \sfrac{1}{2}\,(\mu+3p) > 0$, where everything is
constant in space and time, and there is no redshift. This model
is unstable (see above).\\

\noindent
b) {\bf de Sitter model}: $S(t) = S_{\rm unit}\,\exp(H\,t)$, $H =
\mbox{const}$, $k = 0$, a steady state solution in a constant
curvature space-time: it is empty, because $(\mu+p) = 0$, i.e., it
does not contain ordinary matter, but rather a cosmological
constant,\footnote{A fluid with $(\mu+p) = 0$ is equivalent to a
cosmological constant.} or a scalar field in the strict
`no-rolling' case. It has ambiguous redshift because the choice of
families of worldlines and space sections is not unique in this
case; see \ct{sch55}.\\

\noindent
c) {\bf Milne model}: $S(t) = t$, $k = -\,1$. This is flat, empty
space-time in expanding coordinates (again $(\mu+p) = 0$). \\

\noindent
d) {\bf Einstein--de Sitter model}: the simplest non-empty
expanding model, with $$k = 0 = \Lambda: ~~S(t) = a\,t^{2/3} \ ,
~a = \mbox{const} ~~\mbox{if} ~~p = 0 \ . $$ $\Omega = 1$ is always
identically true in this case (this is the critical density case
that just manages to expand forever). The age of such a model is
$t_0 = 2/(3H_0)$; if the cosmological constant vanishes, higher
density models ($\Omega_0 > 1)$ will have ages less than this,
and lower density models ($0 < \Omega_0 < 1$) ages between this
value and (\r{eq:age}). This is the present state of the Universe
if the standard inflationary universe theory is correct, the high
value of $\Omega$ then implying that most of the matter in the
Universe is invisible (the `dark matter' issue; see \ct{colell97}
for a summary of ways of estimating the matter content of the
Universe, leading to estimates that the detected matter in the
Universe in fact corresponds to $\Omega_0 \approx 0.2$ to
$0.3$). It is thus difficult to reconcile this model with
observations (the Universe could have flat space sections and a
large cosmological constant; but then that is not the Einstein--de
Sitter model).

\subsubsection{Parametric solutions}
Use dimensionless {\bf conformal time} $\tau = \int{dt/S(t)}$ and
rescale $S \rightarrow y = S(t)/S_0$. Then, for a non-interacting
mixture of pressure-free matter and radiation, we find in the
three cases $k = +\,1,\,0,\,-\,1$,
\bea
k = +\,1: &  y = &\alpha\,(1 - \cos\tau) + \beta\,\sin\tau \ , \\
k = 0:    &  y = &\alpha\,\tau^2/2 + \beta\,\tau \ , \\
k = -\,1: &  y = &\alpha\,(\cosh\tau - 1) + \beta\,\sinh\tau \ ,
\eea
where $\alpha = S_0^2\,H_0^2\,\Omega_m/2$, $\beta =
(S_0^2\,H_0^2\,\Omega_r)^{1/2}$, and, on setting $t = \tau = 0$
when $S = 0$,
\bea
k = +\,1: &  t = &S_0\,[\ \alpha\,(\tau - \sin\tau) + \beta\,
                         (1 - \cos\tau)\ ] \ , \\
k = 0:    &  t = &S_0\,[\ \alpha\,\tau^3/6 + \beta\,\tau^2/2\ ] \ ,
                                               \\
k = -\,1: &  t = &S_0\,[\ \alpha\,(\sinh\tau - \tau) + \beta\,
                         (\cosh\tau-1)\ ] \ .
\eea
It is interesting how in this parametrization the dust and
radiation terms decouple; this solution includes as special cases
the pure dust solutions, $\beta = 0$, and the pure
radiation solution, $\alpha = 0$. The general case represents a
smooth transition from a radiation dominated early era to a matter
dominated later era, and (if $k \neq 0$) on to a curvature
dominated era, recollapsing if $k = +1$.

\subsubsection{Early-time solutions}
At early times, when matter is relativistic or negligible compared
with radiation, the equation of state is $p = \sfrac13\,\mu$ 
and the curvature term can be ignored. The solution is
\be
\l{eq:early}
S(t) = c\,t^{1/2} \ , ~c = \mbox{const} \ ,
~\mu = \sfrac34\,t^{-2} \ , ~T = \left({3 \over 4a}\right)^{1/4}\,
{1\over t^{1/2}} \ ,
\ee
which determines the expansion time scale during nucleosynthesis
and so the way the temperature $T$ varies with time (and hence
determines the element fractions produced), and has no adjustable
parameters. Consequently, the degree of agreement attained between
nucleosynthesis theory based on this time scale and element
abundance observations \ct{wei72}--\ct{schtur} may be taken as
supporting both a FLRW geometry and the validity of the EFE at that
epoch.\\

The standard thermal history of the hot early Universe
(e.g. \ct{wei72}) follows; going back in time, the temperature
rises indefinitely (at least until an inflationary or
quantum-dominated epoch occurs), so that the very early Universe is
an opaque near-equilibrium mixture of elementary particles that
combine to form nuclei, atoms, and then molecules after pair
production ends and the mix cools down as the Universe expands,
while various forms of radiation (gravitational radiation,
neutrinos, electromagnetic radiation) successively decouple and
travel freely through the Universe that has become transparent to
them. This picture is very well supported by the detection of the
extremely accurate black body spectrum of the CBR, together with
the good agreement of nucleosynthesis observations with predictions
based on the FLRW time scales (\r{eq:early}) for the early
Universe.\\

{\it Exercise}: The early Universe was radiation dominated but
later became matter dominated (as at the present day).  Determine
at what values $S_{\rm equ}$ of the scale factor $S(t)$
matter--radiation equality occurs, as a function of $\Omega_0$. For
what values of $\Omega_0$ does this occur before decoupling of
matter and radiation?  (Note that if the Universe is dominated by
Cold Dark Matter (`CDM'), then equality of baryon and radiation
density occurs after this time.) When does the Universe become
curvature dominated?

\subsubsection{Scalar field}
The inflationary universe models use many approximations to model a
FLRW universe with a scalar field $\phi$ as the dominant
contribution to the dynamics, so allowing accelerating models that
expand quasi-exponentially through many efoldings at a very early
time \ct{gut80,koltur90}, possibly leading to a very inhomogeneous
structure on very large (super-particle-horizon) scales
\ct{lin90}. This then leads to important links between particle
physics and cosmology, and there is a very large literature on this
subject. If an inflationary period occurs in the very early
Universe, the matter and radiation densities drop very close to
zero while the inflaton field dominates, but is restored during
`reheating' at the end of inflation when the scalar field energy
converts to radiation.\\
 
This will not be pursued further here, except to make one point:
because the potential $V(\phi)$ is unspecified (the nature of the
inflaton is not known) and the initial value of the `rolling rate'
$\dot{\phi}$ can be chosen at will, it is possible to specify a
precise procedure whereby any desired evolutionary history $S(t)$
is attained by appropriate choice of the potential $V(\phi)$ and
the initial `rolling rate' (see \ct{ellmad91} for details). Thus,
inflationary models may be adjusted to give essentially any desired
results in terms of expansion history.

\subsubsection{Kinetic theory}
While a fluid description is used most often, it is also of
interest to use a kinetic theory description of the matter in the
Universe \ct{ehl71}. The details of collisionless isotropic kinetic
models in a FLRW geometry are given by Ehlers, Geren and Sachs
\cite{egs68}; this is extended to collisions in
\ct{treell71}. Curiously, it is also possible to obtain exact
anisotropic collisionless solutions in FLRW geometries; details are
given in \ct{ellmattre83}.

\subsection{Phase planes}
{}From these equations, as well as finding simple exact solutions,
one can determine evolutionary phase planes for this family of
models; see Stabell and Refsdal \ct{refsta66} for $(\Omega_m,q_0$),
Ehlers and Rindler \ct{er89} for $(\Omega_m,\Omega_r,q_0$),
Wainwright and Ellis \ct{waiell97} for $(\Omega_0,H_0)$, and Madsen
and Ellis \ct{me88} for $(\Omega,S)$. The latter are based on the
phase plane equation
\be
\l{eq:phase}
{d\Omega \over dS} = -\,(3\gam-2)\,{\Omega \over S}\,(1 -\Omega)
\ .
\ee
This equation is valid for any $\gamma$, i.e., for arbitrary
relations between $\mu$ and $p$, but gives a $(\Omega,S)$ phase
plane flow if $\gamma = \gamma(\Omega,S)$, and in particular if
$\gamma = \gamma(S)$ or $\gamma = \mbox{const}$. Non-static
solutions can be followed through turnaround points where $\dot{S}
= 0$ (and so $\Omega$ is infinite). This enables one to attain
complete (time-symmetric) phase planes for models with and without
inflation; see \ct{me88} and \ct{ell91} for details.

\subsection{Observations}
Astronomical observations are based on radiation travelling to us
on the {\bf geodesic null rays} that generate our {\bf past light
cone}. In the case of a FLRW model, we may consider only
{\bf radial} null rays as these are generic (because of spatial
homogeneity, we can choose the origin of coordinates on any light
ray of interest; because of isotropy, light rays travelling in any
direction are equivalent to those travelling in any other
direction). Thus, we may consider geodesic null rays travelling in
the FLRW metric (\r{eq:frw1}) such that $ds^2 = 0 = d\theta =
d\phi$; then it follows that $0 = -\,dt^2 + S^2(t)\,dr^2$ on these
geodesics. Hence, radiation emitted at $E$ and received at $O$
obeys the basic relations
\be
\l{eq:o1}
r = \int_E^O dr = \int_{t_E}^{t_0} {dt \over S(t)}
= \int_{S_E}^{S_0} {dS \over S(t)\,\dot{S}(t)}\ ,
\ee
yielding the dimensionless matter-comoving radial coordinate
distance, where the term $\dot{S}$ may be found from the
Friedmann equation (\r{friedmann}), once a suitable matter
description has been chosen.

\subsubsection{Redshift}
The first fundamental quantity is {\bf redshift}. Considering two
successive pulses sent from $E$ to $O$, each remaining at the same
matter-comoving coordinate position, it follows from (\r{eq:o1}
that the cosmological redshift in a FLRW model is given by
\be
\l{eq:z}
(1+z_c) = {\lambda_0 \over \lambda_E}
= {\Delta T_0 \over \Delta T_E} 
= {S(t_0) \over S(t_E)} \ ,
\ee
and so directly measures the expansion of the model between when
light was emitted and when it is received. Two comments are in
order. First, redshift is essentially a time-dilation effect, and
will be apparent in all observations of a source, not just in its
spectra; this characterisation has the important consequences that
(i) redshift is achromatic --- the fractional shift in wavelength
is independent of wavelength, (ii) the width of any emitted
frequency band $d\nu_E$ is altered proportional to the redshift
when it reaches the observer, i.e., the observed width of the band
is $d\nu_0 = (1+z)\,d\nu_E$, and (iii) the observed rate of
emission of radiation and the rate of any time variation in its
intensity will both also be proportional to $(1+z)$. Second, there
can be local gravitational and Doppler contributions $z_0$ at the
observer, and $z_E$ at the emitter; observations of spectra tell us
the overall redshift $z$, given by
\be
(1+z) = (1+z_0)\,(1+z_c)\,(1+z_E) \ ,
\ee
but cannot tell us what part is cosmological and what part is due
to local effects at the source and the observer. The latter can be
determined from the CBR anisotropy, but the former can only be
estimated by identifying cluster members and subtracting off the
mean cluster motion. The essential problem is in identifying which
sources should be considered members of the same cluster.  This is
the source of the controversies between Arp {\em et al\/} and the
rest of the observational community (see, e.g., Field {\em et al\/}
\ct{fieetal73}).

\subsubsection{Areas}
The second fundamental issue is {\bf apparent size}. Considering light
rays converging to the observer at time $t_0$ in a {\bf solid angle}
$d\Omega = \sin\theta\,d\theta\,d\phi$, from the metric form
(\r{eq:frw1}) the corresponding null rays\footnote{Bounded by
geodesics located at ($\phi_0,\theta_0)$,
$(\phi_0+d\phi,\theta_0)$, $(\phi_0,\theta_0+d\theta)$,
$(\phi_0+d\phi,\theta_0+d\theta)$.} will be described by constant
values of $\theta$ and $\phi$ and at the time $t_E$ will encompass
an {\bf area} $dA = S^2(t_E) f^2(r) d\Omega$ orthogonal to the light
rays, where $r$ is given by (\r{eq:o1}). Thus, on defining the {\bf
observer area distance} $r_0$ by the standard area relation, we
find
\be
dA = r_0^2\,d\Omega \hsp5 \Rightarrow \hsp5
r_0^2 = S^2(t_E)\,f^2(r) \ .
\l{area-dist}
\ee
Because these models are isotropic about each point, the {\it same}
distance will relate the observed {\bf angle $\alpha$} corresponding
to a {\bf linear length scale} $\ell$ orthogonal to the light rays:
\be
\l{eq:ang}
\ell = r_0\,\alpha \ .
\ee
One can now calculate $r_0$ from this formula together with
(\r{eq:o1}) and the Friedmann equation, or from the {\bf geodesic
deviation equation} (see \ct{hveell98b}), to obtain for a
non-interacting mixture of matter and radiation \ct{matazi88},
\be
\l{eq:ro}
r_0(z) = {1 \over H_0 q_0 (q_0 + \beta -1)}\,
{\left[\ (q_0-1)\left\{1 + 2q_0z + q_0z^2(1-\beta)\right\}^{1/2} 
- (q_0-q_0\beta z-1)\ \right]\over (1+z)^2} \ ,
\ee
where $\beta$ represents the matter to radiation ratio:
$(1-\beta)\,\rho_{m0} = 2\,\beta\,\rho_{r0}$\,. The
standard {\bf Mattig relation} for pressure-free matter is obtained
for $\beta = 1$ \ct{mat58}, and the corresponding radiation result
for $\beta = 0$.\\

An important consequence of this relation is {\bf refocusing of the
past light cone}: the Universe as a whole acts as a gravitational lens,
so that there is a redshift $z_*$ such that the observer area distance
reaches a maximum there and then decreases for larger $z$;
correspondingly, the apparent size of an object of fixed size would
reach a minimum there and then increase as the object was moved
further away \ct{san61}. As a specific example, in the simplest
(Einstein--de Sitter) case with $p = \Lambda = k = 0$, we find
\be
~\beta = 1 \ , ~q_0 = \sfrac12 \hsp5 \Rightarrow \hsp5
r_0(z) = {2 \over H_0}\,{1 \over (1+z)^{3/2}}\,(\,\sqrt{1+z}-1\,)
\ , 
\ee
which refocuses at $z_* = 5/4$ \ct{rotell93}; objects further away
will look the same size as much closer objects. For example, an
object at a redshift $z_1 = 1023$ (i.e., at about last scattering)
will appear the same angular size as an object of identical size at
redshift $z_2 = 0.0019$ (which is very close --- it corresponds to
a speed of recession of about 570\,{\rm km}/{\rm sec}). In a low
density model, refocusing takes place further out, at redshifts
up to $z \approx 4$, depending on the density, and with apparent
sizes depending on possible source size evolution \ct{elltiv85}.\\

The predicted {\bf (angular size, distance)--relations} are
difficult to test observationally because objects of more or less
fixed size (such as spherical galaxies) do not have sharp edges
that can be used for measuring angular size and so one has rather
to measure isophotal diameters (see e.g. \ct{ellperxx}), while
objects with well-defined linear dimensions, such as double radio
sources, are usually rapidly evolving and so one does not know
their intrinsic size. Thus, these tests, while in principle clean,
are in fact difficult to use in practice.

\subsubsection{Luminosity and reciprocity theorem}
There is a remarkable relation between upgoing and downgoing
bundles of null geodesics connecting the source at $t_E$ and the
observer at $t_0$. Define {\bf galaxy area distance} $r_G$ as
above for observer area distance, but for the upgoing rather than
downgoing bundle of null geodesics. The expression for this
distance will be exactly the same as (\r{area-dist}) except that
the times $t_E$ and $t_0$ will be interchanged. Consequently, on
using the redshift relation (\r{eq:z}),
\begin{quotation}
{\bf Reciprocity Theorem}:
The observer area distance and galaxy area distance are identical
up to redshift factors:
\be
{r_0^2 \over r_G^2} 
= {1 \over (1+z)^2} \ .
\ee
\end{quotation}
This is true in {\em any\/} space-time as a consequence of the
standard first integral of the geodesic deviation equation
\ct{eth33,ell71}. \\

Now from photon conservation, the {\bf flux of light} received
from a source of {\bf luminosity} $L$ at time $t_E$ will be
measured to be
$$F ={L(t_E) \over 4\pi}\,{1\over (1+z)^2}\,
{1 \over r_G^2\,}
\ ,$$
with $r_{G}^{2}=S^{2}(t_{0})\,f^2(r)$ and $r$ given by 
(\r{eq:o1}), and the two factors $(1+z)$ coming
from photon redshift and time dilation of the emission rate,
respectively. On using the reciprocity result, this becomes
\be
F = {L(t_E) \over 4 \pi}\,{1 \over (1+z)^4}\,{1 \over r_0^2} \ ,
\ee
where $r_0$ is given by (\r{eq:ro}). On taking logarithms,
this gives the standard {\bf (luminosity, redshift)--relation} of
observational cosmology \ct{san61}. Observations of this Hubble
relation basically agree with these predictions, but are not
accurate enough to distinguish between the various FLRW models. The
hopes that this relation would determine $q_0$ from galaxy
observations have faded away because of the major problem of {\bf
source evolution}: we do not know what the source luminosity
would have been at the time of emission. We lack {\bf standard
candles} of known luminosity (or equivalently, rigid objects of
known linear size, from which apparent size measurements would
give the answer). Various other distance estimators such as the
{\bf Tully--Fisher relation} have helped considerably, but not
enough to give a definitive answer. Happily it now seems that
{\bf Type Ia supernovae} may provide the answer in the next
decade, because their luminosity can
be determined from their light curves, which should depend only on
local physics rather than their evolutionary history. This is an
extremely promising development at the present time (see
e.g. \ct{per88}).

\subsubsection{Specific intensity}
In practice, we measure (a) in a limited waveband rather than over
all wavelengths, as the `bolometric' calculation above suggests; and
(b) real detectors measure specific intensity (radiation received
per unit solid angle) at each point of an image, rather than total
source luminosity. Putting these together, we see that if the {\bf
source spectrum} is ${\cal I}(\nu_E)$, i.e., a fraction ${\cal
I}(\nu_E)\,d\nu_E$ of the source radiation is emitted in the
frequency range $d\nu_E$, then the observed {\bf specific
intensity} at each image point is given by\footnote{Absorption
effects will modify this if there is sufficient absorbing matter
present; see \ct{ell71} for relevant formulae.}
\be
\l{eq:int}
I_\nu\,d_\nu = {B_E \over (1+z)^3}\,{\cal I}(\nu(1+z))\,d\nu \ ,
\ee
where $B_E$ is the {\bf surface brightness} of the emitting
object, and the observer area distance $r_0$ has canceled out
(because of the reciprocity theorem). This tells us the apparent
intensity of radiation detected in each direction --- which is
independent of (area) distance, and dependent only on the source
redshift, spectrum, and surface brightness. Together with the
{\bf angular diameter relation} (\r{eq:ang}), this determines what
is actually measured by a detector \ct{ellperxx}.\\

An immediate application is {\bf black body radiation}: if any
radiation is emitted as black body radiation at temperature $T_E$,
it follows ({\it Exercise!}) from the black body expression
${\cal I}_\nu = \nu^3\,b(\nu/T_E)$ that the received radiation
will also be black body (i.e., have this same black body form)
but with a measured temperature of
\be
T_0 = {T_E \over (1+z)} \ .
\ee
Note this is true in {\em all\/} cosmologies: the result does
not depend on the FLRW symmetries. The importance of this, of
course, is that it applies to the observed CBR.

\subsubsection{Number counts}
If we observe sources in a given solid angle $d\Omega$ in a
matter-comoving radial coordinate range $(r,r+dr)$, the
corresponding volume is $dV = S^3(t_E)\,r_0^2\,dr\,d\Omega$,
so if the source density is $n(t_E)$
and the probability of detection is $p$, the number of sources
observed will be
\be
dN = p\,n(t_E)\,dV = p\,\left[\,\frac{n(t_E)}{(1+z)^3}\,\right]
\,S^3(t_0)\,f^2(r)\,dr\,d\Omega \ ,
\ee
with $r$ given by (\r{eq:o1}). This is the basic {\bf number count
relation}, where $dr$ can be expressed in terms of observable
quantities such as $dz$; the quantity in brackets is constant if
source numbers are conserved in a FLRW model, that is
\be \l{eq:num}
n(t_E) = n(t_0) (1+z)^3\,. 
\ee
The FLRW predictions agree with
observations only if we allow for source number and/or luminosity
evolution (cf. the discussion of spherically symmetric models in
the next section); but we have no good theory for source
evolution. \\

The additional problem is that there are many undetectable objects
in the sky, including entire galaxies, because they lie below the
detection threshold; thus we face the problem of {\bf dark matter},
which is very difficult to detect by cosmological observations
except by its lensing effects (if it is clustered) and its effects
on the age of the Universe (if it is smoothly distributed). The
current view is that there is indeed such dark matter, detected
particularly through its dynamical effects in galaxies and clusters
of galaxies (see \cite{colell97} for a summary), with the present
day total matter density most probably in the range $0.1 \leq
\Omega_0 \leq 0.3$, while the baryon density is of the order of
$0.01 \leq \Omega_0^{\rm baryons} \leq 0.03$ (from nucleosynthesis
arguments). Thus, most of the dark matter is probably
non-baryonic. \\

To properly deal with source statistics in general, and number
counts in particular, one should have a reasonably good model of
detection limits. It is highly misleading to represent such limits
as depending on source apparent magnitude alone (see Disney
\ct{dis76}); this does not take into account the possible occurrence
of low surface brightness galaxies. A useful model based on both
source apparent size and magnitude is presented in Ellis, Sievers
and Perry \ct{ellpersie84}, summarized in \ct{ell87}. One should
note from this particularly that if there is an evolution in source
size, this has a more important effect on source detectability than
an evolution in surface brightness.\\

{\it Exercise}: Explain why an observer in a FLRW model may at late
times of its evolution see a situation that looks like an island
universe (see \ct{rotell99}).

\subsection{Observational limits}
The first basic observational limit is that we cannot observe
anything outside our past light cone, given by
(\r{eq:o1}). Combined with the finite age of the Universe, this
leads to a maximum matter-comoving radial coordinate distance from
the origin for matter with which we can have had any causal
connection: namely
\be
\l{eq:o2}
r_{ph}(t_0) = \int_{0}^{t_0} {dt \over S(t)} \ ,
\ee
which converges for any ordinary matter. Matter outside is not
visible to us; indeed, we cannot have had any causal contact with
it. Consequently (see Rindler \ct{rin57}), the particles at this
matter-comoving coordinate value define the {\bf particle
horizon}: they separate that matter which can have had any
causal contact with us since the origin of the Universe from
that which cannot. This is most clearly seen by using
Penrose's conformal diagrams, obtained on using as coordinates
the matter-comoving radius and conformal time; see
Penrose \ct{pen83} and Tipler, Clarke and Ellis \ct{tce89}. The
present day distance to the particle horizon is
\be
\l{eq:o3}
D_{ph}(t_0) = S(t_0)\,r_{ph} 
= S(t_0)\,\int_{0}^{t_0} {dt \over S(t)} \ .
\ee
{}From (\r{eq:o1}), this is a sphere corresponding to infinite
measured redshift (because $S(t) \rightarrow 0$ as $t \rightarrow
0$).\\

{\it Exercise}: Show that once comoving matter has entered the
particle horizon, it cannot leave it (i.e., once causal contact has
been established in a FLRW universe, it cannot cease).\\

Actually we cannot even see as far as the particle horizon: on our
past light cone information rapidly fades with redshift (because of
(\r{eq:int})); and because the early Universe is opaque, we can
only see (by means of any kind of electromagnetic radiation) to the
{\bf visual horizon} (Ellis and Stoeger \ct{ellsto88}), which is
the sphere at matter-comoving radial coordinate distance
\be
\l{eq:o4}
r_{vh}(t_0) = \int_{t_d}^{t_0} {dt \over S(t)} \ ,
\ee
where $t_d$ is the time of {\bf decoupling of matter and
radiation}, when the Universe became transparent (at about a
redshift of $z= 1100$). The matter we see at that time is the
matter which emitted the CBR we measure today with a present
temperature of $2.73\,{\rm K}$; its present distance
from us is
\be
D_{vh}(t_0) = S(t_0)\,r_{vh} \ .
\ee
If we evaluate these quantities in an Einstein--de Sitter model,
we find an interesting paradox: (re-establishing the fundamental
constant $c$,) the present day distance to the particle horizon is
$D_{ph}(t_0) = 3c t_0 (= 2c/H_{0})$. The question is how can this
be bigger than $c t_0$ (the distance corresponding to the age of the
Universe). This suggests that the matter comprising the particle
horizon has been moving away from us at {\em faster\/} than the
speed of light in order to reach that distance. How can this be? To
investigate this (Ellis and Rothman \ct{rotell93}), note that the
{\bf proper distance} from the origin to a galaxy at matter-comoving
radial coordinate $r$ at time $t$ is $D(t,r) = S(t)\,r$. Its
velocity away from us is thus given by {\bf Hubble's law}
\be
v = \dot{D} = \dot{S}\,r = \frac{\dot{S}}{S}\,D
= H\,D \ .
\ee
Thus, at any time $t$, $v = c$ when $D_c(t) = {c/H(t)}$; this is
the {\bf speed of light sphere}, where galaxies are (at the
present time) receding away from us at the speed of light; those
galaxies at a larger distance will be (instantaneously) moving away
at a speed greater than $c$. In the case of an Einstein--de Sitter
universe, it occurs when $D_c = 3c t_0/2 (= c/H_{0})$. This is
precisely half the present distance to the particle horizon; the
latter is thus {\it not} the distance where points are moving away
from us at the speed of light (however, it {\it is} the surface of
infinite redshift).\\

To see that this is compatible with local causality, change from
Lagrangian coordinates $(t,r)$ to Eulerian coordinates $(t,D)$,
where $D$ is the instantaneous proper distance, as above.  Then we
find the (non-comoving) metric form
\be
ds^{2} = -\,[\,1 - \frac{(\dot{S}/c)^{2}}{S^{2}}\,D^{2}\,]\,
(dct)^{2} - 2\,\frac{\dot{S}/c}{S}\,D\,dct\,dD
+ dD^{2} + S^{2}(t)\,f^{2}(r(t,D))\,d\Omega^{2} \ .
\ee
It follows that the local light cones are given by
\be
{dD_\pm \over dt} = {\dot{S} \over S}\,D \pm c \ .
\ee
It is easily seen then that there is no violation of local
causality. We also find from this that the past light cone of $t =
t_0$ intersects the family of speed of light spheres at its {\bf
maximum distance} from the origin (the place where the past light
cone starts refocusing), i.e., at
\be
\l{max}
t_* = {8\over 27}\,t_0 \ , \hsp5 D_* = {4\over 9}\,c t_0
= {8\over 27}\,{c \over H_0} \ , \hsp5 S(t_*)= {4\over 9}\,S(t_0)
\ , \hsp5 z_* = 1.25 \ .
\ee
At that intersection, $dD_-/dt = 0$ (maximum distance!), $dD_+/dt =
2c$, so there is no causality violation by the matter moving at
speed $c$ relative to the central worldline. That matter is
presently at a distance $ct_0$ from us. By contrast, the matter
comprising the visual horizon was moving away from us at a speed $v
= 61 c$ when it emitted the CBR, and was at a distance of about
$10^7$ light years from our past worldline at that time. Hence, it
is the {\bf fastest moving matter} we shall ever see, but was
{\em not\/} at the greatest proper distance to which we can see
(which is $D_*$, see (\r{max})). For a full investigation of these
matters see \ct{rotell93}.\\

Finally it should be noted that an early inflationary era will move
the particle horizon out to very large distances, thus
\ct{gut80,koltur90} solving the causal problem presented by the
isotropy of CBR arriving here from causally disconnected regions
(see \ct{ellsto88,rotell93} for the relevant causal diagrams), but
it will have no effect on the visual horizon. Thus, it changes the
causal limitations, but does not affect the visual limits on the
part of the Universe we can see.\\

{\it Exercise}: Determine the angular size seen today for the
horizon distance $D_{ph}(t_d)$ at the time of decoupling. What is
the physical significance of this distance? How might this relate
to CBR anisotropies?

\subsubsection{Small universes}
The existence of visual horizons represent absolute limits on what
we can ever know; because of them, we can only hope to investigate
a small fraction of all the matter in the Universe. Furthermore,
they imply we do not in fact have the data needed to predict to the
future, for at any time gravitational radiation from as yet unseen
objects (e.g., domain walls in a chaotic inflationary universe) may
cross the visual horizon and undermine any predictions we may have
made \ct{ell84}. However, there is one exceptional situation: it is
possible we live in a {\bf small universe}, with a spatially closed
topology on such a length scale (say, $300$ to $800\,{\rm Mpc}$)
that we have already seen around the universe many times, thus
already having seen all the matter there is in the universe. The
effect is like being in a room with mirrors on the floor, ceiling,
and all walls; images from a finite number of objects seem to
stretch to infinity. There are many possible topologies, whatever
the sign of $k$ \ct{ell71a}; the observational result --- best
modelled by considering many identical copies of a basic cell
attached to each other in an infinitely repeating
pattern\footnote{In mathematical terms, the universal covering
space.} --- can be very like the real Universe (Ellis and Schreiber
\ct{ellsch86}). In this case we would be able to see our own galaxy
many times over, thus being able to observationally examine its
historical evolution once we had identified which images of distant
galaxies were in fact repeated images of our own galaxy. \\

It is possible the real Universe is like this. Observational tests
can be carried out by trying to identify the same cluster of
galaxies, QSO's \ct{rou96}, or X-ray sources in different
directions in the sky \ct{rou98}; or by detecting circles of
identical temperature variation in the CBR sky (Cornish et al
\ct{cor97}). If no such circles are detected, this will be a
reasonably convincing proof that we do not live in such a small
universe --- which has various philosophical advantages over the
more conventional models with infinite spatial sections \ct{ell84}.
Inter alia they give some degree of mixing of CBR modes so giving a
potentially powerful explanation of the low degree of CBR
anisotropy (but this effect is not as strong as some have claimed;
see \ct{elltav94}).

\subsection{FLRW universes as cosmological models}
These models are very successful in explaining the major features
of the observed Universe --- its expansion from a hot big bang
leading to the observed galactic redshifts and remnant black body
radiation, tied in well with element abundance predictions and
observations (Peebles et al \ct{pee88}). However, these models do
not describe the real Universe well in an essential way, in that
the highly idealized degree of symmetry does not correspond to the
lumpy real Universe. Thus, they can serve as basic models giving
the largest-scale smoothed out features of the observable physical
Universe, but one needs to perturb them to get realistic
(`almost-FLRW') Universe models that can be used to examine the
inhomogeneities and anisotropies arising during structure
formation, and that can be compared in detail with observations.
This is the topic of the last sections. \\
 
However, there is a major underlying issue: because of their high
symmetry, these models are infinitely improbable in the space of
all possible cosmologies. This high symmetry represents a very high
degree of fine tuning of initial conditions, which is
extraordinarily improbable, unless we can show physical reasons why
it should develop from much more general conditions. In order to
examine that question, one needs to look at much more general
models and see if they do indeed evolve towards the FLRW models
because of physical processes (this is the chaotic cosmology
programme, initiated by Misner \ct{mis68,mis69}, and taken up much
later by the inflationary universe proposal of Guth \ct{gut80}).
Additionally, while the FLRW models seem good models for the
observed Universe at the present time, one can ask (a) are they the
only possible models that will fit the observations? (b) does the
Universe necessarily have the same symmetries on very large scales
(outside the particle horizon), or at very early and/or very late
times? \\

To study these issues, we need to look at more general models,
developing some understanding of their geometry and dynamics. This
is the topic of the next section. 
We will find there is a range of models in addition to the FLRW
models that can fulfill all present day observational
requirements. Nevertheless, it is important to state that the
family of perturbed FLRW models can meet all present observational
requirements, {\it provided} we allow suitable evolution of source
properties back in the past. They also provide a powerful
theoretical framework for considering the nature of and effects of
cosmic evolution. Hence, they are justifiably the {\bf standard
models of cosmology}. No evidence stands solidly against them.\\

{\it Exercise}: Apart from the detection of major anisotropy, there
are a series of other observations which could, if they were ever
observed, decisively disprove this family of standard models. What
are these observations? (See \cite{ree95} for some.)

\subsection{General observational relations}
Before moving to that section, we briefly consider general
observational relations. The present section has focussed on the
observational relations holding in FLRW models. However,
corresponding generic relations can be found determining
observations in arbitrary cosmologies (see Kristian and Sachs
\ct{krisac66} and Ellis \ct{ell71}). The essential points are as
follows.\\

In a general model, observations take place on our {\bf past light
cone}, which will develop many cusps and caustics at early times
because of {\bf gravitational lensing}, but is still locally
generated by {\bf geodesic null rays}. The information we receive
comes to us along these null rays, with tangent vector
\be
k^{a} = {dx^{a} \over dv} \ , \hsp5k_ak^a = 0 \ , \hsp5
k_a = \nabla_{a}\phi \hsp5 \Rightarrow \hsp5
k^{b}\nabla_{b}k^{a} = 0 \ , \hsp5 k^{a}\nabla_{a}\phi = 0 \ .
\ee
The phase factor $\phi$ determines the local light cone $\{\phi =
\mbox{const}\}$. Relative to an observer with 4-velocity $u^a$, the
null vector $k^a$ determines a redshift factor $(-k_au^a)$ and a
direction $e_a$:
\be
k^a = (-k_bu^b)\,(u^a + e^a) \ , ~~e_au^a = 0 \ , ~~e_ae^a = 1 \ .
\ee
Considering the observed variation $\dot{\phi} = u^a\nabla_a\phi$
of the phase $\phi$, we see that the observed cosmological
{\bf redshift} $z$ for comoving matter\footnote{Cf. the comment
on cosmological and local sources of redshift above.} is given by
\be
\l{eq:red2}
(1+z) = {\lambda_{O} \over \lambda_{E}}
= {(k_au^a)_{E} \over (k_bu^b)_{O}} \ . 
\ee
Taking the derivative of this equation along $k^a$, we get the
fundamental equation \ct{ehl61,ell71}
\be
{d\lambda \over \lambda}
= - \,{d(k_au^a) \over (k_bu^b)}
= \left[\ \sfrac{1}{3}\,\Th + (\udot_{a}e^{a})
+ (\sig_{ab}e^{a}e^{b})\ \right] dl \ ,
\ee
where $dl = (-k_au^a)\,dv$. This shows directly the isotropic and
anisotropic contributions to redshift from the expansion and shear,
respectively, and the gravitational redshift contribution from the
acceleration.\footnote{In a static gravitational field, this will
be given by an acceleration potential: $\udot_{a} = \3nab_{a}\Phi$;
see \ct{ehl61}.} In a FLRW model, the last
two contributions will vanish.

Area distances are defined as before, and because of the geodesic
deviation equation, the reciprocity theorem holds unchanged
\ct{ell71}.\footnote{Because of the first integrals of the geodesic
deviation equation; this result can also be shown from use of
Liouville's theorem in kinetic theory.} Consequently, the same
surface brightness results as discussed above hold generically;
specifically, Eq. (\r{eq:int}) holds in any anisotropic or
inhomogeneous cosmology. The major difference from the isotropic
case is that due to the effect of the electric and magnetic Weyl
curvatures in the {\bf geodesic deviation equation}, distortions
occur in bundles of null geodesics which then cause focusing,
resulting both in {\bf strong lensing} (multiple images, Einstein
rings, and arcs related to cusps and caustics in the past light
cone) and {\bf weak lensing} (systematic distortion of images in
an observed area); see the lectures by Y Mellier and F Bernardeau,
the book by Schneider, Ehlers and Falco \ct{sef}, and the work
by Holz and Wald \ct{holwal}.\\

Power series equations showing how the kinematical quantities and
the electric and magnetic Weyl curvatures affect cosmological
observations have been given in a beautiful paper by Kristian and
Sachs \ct{krisac66}. The generalisation of those relations to
generic cosmologies has been investigated by Ellis {\em et al\/}
\ct{ensmw85}, showing how in principle cosmological observations can
directly determine the space-time structure on the past null cone,
and thence off it. Needless to say, major practical observational
difficulties make this a formidable task, but some progress in this
direction is possible (see e.g. \ct{ell84}, \ct{ell95} and
\ct{sto87}).\\

{\it Exercise}: Apart from area distances, distortions, matter
densities, and redshifts, a crucial data set needed to completely
determine the space-time geometry from the EFE is the transverse
velocities of the matter we observe on the past light cone
\ct{ensmw85}. Consider how one might try to measure these velocity
components, and what are the best limits one might place on them by
practical measurement techniques. [Hint: One possible route is by
solar system interferometry. Another is by the Sunyaev--Zel'dovich
effect \ct{sunzel70}.]

\section{Solutions with symmetries} 
\subsection{Symmetries of cosmologies}
Symmetries of a space or a space-time (generically, `space') are
transformations of the space into itself that leave the metric
tensor and all physical and geometrical properties invariant. We
deal here only with {\em continuous\/} symmetries, characterised
by a continuous group of transformations and associated vector
fields \ct{eis33}.

\subsubsection{Killing vector fields}
A space or space-time {\bf symmetry}, or {\bf isometry}, is a
transformation that drags the metric along a certain congruence of
curves into itself. The generating vector field $\xi_{i}$ of such
curves is called a {\bf Killing vector field} (or `KV'), and
obeys {\bf Killing's equations},
\be
(L_\xi g)_{ij} = 0 \hsp5 \Leftrightarrow \hsp5
\nabla_{(i}\xi_{j)} = 0 \hsp5 \Leftrightarrow \hsp5
\nabla_{i}\xi_{j} = -\,\nabla_{j}\xi_{i} \ ,
\ee
where $L_{\xi}$ is the {\bf Lie derivative} along $\xi_{i}$. By the
Ricci identities for a KV, this implies the curvature equation:
\be
\l{eq:int3}
\nabla_{i}\nabla_{j}\xi_{k} = R^{m}{}_{ijk}\,\xi_{m} \ ,
\ee
and so the infinite series of further equations that follows by
taking covariant derivatives of this one, e.g.,
\be
\nabla_{l}\nabla_{i}\nabla_{j}\xi_{k}
= (\nabla_{l}R^{m}{}_{ijk})\,\xi_{m}
+ R^{m}{}_{ijk}\,\nabla_{l}\xi_{m} \ .
\ee
The set of all KV's forms a Lie algebra with a basis
$\{\,\xi_a\,\}_{a = 1, 2, \dots, r}$, of dimension $r\leq
\sfrac{1}{2}\,n\,(n-1)$. $\xi_a^i$ denote the components
with respect to a local coordinate basis; $a, b, c$ label the KV
basis, and $i, j, k$ the coordinate components. Any KV can be
written in terms of this basis, with {\it constant
coefficients}. Hence: if we take the commutator
$[\,\xi_a,\,\xi_b\,]$ of two of the basis KV's, this is also a KV,
and so can be written in terms of its components relative to the KV
basis, which will be constants. We can write the constants as
$C^c{}_{ab}$, obtaining\footnote{Cf. equation (\r{eq:55}).}
\be
[\,\xi_a,\,\xi_b\,] = C^c{}_{ab}\,\xi_c \ , \hsp5
C^a{}_{bc} = C^a{}_{[bc]} \ .
\ee
By the Jacobi identities for the basis vectors, these {\bf structure
constants} must satisfy
\be
\l{eq:jac3}
C^a{}_{e[b}C^e{}_{cd]} = 0 \ ,
\ee
(which is just equation (\r{eq:jac}) specialized to the case of a
set of vectors with constant commutation functions). These are the
integrability conditions that must be satisfied in order that the
Lie algebra exist in a consistent way. The transformations
generated by the Lie algebra form a Lie group of the same dimension
(see Eisenhart \ct{eis33} or Cohn \ct{coh61}).\\

{\bf Arbitrariness of the basis}: We can change the basis of KV's
in the usual way;
\be
\l{eq:trans}
\xi_{a'} = \Lambda_{a'}{}^a\,\xi_a \hsp5 \Leftrightarrow \hsp5
\xi_{a'}^i = \Lambda_{a'}{}^a\,\xi_a^i \ ,
\ee 
where the $\Lambda_{a'}{}^a$ are constants with
$\det\,(\Lambda_{a'}{}^a) \neq 0$, so unique inverse matrices
$\Lambda^{a'}{}_{a}$ exist. Then the structure constants transform
as tensors:
\be
\l{eq:trans1}
C^{c'}{}_{a'b'} = \Lambda^{c'}{}_c\,\Lambda_{a'}{}^a\,
\Lambda_{b'}{}^b\,C^c{}_{ab} \ .
\ee
Thus the possible equivalence of two Lie algebras is not obvious,
as they may be given in quite different bases.

\subsubsection{Groups of isometries}
The isometries of a space of dimension $n$ must be a group, as the
identity is an isometry, the inverse of an isometry is an isometry,
and the composition of two isometries is an isometry. Continuous
isometries are generated by the Lie algebra of KV's. The group
structure is determined locally by the Lie algebra, in turn
characterised by the structure constants \ct{coh61}. The action of
the group is characterised by the nature of its orbits in space;
this is only partially determined by the group structure (indeed
the same group can act as a space-time symmetry group in quite
different ways).

\subsubsection{Dimensionality of groups and orbits}
Most spaces have no KV's, but special spaces (with symmetries) have
some. The group action defines orbits in the space where it acts,
and the dimensionality of these orbits determines the kind of
symmetry that is present.

The {\bf orbit} of a point $p$ is the set of all points into which
$p$ can be moved by the action of the isometries of a space. Orbits
are necessarily homogeneous (all physical quantities are the same
at each point). An {\bf invariant variety} is a set of points moved
into itself by the group. This will be bigger than (or equal to)
all orbits it contains. The orbits are necessarily invariant
varieties; indeed they are sometimes called minimum invariant
varieties, because they are the smallest subspaces that are always
moved into themselves by all the isometries in the group.
{\bf Fixed points} of a group of isometries are those points which
are left invariant by the isometries (thus the orbit of such a
point is just the point itself). These are the points where all
KV's vanish (however, the derivatives of the KV's there are
non-zero; the KV's generate isotropies about these points).
{\bf General points} are those where the dimension of the space
spanned by the KV's (i.e., the dimension of the orbit through
the point) takes the value it has almost everywhere; {\bf special
points} are those where it has a lower dimension (e.g., fixed
points). Consequently, the dimension of the orbits through special
points is lower than that of orbits through general points. The
dimension of the orbit and isotropy group
 is the same at each point of an orbit,
because of the equivalence of the group action at all points on
each orbit. \\

The group is {\bf transitive on a surface} $S$ (of whatever
dimension) if it can move any point of $S$ into any other point of
$S$. Orbits are the largest surfaces through each point on which
the group is transitive; they are therefore sometimes referred to
as {\bf surfaces of transitivity}. We define their dimension as
follows, and determine limits from the maximal possible initial
data for KV's:

{\bf dim surface of transitivity} $= s $, where in a space of
dimension $n$, $s \leq n$. \\

At each point we can also consider the dimension of the {\bf isotropy
group} (the group of isometries leaving that point fixed), generated
by all those KV's that vanish at that point:

{\bf dim of isotropy group} $= q$, where in a space of
dimension $n$, $q \leq \sfrac{1}{2}\,n\,(n-1)$. \\

The {\bf dimension $r$ of the group of isometries} of a space of
dimension $n$ is $r = s + q$ (translations plus rotations). From
the above limits , $0 \leq r \leq n + \sfrac{1}{2}\,n\,(n-1) =
\sfrac{1}{2}\,n\,(n+1)$ (the maximal number of translations and of
rotations). This shows the Lie algebra of KV's is finite
dimensional.\\

{\bf Maximal dimensions}: If $r = \sfrac{1}{2}\,n\,(n+1)$, we have
a space(-time) of constant curvature (maximal symmetry for a space
of dimension $n$). In this case,
\be
R_{ijkl} = K\,(\,g_{ik}\,g_{jl} - g_{il}\,g_{jk}\,) \ ,
\ee
with $K$ a constant; and $K$ necessarily {\it is} a constant if
this equation is true and $n \geq 3$. One cannot get $q =
\sfrac{1}{2}\,n\,(n-1)-1$ so $ r \neq \sfrac{1}{2}\,n\,(n+1)-1$.\\

A group is {\bf simply transitive} if $r = s ~\Leftrightarrow ~q =
0 $ (no redundancy: dimensionality of group of isometries is just
sufficient to move each point in a surface of transitivity into
each other point). There is no continuous isotropy group.

A group is {\bf multiply transitive} if $r > s ~\Leftrightarrow ~q
> 0 $ (there is redundancy in that the dimension of the group of
isometries is larger than is needed to move each point in an orbit
into each other point). There exist non-trivial isotropies.

\subsection{Classification of cosmological symmetries}
We consider non-empty perfect fluid models, i.e., (\r{eq:pf}) holds
with $(\mu+p) > 0$.
 
\begin{figure}
\begin{verbatim}
---------------------------------------------------------------------
                             Dim invariant variety
Dimension
Isotropy      s = 2                   s = 3                s = 4
Group
             inhomogeneous           spatially            space-time
                                     homogeneous          homogeneous

---------------------------------------------------------------------
q = 0     generic metric form known.    Bianchi:          Osvath/Kerr
            Spatially self-similar,      orthogonal,
aniso-      Abelian G_2 on 2-d           tilted
tropic       spacelike surfaces,
            non-Abelian G_2             

--------   -----------------------   ---------------    -------------
q = 1        Lemaitre-Tolman-        Kantowski-Sachs,      G"odel
LRS            Bondi family           LRS Bianchi

--------   -----------------------   ---------------    -------------
q = 3         none                      Friedmann          Einstein
isotropic     (cannot happen)                               static

---------------------------------------------------------------------
	            two non-ignorable      one non-ignorable   algebraic EFE
	            coordinates            coordinate          (no redshift)
---------------------------------------------------------------------
        
---------------------------------------------------------------------
                             Dim invariant variety
Dimension 
Isotropy      s = 0                   s = 1
Group
             inhomogeneous           inhomogeneous/no isotropy group    
---------------------------------------------------------------------
q = 0       Szekeres-Szafron,         General metric
            Stephani-Barnes,          form independent
            Oleson type N             of one coord;
                                       KV h.s.o./not h.s.o.
           The real universe!
 
---------------------------------------------------------------------
\end{verbatim}
\caption{Classification of cosmological models (with $(\mu+p) > 0)$ 
by isotropy and homogeneity.} 
\end{figure}
 
For a cosmological model, because space-time is 4-dimensional, the
possibilities for the dimension of the surface of transitivity are
$s = 0, 1, 2, 3, 4$. As to isotropy, we assume $(\mu+p) \neq 0$;
then $q = 3$, $1$, or $0$ because $u^a$ is invariant and so the
isotropy group at each point has to be a sub-group of the rotations
acting orthogonally to $u^a$ (and there is no 2-dimensional
subgroup of $O(3)$.) The dimension $q$ of the isotropy group can
vary over the space (but not over an orbit): it can be greater at
special points (e.g., an axis centre of symmetry) where the
dimension $s$ of the orbit is less, but $r$ (the dimension of the
total symmetry group) must stay the same everywhere. Thus the
possibilities for isotropy at a general point are:\\

a) {\bf Isotropic}: $q =3$, the Weyl curvature tensor vanishes,
 kinematical quantities vanish except $\Theta$. All observations
 (at every point) are isotropic. This is the FLRW family of
 space-time geometries;\\

b) {\bf Local Rotational Symmetry} (`LRS'): $q = 1$, the Weyl
 curvature tensor is of algebraic Petrov type D, kinematical
 quantities are rotationally symmetric about a preferred spatial
 direction. All observations at every general point are rotationally
 symmetric about this direction. All metrics are known in the case
 of dust \ct{ell67} and a perfect fluid (see \ct{steell68} and also
 \ct{veel96}). \\

c) {\bf Anisotropic}: $q=0$; there are no rotational
 symmetries. Observations in each direction are different from
 observations in each other direction.\\

Putting this together with the possibilities for the dimensions of
the surfaces of transitivity, we have the following possibilities
(see Figure 1):

\subsubsection{Space-time homogeneous models}
These models with $s = 4$ are unchanging in space and time, hence
$\mu$ is a constant, so by the energy conservation equation
(\r{eq:en}) they cannot expand: $\Th = 0$. Thus by (\r{eq:red2})
they cannot produce an almost isotropic redshift, and are not
useful as models of the real Universe. Nevertheless, they are of
some interest.

The isotropic case $q = 3$ ($\Rightarrow r = 7)$ is the
{\bf Einstein static universe}, the non-expanding FLRW model (briefly
mentioned above) that was the first relativistic cosmological model
found. It is not a viable cosmology inter alia because it has no
redshifts, but it laid the foundation for the discovery of the
expanding FLRW models. \\

{\it Exercise}: What other features make this space-time
problematic as a cosmological model?\\

The LRS case $q = 1$ ($\Rightarrow r = 5$) is the {\bf G\"{o}del
stationary rotating universe} \ct{god49}, also with no
redshifts. This model was important because of the new
understanding it brought as to the nature of time in General
Relativity (see \ct{he73,tce89,ell97}). Inter alia, it is a model
in which causality is violated (there exist closed timelike curves
through each space-time point) and there exists no cosmic time
function whatsoever.

The anisotropic models $q = 0$ ($\Rightarrow r = 4$) are all known,
\ct{osz65}, but are interesting only for the light they 
shed on Mach's principle; see \ct{osz62}.

\subsubsection{Spatially homogeneous universes}
These models with $s=3$ are the major models of theoretical
cosmology, because they express mathematically the idea of the
`cosmological principle': all points of space at the same time are
equivalent to each other \ct{bon60}.

The isotropic case $q = 3$ ($\Rightarrow r = 6)$ is the
family of {\bf FLRW models}, the standard models of cosmology
discussed above that have the matter-comoving metric
form (\r{eq:frw1}).

The LRS case $q = 1$ ($\Rightarrow r = 4)$ is the family of
{\bf Kantowski--Sachs universes} \ct{Kc65}--\ct{col77} plus the {\bf LRS
orthogonal} \ct{ellmac69} and {\bf LRS tilted} \ct{ke73} {\bf Bianchi
models}. The simplest are the Kantowski--Sachs family, with
matter-comoving metric form
\be
ds^2 = -\,dt^2 + A^2(t)\,dr^2 + B^2(t)\,
(\,d\theta^2 + f^2(\theta)\,d\phi^2\,) \ ,
\ee
where $f(\theta)$ is given by (\r{eq:fr}).

The anisotropic case $q = 0$ ($\Rightarrow r = 3)$ is the
family of {\bf orthogonal} and {\bf tilted Bianchi models} with
a group of isometries $G_{3}$ acting
simply transitively on spacelike surfaces.
The simplest class is the Bianchi Type I family,
discussed later in this section. The family as a whole has quite
complex properties; these models are discussed in the following
section.

\subsubsection{Spatially inhomogeneous universes}
These models have $s \leq 2$. 

The LRS cases ($q = 1 \Rightarrow s = 2, r = 3)$ are the 
spherically symmetric family with matter-comoving metric form
\be
ds^2 = -\,C^2(t,r)\,dt^2 + A^2(t,r)\,dr^2 + B^2(t,r)\,(\,d\theta^2
+ f^2(\theta)\,d\phi^2\,) \ ,
\ee
where $f(\theta)$ is given by (\r{eq:fr}). In the dust case, we can
set $C(t,r) = 1$ and can integrate Einstein's field equations
analytically; for $k = +\,1$, these are the spherically symmetric
{\bf Lema\^{\i}tre--Tolman--Bondi models} (`LTB')
\ct{lem33}--\ct{bon47}, discussed later in this section. They may
have a centre of symmetry (a timelike worldline), and can even
allow two such centres, but they
cannot be isotropic about a general point (because isotropy
everywhere implies spatial homogeneity; see the discussion of FLRW
models).

The anisotropic cases ($q = 0 \Rightarrow s \leq 2, r \leq
2$) include solutions admitting an Abelian or non-Abelian group of
isometries $G_{2}$, the $\boldsymbol{G_{2}}$ {\bf cosmologies},
and spatially self-similar models (see e.g. \ct{waiell97}).

Solutions with no symmetries at all have $r = 0 \Rightarrow s = 0,
q = 0$. The real Universe, of course, belongs to this class; all
the others are intended as approximations to this unique
Universe. Remarkably, we know some exact solutions without
symmetries, specifically (a) {\bf Szekeres' quasi-spherical
models} \ct{sze75a,sze75b}, that are in a sense non-linear FLRW
perturbations \ct{goowai82}, with matter-comoving metric form
\be
ds^2 = -\,dt^2 + e^{2A}\,dx^2 + e^{2B}(dy^2 + dz^2) \ ,
\hsp5 A = A(t,x,y,z) \ , \hsp5 B = B(t,x,y,z) \ ,
\ee
(b) {\bf Stephani's conformally flat models} \ct{ste87,kra83}, and (c)
{\bf Oleson's type N solutions} (for a discussion of these and all the
other inhomogeneous models, see Krasi\'{n}ski \ct{kra93} and Kramer
{\em et al\/} \ct{macetal}). One further interesting family without
global symmetries are the {\bf Swiss-Cheese models} made by cutting
and pasting segments of spherically symmetric models. These are
discussed below.\\

We now discuss the simplest useful anisotropic and inhomogeneous
models, before turning to the Bianchi models in the next section.

\subsection{Bianchi Type I universes ($s = 3$)}
These are the simplest anisotropically expanding Universe models.
The metric can be given in matter-comoving coordinates in the form
\ct{hs62}
\be
ds^{2} = -\,dt^2 + X^2(t)\,dx^2 + Y^2(t)\,dy^2 + Z^2(t)\,dz^2 \ , 
\hsp5 u^a = \delta^a{}_0 \ . 
\ee
This is the simplest generalisation of the spatially flat FLRW
models to allow for different expansion factors in three orthogonal
directions; the corresponding {\bf average expansion scale
factor} is $S(t) = (XY\!Z)^{1/3}$.  They are spatially
homogeneous, being invariant under an Abelian group of isometries
$G_{3}$, simply transitive on spacelike surfaces $\{t =
\mbox{const}\}$, so $s = 3$; in general $q = 0 \Rightarrow r =3$,
but there are LRS and isotropic subcases (the latter being the
Einstein--de Sitter universe). The space sections $\{t =
\mbox{const}\}$ are flat (when $t = t_0$, all the metric
coefficients are constant), and all invariants depend only on the
time coordinate $t$. The fluid flow (orthogonal to these
homogeneous surfaces) is necessarily geodesic and
irrotational. Thus these models obey the restrictions
\be
0 = \udot^a = \om^a \ , \hsp5 0 = X_a = Z_a = \3nab_{a}p \ ,
\hsp5 0 = {}^3\!R_{ab}\ .
\ee
The $1+3$ covariant equations obeyed by these models follow from
the $1+3$ covariant equations in subsection \r{subsec:13ceqs} on
making these restrictions. We can find a tetrad in the obvious way
from the above coordinates ($e_1{}^i = X(t)^{-1}\,\delta_1{}^i$,
etc.); then the tetrad equations of the subsection
\r{subsec:13teqs} hold with
\be
0 = \udot^{\alpha} = \om^{\alpha} = \Omega^{\alpha} \ , \hsp5
0 = a^{\alpha} = n_{\alpha\beta} \ , \hsp5
0 = \vec{e}_{\alpha}(\Th) = \vec{e}_{\alpha}(\sig_{\beta\gam}) \ , \hsp5
0 = \vec{e}_{\alpha}(\mu) = \vec{e}_{\alpha}(p) \ .
\ee
It follows that the $(0\alpha)$-equation (\r{eq:onu}), which is
$(C_1)^\alpha$ in the tetrad form, is identically satisfied, and
also that $H_{ab} = 0$ and $\3nab_{b}E^{ab} = 0$. From the Gauss
embedding equation (\r{eq:3rab}), the shear obeys
\be
\l{eq:sig}
(S^3\sigma_{ab})\,\dot{} = 0 \hsp5 \Rightarrow \hsp5
\sigma_{ab} = {\Sigma_{ab} \over
S^3} \ , \hsp5 (\Sigma_{ab})\,\dot{} = 0 \ ,
\ee
which implies 
\be
\l{eq:sig2}
\sigma^2 = {\Sigma^2 \over S^6} \ , \hsp5
\Sigma^2 = \sfrac12\,\Sigma_{ab}\Sigma^{ab} \ , \hsp5
(\Sigma^2)\,\dot{} = 0 \ .
\ee
All of Einstein's field equations will then be satisfied if the
conservation equation (\r{eq:en}), the Raychaudhuri equation
(\r{eq:ray}), and the Friedmann-like equation (\r{eq:3R}) are
satisfied. As in the
FLRW case, the latter is the first integral of the other two.
Assuming a $\gamma$-law equation of state, (\r{eq:en2}) will be
satisfied and, using (\r{eq:sig2}), equation (\r{eq:3R}) becomes
the generalised Friedmann equation,
\be
\l{eq:fried2}
3\,{\dot{S}^2 \over S^2} = {\Sigma^2 \over S^6}
+ {M \over S^{3\gamma}} \ .
\ee 
This shows that no matter how small the shear today, it will (for
ordinary matter) dominate the very early evolution of the Universe
model, which will then approximate {\bf Kasner's vacuum solution}
\ct{macetal}. \\

On writing out the tetrad components of the shear equation (\r{eq:sig}), 
using the commutator relations (\r{onfcomts}) to determine the shear 
components, one finds that the individual length scales are given by,
$$X(t) = S(t)\,\exp(\Sigma_{1}\,W(t)) \ , \hsp5
Y(t) = S(t)\,\exp(\Sigma_{2}\,W(t)) \ , \hsp5
Z(t) = S(t)\,\exp(\Sigma_{3}\,W(t)) \ ,$$
where
\be
\l{eq:w}
W(t) = \int {dt \over S^3(t)} \ ,
\ee
and the constants $\Sigma_\alpha$ satisfy
$$
\Sigma_1 + \Sigma_2 + \Sigma_3 = 0 \ , \hsp5
\Sigma_1^2 + \Sigma_2^2 + \Sigma_3^2 = 2 \Sigma^2 \ .
$$
These relations can be satisfied by setting
\be
\Sigma_\alpha = (2\Sigma/3) \sin\,\alpha_\alpha \ , \hsp5
\alpha_1 = \alpha \ , \hsp5
\alpha_2 = \alpha + {2\pi \over 3} \ , \hsp5
\alpha_3 = \alpha + {4\pi \over 3} \ ,
\l{alpha}
\ee
and $\alpha$ is a constant. Thus, the solution is given by choosing
a value for $\gamma$, and then integrating successively
(\r{eq:fried2}) and (\r{eq:w}). \\

{\it Exercise}: Show that, on using the obvious tetrad associated
with the coordinates above, all the tetrad (and $1+3$ covariant)
equations are then satisfied.\\

For example, in the case of dust
($\gamma = 1$) we have:
$$S(t) = (\,{9 \over 2}\,M\,t^2 + \sqrt{3}\Sigma\,t\,)^{1/3} \ ,
\hsp5
W(t) = {1 \over \sqrt{3}\Sigma}\,\ln\left(\,{t \over {3\over 4}Mt 
+ \sqrt{3}\Sigma}\,\right) \ ,$$
so
$$X(t) = S(t) \left({t^2 \over S(t)^3}\right)^{{2\over
3}\sin\alpha_1} \ , \hsp5
Y(t) = S(t) \left({t^2 \over S(t)^3}\right)^{{2\over
3}\sin\alpha_2} \ , \hsp5
Z(t) = S(t) \left({t^2 \over S(t)^3}\right)^{{2\over
3}\sin\alpha_ 3} \ .$$
The generic case is anisotropic; LRS cases occur when $\alpha =
\pi/6$ and $\alpha =\pi/2$ in (\r{alpha}), and isotropy when
$\Sigma = 0$. 

At late times this isotropizes to give the Einstein--de Sitter
model, and, hence, as mentioned above, can be a good model of the
real Universe if $\Sigma$ is chosen appropriately. However, at
early times, the situation is quite different. As $t \rightarrow
0$, provided $\Sigma \neq 0$, then
$S(t) \rightarrow (\sqrt{3}\Sigma)^{1/3}\,t^{1/3}$ and
$$
X(t) \rightarrow X_0\,t^{{1\over 3}(1 + 2\sin\alpha_1)} \ ,
\hsp5
Y(t) \rightarrow Y_0\,t^{{1 \over 3}(1 + 2\sin\alpha_2)} \ ,
\hsp5
Z(t) \rightarrow Z_0\,t^{{1 \over 3}(1 + 2\sin\alpha_3)} \ .
$$
Plotting the function $f(\alpha) = {2 \over 3}\,(\sfrac12 +
\sin\alpha)$, we see that the generic behaviour occurs for $\alpha
\neq \pi/2$; in this case two of the powers are positive but one is
negative, so going backwards in time, the collapse along the
preferred axis reverses and changes to a (divergent) expansion,
while collapse continues (divergently) along the two orthogonal
direction; the singularity is a {\bf cigar singularity}. Going
forward in time, a collapse along the preferred axis stops and
reverses to become an expansion. 
However when $\alpha = \pi/2$, one exponent is positive but the
other two are zero. Hence, going back in time, collapse continues
divergently along the preferred direction in these LRS solutions
back to the singularity, but in the orthogonal directions it slows
down and halts; this is a {\bf pancake singularity}.  An important
consequence in this special case is that particle horizons are
broken in the preferred direction --- communication is possible
to arbitrary distance in a cylinder around this axis \ct{he73}. \\

One can work out detailed observational relations in these models.
Because of the high symmetry, the null geodesics can be found
explicitly; those along the three preferred axes are particularly
simple. Redshift along each of these axes simply scales with the
expansion ratio in that direction. Area distances can be found
explicitly \ct{tom68,macell70}.  An interesting feature is that all
observations will show an eight-fold discrete isotropy symmetry
about the preferred axes. One can also work out helium production
and CBR anisotropy in these models, following the pioneering paper
by Thorne \ct{tho67}. Because the shear can dominate the dynamics
at nucleosynthesis or baryosynthesis time, causing a speeding up of
the expansion, one can get quite different results than in the FLRW
models. Consequently, one can use the nucleosynthesis observations
to limit the shear constant $\Sigma$, but still allowing extra
freedom at the time of baryosynthesis. The CBR quadrupole
anisotropy will directly measure the difference in expansion along
the three principal axes since last scattering, and, hence, may
also be used to limit the anisotropy parameter $\Sigma$.
Nucleosynthesis gives stronger limits, because it probes to earlier
times. These models have also been investigated in the case of
viscous fluids and kinetic theory solutions (Misner \ct{mis67}),
with electromagnetic fields, and also the effects of `reheating' on
the CBR anisotropy and spectrum have been examined; see Rees
\ct{ree67}. \\

Thus, these models can have arbitrarily small shear at the present
day, thus can be arbitrarily close to an Einstein--de Sitter
universe since decoupling, but can be quite different early on.\\

{\it Exercise}: Show how the solutions will be altered by (i) a
fluid with simple viscosity: $\pi_{ab} = -\,\eta\,\sigma_{ab}$ with
constant viscocity coefficient $\eta$, (ii) freely propagating
neutrinos \ct{mis67}.
 
\subsection{Lema\^{\i}tre--Tolman--Bondi family ($s = 2$)}
The simplest inhomogeneous models are those that are spherically
symmetric. In general they are time-dependent, with 2-dimensional
spherical surfaces of symmetry: $s =2$, $q = 1$ $\Rightarrow r=3$.
The geometry of this family (including the closely related models
with flat and negatively curved 2-surfaces of symmetry), is
examined in a $1+3$ covariant way by van Elst and Ellis
\ct{veel96}, and a tetrad analysis is given by Ellis \ct{ell67}
(the pressure-free case) and Stewart and Ellis \ct{steell68} (for
perfect fluids). Here we only consider the dust case, because then
a simple analytic solution is possible; the perfect fluid case
includes spherical stellar models and collapse solutions (see,
e.g., Misner, Thorne and Wheeler \ct{mtw73}). \\

The general spherically symmetric metric for an irrotational dust
matter source in synchronous matter-comoving coordinates is the
Lema\^{\i}tre--Tolman--Bondi (`LTB') metric \ct{lem33}--\ct{bon47}
\be
\l{eq:met1}
ds^2 = -\,dt^2 + X^2(t,r)\,dr^2 + Y^2(t,r)\,d\Omega^2 \ , \hsp5
u^a = \delta^a{}_0 \ .
\ee
The function $Y = Y(t,r)$ is the {\bf areal radius}, since the
proper area of a sphere of coordinate radius $r$ on a time
slice of constant $t$ is $4\pi Y^2$ (upon re-establishing
factors of $\pi$). Solving Einstein's field equations \ct{bon47}
shows
\be 
\l{eq:met} 
ds^2 = -\,dt^2 + \frac{[\,Y'(t,r)\,]^2}{1 + 2E(r)}\,dr^2
+ Y^2(t,r)\,d\Omega^2 \ ,
\ee 
where $Y^{\prime}(t,r) = \p Y(t,r)/\p r$, and
$d\Omega^2 = d\theta^2 + \sin^2\theta\,d\phi^2$, with
$Y(t,r)$ obeying a generalised Friedmann equation,
\be 
\l{eq:Rdot} 
\dot{Y}(t,r) = \pm\,\sqrt{\,{2\,M(r) \over Y(t,r)} + 2\,E(r)\,} \ ,
\ee 
and the energy density given by
\be
\l{eq:density}
4\pi\mu(t,r) = {M^{\prime}(r) \over {Y^2(t,r)\,Y'(t,r)}} \ .
\ee
Equation (\r{eq:Rdot}) can be solved in terms of a parameter $\eta = 
\eta(t,r)$:
\be
\l{eq:R}
Y(t,r) = \frac{M(r)}{{\cal E}(r)}\,\phi_0(t,r) \ , \hsp5 
\xi(t,r) = \frac{[\,{\cal E}(r)\,]^{3/2}\,(t - t_B(r))}{M(r)} \ ,
\ee
where\footnote{Strictly speaking, the hyperbolic, parabolic and
elliptic solutions obtain when $YE/M$ $> 0$, $=0$ and $<0$,
respectively, since $E=0$ at a spherical origin in both hyperbolic
and elliptic models.}
\be
\l{eq:xi_phi0}
    {\cal E}(r) = \left\{
      \begin{array}{l}
               2 E(r), \\
               1, \\
               - 2 E(r), 
      \end{array}
             \right.
   ~~~ \phi_0 = \left\{
      \begin{array}{l}
               \cosh \eta - 1, \\
               (1/2) \eta^2, \\
               1 - \cos \eta, 
      \end{array}
             \right.
   ~~~ \xi = \left\{
         \begin{array}{l}
               \sinh \eta - \eta, \\
               (1/6) \eta^3, \\
               \eta - \sin \eta,
         \end{array}
             \right.
   ~~~ \mbox{when~} \left\{
         \begin{array}{l}
               E > 0 \\
               E = 0 \\
               E < 0
         \end{array}
             \right. \ ,
\ee
for hyperbolic, parabolic and elliptic solutions, respectively.\\

The LTB model is characterised by {\em three arbitrary functions\/}
of the matter-comoving coordinate radius $r$. $E = E(r) \geq -1$
has a geometrical role, determining the local `embedding angle' of
spatial slices, and also a dynamical role, determining the local
energy per unit mass of the dust particles, and, hence, the type of
evolution of $Y$. $M = M(r)$ is the effective gravitational mass
within coordinate radius $r$. $t_B = t_B(r)$ is the local time at
which $Y = 0$, i.e., the local time of the big bang --- we have a
non-simultaneous bang surface.  Specification of these three
arbitrary functions --- $M(r)$, $E(r)$ and $t_B(r)$ --- fully
determines the model, and whilst all have some type of physical or
geometrical interpretation, they admit a freedom to choose the
radial coordinate, leaving {\em two physically meaningful
choices\/}, e.g., $r = r(M)$, $E = E(M)$, $t_B = t_B(M)$. For
particular choices of this initial data, one obtains FLRW models,
which, of course, are special cases of these spherical models
with very specific initial data. In fact, the FLRW models are
obtained if one sets
\be
2E(r) = -\,k\,r^2 \ , \hsp5 Y(t,r) = S(t)\,r \ , \hsp5
M(r) = {4\pi \over 3}\,\mu(t)\,Y^3 \ .
\ee

The LTB models have been used in a number of interesting ways in
cosmology: \\

* to give simple models of structure formation \ct{bon72,bon74},
 e.g., by looking at evolution of a locally open region in a
 closed universe \ct{zel84} and evolution of density
 contrast \ct{mentav98},

* to give Universe models that are inhomogeneous on a cosmological
  scale \ct{sil77,kan69b},

* to examine inhomogeneous big bang structures \ct{hellak84},

* to examine CBR anisotropies \ct{rai81}--\ct{pan92},

* to investigate observational conditions for spatial homogeneity
  \ct{bonell86}--\ct{humetal98},

* to trace the effect of averaging on spatial inhomogeneities
  \ct{hel88}, and

* to look at the relationship between cosmic evolution and closure
  of the Universe \ct{hellak85}. \\

These aspects are discussed in Krasi\'{n}ski's book \ct{kra93},
Part III. Here we will only summarize one interesting result:
namely, regarding observational tests of whether the real Universe
is more like a LTB inhomogeneous model, or a FLRW model. In
Mustapha, Hellaby and Ellis \ct{muheel98}, the following result is
shown:

\begin{quotation}
{\bf Isotropic Observations Theorem (1)}: Any given isotropic set
of source observations $n(z)$ and $m(z)$, together with any given
source luminosity and number evolution functions $L(z)$ and $N(z)$,
can be fitted by a spherically symmetric dust cosmology --- a LTB
model --- in which observations are spherically symmetric about us
because we are located near the central worldline.
\end{quotation}

This shows that any spherically symmetric observations we may
eventually make can be accommodated by appropriate inhomogeneities
in a LTB model --- irrespective of what source evolution may occur.
In particular, one can find such a model that will fit the
observations if there is no source evolution. The following result
also holds:\\

\begin{quotation}
{\bf Isotropic Observations Theorem (2)}: Given any spherically
symmetric geometry and any spherically symmetric set of
observations, we can find evolution functions that will make the
model compatible with the observations. This applies in particular
if we want to fit observations to a FLRW model.
\end{quotation}

The point of the first result is that it shows that these models
--- spherical inhomogeneous generalisations of the FLRW models ---
are viable models of the real Universe since decoupling, because
they cannot be observationally disproved (at least not in any
simple way). The usual response to this is: but the FLRW models are
confirmed by observations. The second result clarifies this: yes
they are, provided you allow an evolution function to be chosen
specifically so that the initially discrepant number counts fit the
FLRW model predictions. That is, we {\it assume} the FLRW geometry,
and then determine what source evolution makes this assumption
compatible with observations \ct{ell75}. Without this freedom, the
FLRW models are contradicted by, e.g., radio source observations,
which without evolution are better fitted by a spatially flat
(Euclidean) model. Thus, the FLRW model fit is obtained only because
of this freedom, allowed because we do not understand source
luminosity and number evolution. The inhomogeneous LTB models
provide an alternative understanding of the data; the observations
do not contradict them. \\

{\it Exercise}: Suppose (a) observations are isotropic, and (b) we
knew the source evolution function and selection function, and (c)
we were to observationally show that after taking them into
account, the observer area distance relation has precisely the
FLRW form (\r{eq:ro}) for $\beta=1$ and the number count relation
implies the FLRW form (\r{eq:num}). Assuming the space-time matter
content is dust, (i) prove from this that the space-time is a
FLRW space-time. Now (ii) explain the observational difficulties
that prevent us using this exact result to prove spatial
homogeneity in practice.\\

An alternative approach to proving homogeneity is via the
{\bf Postulate of Uniform Thermal Histories} (`PUTH') --- i.e.,
the assumption that because we see similar kinds of objects at
great distances and nearby, they must have had similar thermal
histories. One might then hope that from this one could deduce
spatial homogeneity of the space-time geometry (for otherwise
the thermal histories would have been different). Unfortunately,
the argument here is not watertight, as can be shown by a
counterexample based on the LTB models (Bonnor and Ellis
\ct{bonell86}). Proving --- rather than assuming --- spatial
homogeneity remains elusive. We cannot observationally disprove
spatial inhomogeneity. However, we can give a solid argument for
it via the {\bf EGS theorem} discussed below.

\subsection{Swiss-Cheese models}
Finally, an interesting family of inhomogeneous models is the
Swiss-Cheese family of models, obtained by repeatedly cutting out a
spherical region from a FLRW model and filling it in with another
spherical model: Schwarzschild or LTB, for example. This
requires: \\

(i) locating the 3-dimensional timelike {\bf junction surfaces}
$\Sigma_\pm$ in each of the two models; \\

(ii) defining a proposed {\bf identification} $\Phi$ between
$\Sigma_+$ and $\Sigma_-$; \\

(iii) determining the {\bf junction conditions} that (a) the
3-dimensional metrics of $\Sigma_+$ and $\Sigma_-$ (the {\bf first
fundamental forms} of these surfaces) be isometric under this
identification, so that there be no discontinuity when we glue them
together --- we arrive at the same metric from both sides --- and
(b) the {\bf second fundamental forms} of these surfaces (i.e., the
covariant first derivatives of the 3-dimensional metrics along the
spacelike normal directions) must also be isometric when we make
this identification, so that they too are continuous in the
resultant space-time --- equivalently, there is no discontinuity in
the direction of the spacelike unit normal vector as we cross the
junction surface $\Sigma$ (this is the condition that there be no
surface layer on $\Sigma$ once we make the join; see Israel
\ct{isr66}).

Satisfying these junction conditions involves deciding how the
3-dimensional junction surfaces $\Sigma_\pm$ should be placed in
the respective background space-times. It follows from them that 4
of the 10 components of $T_{ab}$ must be continuous: if $n^{a}$
denotes the spacelike (or, in some other matching problems,
timelike) unit normal to $\Sigma$ and $p_{a}{}^{b}$ the tensor
projecting orthogonal to $n^{a}$, then $(T_{ab}n^{a}n^{b})$ and
$T_{bc}\,n^{b}p_{a}{}^{c}$ must be continuous, but the other 6
components $T_{cd}\,p_{a}{}^{c}p_{b}{}^{d}$ can be discontinuous
(thus at the surface of a star, in which case $n^{a}$ will be
spacelike, the pressure is continuous but the energy density can be
discontinuous). Conservation of the energy-momentum tensor across
the junction surface $\Sigma$ will then be satisfied by the
constraints $0 = (G_{ab}-T_{ab})\,n^{a}n^{b}$ and $0 =
(G_{bc}-T_{bc})\,n^{b}p_{a}{}^{c}$. \\

(iv) Having determined that these junction conditions can be
satisfied for some particular identification of points, one can
then proceed to identify these corresponding points in the two
surfaces $\Sigma_+$ and $\Sigma_-$, thus gluing an interior
Schwarzschild part to an exterior FLRW part, for example. Because
of the reciprocal nature of the junction conditions, it is then
clear we could have joined them the other way also, obtaining a
well-matched FLRW interior and Schwarzschild exterior. \\

(v) One can continue in this way, obtaining a family of holes of
 different sizes in a FLRW model with different interior fillings,
 with further FLRW model segments fitted into the interiors of
 some of these regions, obtaining a Swiss-Cheese model.  One can
 even obtain a hierarchically structured family of spherically
 symmetric vacuum and non-vacuum regions in this way.\\

It is important to note that one {\em cannot\/} match arbitrary
masses. It follows from the junction conditions that the
Schwarzschild mass in the interior of a combined
FLRW--Schwarzschild solution must be the same as the mass that has
been removed: $M_{\rm Schw} = (4\pi/3)\, (\mu\,S^3\,r^3)_{\rm
FLRW}$. If the masses were wrongly matched, there would be an
excess gravitational field from the mass in the interior that would
not fit the exterior gravitational field, and the result would be
to distort the FLRW geometry in the exterior region --- which then
would no longer be a FLRW model. Alternatively viewed, the reason
this matching of masses is needed is that otherwise we will have
fitted the wrong background geometry to the inhomogeneous
Swiss-Cheese model --- averaging the masses in that model will not
give the correct background average
\ct{elljak}, and they could not have arisen from rearranging
uniformly distributed masses in an inhomogeneous way (this is the
content of {\bf Traschen's integral constraints}
\ct{tra}). Consequently, there can be {\it no} long-range effects
of such matching: the Schwarzschild mass cannot cause large-scale
motions of matter in the FLRW region.\\

These models were originally developed by Einstein and Straus
\ct{eist47} (see also Sch\"{u}cking \ct{sch54}) to examine the
effect of the expansion of the Universe on the solar system (can we
measure the expansion of the Universe by laser ranging within the
solar system?). Their matching of a Schwarzschild interior to a
FLRW exterior showed that this expansion has no effect on the
motion of planets in the Schwarzschild region. It does not,
however, answer the question as to {\it where} the boundary between
the regions should be placed --- which determines which regions are
affected by the universal expansion. Subsequent uses of these
models have included: \\

* examining Oppenheimer--Snyder collapse in an expanding
 universe \ct{lak80}--\ct{hellak83},

* examining gravitational lensing effects on area distances
 \ct{kan69a},

* investigating CBR anisotropies \ct{reesci68}--\ct{meszmol96},

* modelling voids in large-scale structure \ct{boncho90,cha91},
perhaps using surface-layers \ct{lake87},

* modelling the Universe as a patchwork of domains of different
 curvature $k = 0, \pm\,1$ \ct{har92}. \\

{\it Exercise}: Show how appropriate choice of initial data in a
LTB model can give an effective Swiss-Cheese model with one centre
surrounded by a series of successive FLRW and non-FLRW spherical
regions. Can you include (i) flat, (ii) vacuum (Schwarzschild)
regions in this construction?\\

One of the most intriguing questions is what non-spherically
symmetric models can be joined regularly onto a FLRW model. Bonnor
has shown that some Szekeres anisotropic and inhomogeneous models
can be matched to a dust FLRW model across a matter-comoving
spherical junction surface \ct{bonnor76}. Dyer {\em et al\/}
\ct{dye93} have shown that one can match FLRW and LRS Kasner
(anisotropic vacuum Bianchi Type I) models across a flat
3-dimensional timelike junction surface. Optical properties of
these Cheese-slice models have been investigated in depth
\ct{landye97}.

\section{Bianchi models} 
These are the models in which there is a group of isometries $G_3$
acting simply transitively on spacelike surfaces
$\{t = \mbox{const}\}$, so they are spatially homogeneous.
There is only {\em one\/}
essential dynamical coordinate (the time $t$), and Einstein's
field equations reduce to {\bf ordinary differential equations},
because the inhomogeneous degrees of freedom have been `frozen
out'. They are thus quite special in
geometrical terms; nevertheless, they form a rich set of models
where one can study the exact dynamics of the full non-linear field
equations. The solutions to the EFE will depend on the matter in
the space-time. In the case of a fluid (with uniquely defined flow
lines), we have two different kinds of models:\\

{\bf Orthogonal models}, with the fluid flow lines orthogonal to
the surfaces of homogeneity (Ellis and MacCallum \ct{ellmac69}, see
also \ct{waiell97});

{\bf Tilted models}, with the fluid flow lines not orthogonal to
the surfaces of homogeneity; the components of the fluid's
{\bf peculiar velocity} enter as further variables (King and Ellis
\ct{ke73}, see also \ct{colell79}). \\

Rotating models {\it must} be tilted (\,cf. Eq. (\r{eq:tilt})\,),
and are much more complex than non-rotating models.

\subsection{Constructing Bianchi models}
There are essentially three direct ways of constructing the
{\bf orthogonal models}, all based on properties of a tetrad of
vectors $\{\,\vec{e}_{a}\,\}$ that commute with the basis of KV's
$\{\,\xi_\alpha\,\}$, and usually with the timelike basis vector
chosen parallel to the unit normal $n_a = -\,\nabla_{a}t$ to the
surfaces of homogeneity, i.e., $\vec{e}_{0} = \vec{n}$. \\

The {\it first approach} (Taub \ct{tau51}, Heckmann and
Sch\"{u}cking \ct{hs62}) puts all the time variation in the
{\bf metric components}:
\be
ds^2 = -\,dt^2 + \gam_{\alpha\beta}(t)\,
(e^\alpha{}_i(x^\gam)\,dx^i)\,(e^\beta{}_j(x^\delta)\,dx^j) \ ,
\ee
where $e^\alpha{}_i(x^\gam)$ are 1-forms inverse to the spatial
triad vectors $e_\alpha{}^i(x^\gam)$, which have the same
commutators $C^\alpha{}_{\beta\gamma}$, $\alpha,\beta,\gamma, \dots
= 1,2,3$, as the structure constants of the group of isometries and
commute with the unit normal vector $\vec{e}_{0}$ to the surfaces of
homogeneity; i.e., $\vec{e}_{\alpha} = e_{\alpha}{}^{i}\,\p /\p x^i$,
$\vec{e}_{0} = \p /\p t$ obey the commutator relations
\be
[\,\vec{e}_{\alpha},\,\vec{e}_{\beta}\,] = C^\gamma{}_{\alpha\beta}\,
\vec{e}_{\gamma}
\ , \hsp5 [\,\vec{e}_{0},\,\vec{e}_{\alpha}\,] = \vec{0} \ ,
\ee
where the $C^\gamma{}_{\alpha\beta}$ are the Lie algebra structure
constants satisfying the Jacobi identities (\r{eq:jac3}). Einstein's
field equations (\r{eq:efe}) become ordinary differential equations
for the metric functions $\gamma_{\alpha\beta}(t)$.\\

The {\it second approach} is based on use of the {\bf automorphism
group} of the isometry group with time-dependent parameters.
We will not consider it further here (see Collins and Hawking
\ct{colhaw73b}, Jantzen \ct{jan79,jan84} and Wainwright and
Ellis \ct{waiell97} for a discussion).\\

The {\it third approach} (Ellis and MacCallum \ct{ellmac69}), which
is in our view the preferable one, uses an {\bf orthonormal tetrad}
based on the normals to the surfaces of homogeneity (i.e.,
$\vec{e}_{0} = \vec{n}$, the unit normal vector to these surfaces).
The tetrad is chosen to be invariant under the group of isometries,
i.e., the tetrad vectors commute with the KV's, and the metric
components in the tetrad are space-time constants, $g_{ab} =
\eta_{ab}$; now the dynamical time variation is in the commutation
functions for the basis vectors, which then determine the time-(and
space-)dependence in the basis vectors themselves. Thus, we have an
orthonormal basis $\{\,\vec{e}_{a}\,\}_{a = 0,1,2,3}$, such that
\be
[\,\vec{e}_{a},\,\vec{e}_{b}\,] = \gamma^c{}_{ab}(t)\,\vec{e}_{c} \ .
\ee
The commutation functions $\gamma^a{}_{bc}(t)$, together with the
matter variables, are then treated as the dynamical variables.
Einstein's field equations (\r{eq:efe}) are first-order equations
for these quantities, supplemented by the Jacobi identities
for the $\gamma^a{}_{bc}(t)$, which are also first-order equations.
It is sometimes useful to introduce also the Weyl curvature
components as auxiliary variables, but this is not necessary in
order to obtain solutions. Thus the equations needed are just the
tetrad equations given in section \r{sec:onf}, specialised to the
case
\be
\l{eq:bia}
\udot^{\alpha} = \om^{\alpha} = 0
= \vec{e}_{\alpha}(\gamma^a{}_{bc}) \ .
\ee
The spatial commutation functions
$\gamma^\alpha{}_{\beta\gamma}(t)$ can be decomposed into a
time-dependent matrix $n_{\alpha\beta}(t)$ and a vector
$a^{\alpha}(t)$ (see (\r{onfgam4})), and are equivalent to the
structure constants $C^\alpha{}_{\beta\gamma}$ of the symmetry
group at each point.\footnote{That is, they can be brought to the
canonical forms of the $C^\alpha{}_{\beta\gamma}$ by a suitable
change of group-invariant basis (the final normalisation to $\pm1$
may require changing from normalised basis vectors); the
transformation to do so is different at each point and at each
time.} In view of (\r{eq:bia}), the Jacobi identities (\r{onfjac})
now take the simple form
\be
n^{\alpha\beta}\,a_{\beta} = 0 \ .
\ee
The tetrad basis can be chosen to diagonalise $n_{\alpha\beta}$ to
attain $n_{\alpha\beta} = \mbox{diag}\,(n_1, n_2, n_3)$ and to set
$a^\alpha = (a,0,0)$, so that the Jacobi identities are then simply
$n_1\,a = 0$. Consequently, we define two major classes of structure
constants (and so Lie algebras):\\

{\bf Class A}: $a = 0$ \ ,

{\bf Class B}: $a \neq 0$ \ . \\

Following Sch\"{u}cking, the adaptation of the Bianchi
classification of $G_3$ isometry group types used is as in Figure 2.
Given a specific group type at one instant, this type will be
preserved by the evolution equations for the quantities
$n_\alpha(t)$ and $a(t)$. This is a consequence of a generic
property of Einstein's field equations: they will always preserve
symmetries in initial data (within the Cauchy development of that
data); see Hawking and Ellis \ct{he73}.
\\

\begin{figure}
\begin{verbatim}
---------------------------------------------------------------
Class    Type       n_1    n_2    n_3    a
---------------------------------------------------------------
A         I          0      0      0     0         Abelian
        ------------------------------------
         II         +ve     0      0     0
         VI_0        0     +ve    -ve    0
        VII_0        0     +ve    +ve    0
        VIII        -ve    +ve    +ve    0
         IX         +ve    +ve    +ve    0

---------------------------------------------------------------
B         V          0      0      0    +ve
        ------------------------------------
         IV          0      0     +ve   +ve
         VI_h        0     +ve    -ve   +ve       h < 0
         III         0     +ve    -ve   n2n3   same as VI_1
         VII_h       0     +ve    +ve   +ve       h > 0

---------------------------------------------------------------

\end{verbatim}
\caption{Canonical structure constants for different Bianchi types.
The Class B parameter $h$ is defined as $h = a^2/n_2n_3$ (see,
e.g., \ct{waiell97}).}
\end{figure}

In some cases, the Bianchi groups allow higher symmetry subcases:
isotropic (FLRW) or LRS models. Figure 3 gives the
Bianchi isometry groups admitted by FLRW and LRS solutions
\ct{ellmac69}, i.e., these are the simply transitive 3-dimensional
subgroups allowed by the full $G_6$ of isometries (in the FLRW
case) and the $G_4$ of isometries (in the LRS case). The only LRS
models not allowing a simply transitive subgroup $G_3$ are the
Kantowski--Sachs models for $k = 1$.

\begin{figure}
\begin{verbatim}
===========================================================
                       Isotropic Bianchi models

 FLRW k = +1: Bianchi IX [two commuting groups]
 FLRW k =  0: Bianchi  I, Bianchi VII_0
 FLRW k = -1: Bianchi  V, Bianchi VII_h

===========================================================
                          LRS Bianchi models

Orthogonal             c = 0                c \neq 0

   Taub-NUT I        [KS +1: no subgroup]   Bianchi IX
   Taub-NUT 3        Bianchi I, VII_0       Bianchi II
   Taub-NUT 2        Bianchi III [KS -1]    Bianchi VII_h,
                                            III
Tilted

                     Bianchi V, VII_h
                     Farnsworth,
                     Collins-Ellis

===========================================================
\end{verbatim}
\caption{The Bianchi models permitting higher symmetry
subcases. The parameter $c$ is zero iff the preferred spatial
vector is hypersurface-orthogonal.}
\end{figure}

{\bf Tilted models} can be constructed similarly, as discussed
below, or by using non-orthogonal bases in various ways \ct{ke73};
those possibilities will not be pursued further here.

\subsection{Dynamics of Bianchi models}
The set of tetrad equations (section 3) with these restrictions
will determine the evolution of all the commutation functions and
matter variables, and, hence, determine the metric and also the
evolution of the Weyl curvature (these are regarded as auxiliary
variables). In the case of {\bf orthogonal models} --- the fluid
4-velocity $\vec{u}$ is {\em parallel\/} to the normal vectors
$\vec{n}$ --- the matter variables will be just the fluid density
and pressure \ct{ellmac69}; in the case of {\bf tilted models} ---
the fluid 4-velocity $\vec{u}$ is {\em not parallel\/} to the
normal vectors $\vec{n}$ --- we also need the {\bf peculiar
velocity} of the fluid relative to the normal vectors \ct{ke73},
determining the fluid energy--momentum tensor decomposition
relative to the normal vectors (a perfect fluid will
appear as an imperfect fluid in that frame). Various papers relate
these equations to variational principles and {\bf Hamiltonian
formalisms}, thus expressing them in terms of a potential formalism
that gives an intuitive feel for what the evolution will be like
\ct{mm73,mm79}. There have also been many numerical investigations
of these dynamical equations and the resulting solutions. We will
briefly consider three specific aspects here, then the relation to
observations, and finally the related {\bf dynamical systems
approach}.

\subsubsection{Chaos in these universes?}
An ongoing issue since Misner's discovery of the `Mixmaster'
behaviour of the Type IX universes has been whether or not these
solutions show chaotic behaviour as they approach the initial
singularity (see \ct{hobill} and Hobill in \ct{waiell97}). The
potential approach represents these solutions as bouncing in an
expanding approximately triangular shaped potential well, with
three deep troughs attached to the corners. The return map
approximation (a series of Kasner-like epochs, separated by
collisions with the potential walls and consequent change of the
Kasner parameters) suggests the motion is chaotic, but the question
is whether this map represents the solutions of the differential
equations well enough to reach this conclusion (for example, the
potential walls are represented as flat in this approximation; and
there are times when the solution moves up one of the troughs and
then reflects back, but the return map does not represent this part
of the motion). Part of the problem is that the usual definitions
of chaos in terms of a Lyapunov parameter depend on the
definition of time variable used, and there is a good case for
changing to conformal Misner time in these investigations.\\

The issue may have been solved now by an analysis of the motion in
terms of the attractors in phase space given by Cornish and Levin
\ct{cor98}, suggesting that the motion is indeed chaotic,
independent of the definition of time used. There may also be chaos
in Type XIII solutions. Moreover, chaotic behaviour near the
initial singularity was observed in solutions when a source-free
magnetic Maxwell field is coupled to fluid space-times of Type I
\ct{leb97} and Type VI$_{0}$ \ct{lebetal95}.

\subsubsection{Horizons and whimper singularities}
In tilted Class B models, it is possible for there to be a dramatic
change in the nature of the solution. This occurs where the
surfaces of homogeneity change from being spacelike (at late times)
to being timelike (at early times), these regions being separated
by a null surface ${\cal H}$, the horizon associated with this
change of symmetry. At earlier times the solution is no longer
spatially homogeneous --- it is inhomogeneous and
stationary.\footnote{This kind of change happens also in the
maximally extended Schwarzschild solution at the event horizon.}
Associated with the horizon is a singularity where all scalar
quantities are finite but components of the matter energy-momentum
tensor diverge when measured in a parallelly propagated frame as
one approaches the boundary of space-time (this happens because the
parallelly propagated frame gets infinitely rescaled in a finite
proper time relative to a family of KV's, which in the limit have
this singularity as a fixed point). The matter itself originates at
an anisotropic big bang singularity at the origin of the universe
in the stationary inhomogeneous region.\\

Details of how this happens are given in Ellis and King
\ct{ellkin73}, and phase plane diagrams for the simplest models in
which this occurs --- tilted LRS Type V models --- in Collins and
Ellis \ct{colell79}. These models isotropise at late times, and can
be arbitrarily similar to a low density FLRW model at the present
day.

\subsubsection{Isotropisation properties}
An issue of importance is whether these models tend to isotropy at
early or late times. An important paper by Collins and Hawking
\ct{colhaw73b} shows that for ordinary matter many Bianchi models
become anisotropic at very late times, even if they are very nearly
isotropic at present. Thus, isotropy is unstable in this case. On
the other hand, a paper by Wald \ct{wal83} showed that Bianchi
models will tend to isotropise at late times if there is a positive
cosmological constant present, implying that an inflationary era
can cause anisotropies to die away. The latter work, however, while
applicable to models with non-zero tilt angle, did not show this
angle dies away, and indeed it does not do so in general (Goliath
and Ellis \ct{golell99}). Inflation also only occurs in Bianchi
models if there is not too much anisotropy to begin with (Rothman
and Ellis \ct{rotell82}), and it is not clear that shear and
spatial curvature are in fact removed in all cases
\ct{raymod88}. Hence, some Bianchi models isotropise due to
inflation, but not all.\\

To study these kinds of question properly needs the use of phase
planes. These will be discussed after briefly considering
observations.

\subsection{Observational relations}
Observational relations in these models have been examined in
detail.\\

 (a) Redshift, observer area distance, and galaxy observations
 ($(M,z)$ and $(N,z)$ relations) are considered in MacCallum and
 Ellis \ct{macell70}. Anisotropies can occur in all these relations,
 but many of the models will display discrete isotropies in the
 sky.\\

 (b) The effect of tilt is to make the universe look inhomogeneous,
 even though it is spatially homogeneous (King and Ellis
 \ct{ke73}). This will be reflected in particular in a dipole
 anisotropy in number counts, which will thus occur in rotating
 universes \ct{god52}.\footnote{They will also occur in FLRW models
 seen from a reference frame that is not comoving; hence, they
 should occur in the real universe if the standard interpretation
 of the CBR anisotropy as due to our motion relative to a FLRW
 universe is correct; see Ellis and Baldwin \ct{ellbal84}.}\\

 (c) Element formation will be altered primarily through possible
 changes in the expansion time scale at the time of nucleosynthesis
 (\ct{tho67,bar76,rotmat84}). This enables us to put limits on
 anisotropy from measured element abundances in particular Bianchi
 types. This effect could in principle go either way, so a useful
 conjecture \ct{mre} is that in fact the effect of anisotropy will
 always --- despite the possible presence of rotation --- be to
 speed up the expansion time scale in Bianchi models.\\

 (d) CBR anisotropies will result in anisotropic Universe models,
 e.g., many Class B Bianchi models will show a hot-spot and
 associated spiral pattern in the CBR sky
 \ct{colhaw73a}--\ct{batj86}.\footnote{This result is derived in a
 gauge-dependent way; it would be useful to have a gauge-invariant
 version.} This enables us to put limits on anisotropy from
 observed CBR anisotropy limits (Collins and Hawking
 \ct{colhaw73a}, Bunn et al \ct{bunn96}). 
 If `reheating' takes place in an anisotropic
 universe, this will mix anisotropic temperatures from different
 directions, and hence distort the CBR spectrum \ct{ree67}.\\

Limits on present-day anisotropy from the CBR and element abundance
measurements are very stringent: $|\sig_0|/\Th_0 \leq 10^{-6}$ to
$10^{-12}$, depending on the model. However, because of the
anisotropies that can build up in both directions in time, this
does not imply that the very early Universe (before
nucleosynthesis) or late Universe will also be isotropic. The
conclusion applies back to last scattering (CBR measurements) and
to nucleosynthesis (element abundances). In both cases the
conclusion is quite model dependent. Although very strong limits
apply to some Bianchi models, they are much weaker for other
types. Hence, one should be a bit cautious in what one claims in
this regard.

\subsection{Dynamical systems approach}
The most illuminating description of the evolution of families of
Bianchi models is a {\bf dynamical systems approach} based on the
use of orthonormal tetrads, presented in detail in Wainwright and
Ellis \ct{waiell97}. The main variables used are essentially the
commutation functions mentioned above, but rescaled by a common
time dependent factor.

\subsubsection{Reduced differential equations}
The basic idea (Collins \ct{coll71}, Wainwright \ct{wai88}) is to
write the Einstein's field equations in a way that enables one to
study the evolution of the various physical and geometrical
quantities {\it relative to the overall rate of expansion of the
model}, as described by the rate of expansion scalar $\Th$, or,
equivalently, the {\bf Hubble scalar} $H = {\sfrac13}\,\Th$. The
remaining freedom in the choice of orthonormal tetrad needs to be
eliminated by specifying the variables $\Omega^\alpha$ implicitly
or explicitly (e.g., by specifying them as functions of the
$\sig_{\alpha\beta}$). This also simplifies the other quantities
(e.g., choice of a shear eigenframe will result in the tensor
$\sig_{\alpha\beta}$ being represented by two diagonal terms). One
so obtains a reduced set of variables, consisting of $H$ and the
remaining commutation functions, which we denote symbolically by
\be
\l{5.3}
\vec{x} = (\gamma^a{}_{bc}|_{\rm reduced}) \ .
\ee
The {\bf physical state} of the model is thus described by the vector
$(H,\vec{x})$. The details of this reduction differ for the Class A
and B models, and in the latter case there is an algebraic
constraint of the form
\be
\l{5.4}
g(\vec{x}) = 0 \ ,
\ee
where $g$ is a homogeneous polynomial. \\

The idea is now to normalise $\vec{x}$ with the Hubble scalar
$H$. We denote the resulting variables by a vector $\vec{y} \in
\mathbb{R}^{n}$, and write:
\be
\l{5.5}
\vec{y} =  \frac{\vec{x}}{H} \ .
\ee
These new variables are {\bf dimensionless}, and will be referred
to as {\bf Hubble-normalised variables}. It is clear that each
{\bf dynamical state} $\vec{y}$ determines a 1-parameter family of
physical states $(H,\vec{x})$. The evolution equations for the
$\gamma^a{}_{bc}$ lead to evolution equations for $H$ and
$\vec{x}$ and hence for $\vec{y}$. In deriving the evolution
equations for $\vec{y}$ from those for $\vec{x}$, the {\bf
deceleration parameter} $q$ plays an important role. The Hubble
scalar $H$ can be used to define a scale factor $S$ according
to (\r{eq:ell})
\be
\l{5.6}
H = \frac{\dot{S}}{S} \ ,
\ee
where $\cdot$ denotes differentiation with respect to $t$. The
deceleration parameter, defined by $q = -\,\ddot{S}\,S/\dot{S}^{2}$
(see (\r{eq:def_par})), is related to $\dot{H}$ according to
\be
\l{5.8}
\dot{H} = -\,(1+q)\,H^{2} \ .
\ee
In order that the evolution equations define a {\bf flow}, it is
necessary, in conjunction with the rescaling (\r{5.5}), to
introduce a {\bf dimensionless time variable} $\tau$ according to
\be
\l{5.9}
S = S_{0}\,e^{\tau} \ ,
\ee
where $S_{0}$ is the value of the scale factor at some arbitrary
reference time. Since $S$ assumes values $0 < S < +\,\infty$ in an
ever-expanding model, $\tau$ assumes all real values, with $\tau
\rightarrow -\,\infty$ at the initial singularity and $\tau
\rightarrow +\,\infty$ at late times.  It follows from equations
(\r{5.6}) and (\r{5.9}) that
\be
\l{5.10}
\frac{dt}{d \tau} = \frac{1}{H} \ ,
\ee
and the evolution equation (\r{5.8}) for $H$ can be written
\be
\l{5.11}
\frac{dH}{d \tau} = -\,(1+q)\,H \ .
\ee

Since the right-hand sides of the evolution equations for the
$\gam^{a}{}_{bc}$ are homogeneous of degree 2 in the
$\gam^{a}{}_{bc}$, the change (\r{5.10}) of the time variable
results in $H$ canceling out of the evolution equation for
$\vec{y}$, yielding an {\bf autonomous differential equation}
(`DE'):
\be
\l{5.12}
\frac{d\vec{y}}{d \tau} = \vec{f}(\vec{y}) \ ,
\quad \vec{y} \in \mathbb{R}^{n} \ .
\ee
The constraint $g(\vec{x}) = 0$ translates into a constraint
\be
\l{5.13}
g(\vec{y}) = 0 \ ,
\ee
which is preserved by the DE. The functions $\vec{f}:
\mathbb{R}^{n} \rightarrow \mathbb{R}^{n}$ and $g: \mathbb{R}^{n}
\rightarrow \mathbb{R}$ are polynomial functions in $\vec{y}$.
An essential feature of this process is
that the evolution equation for $H$, namely (\r{5.11}), decouples
from the remaining equations (\r{5.12}) and (\r{5.13}). In other
words, the DE (\r{5.12}) describes the evolution of the non-tilted
Bianchi cosmologies, the transformation (\r{5.5}) essentially
scaling away the effects of the overall expansion. An important
consequence is that the new variables are bounded near the initial
singularity.

\subsubsection{Equations and orbits}
The first step in the analysis is to formulate Einstein's
field equations, using Hubble-normalised variables, as a
DE (\r{5.12}) in $\mathbb{R}^{n}$, possibly subject to a
constraint (\r{5.13}). Thus one uses the
tetrad equations presented above, now adapted to apply to the
variables rescaled in this way.  Since $\tau$ assumes all real
values (for models which expand indefinitely), the solutions of
(\r{5.12}) are defined for all $\tau$ and hence define a {\bf flow}
$\{\phi_{\tau} \}$ on $\mathbb{R}^{n}$. The evolution of the
cosmological models can thus be analyzed by studying the orbits
of this flow in the physical region of the state space, which is
a subset of $\mathbb{R}^{n}$ defined by the requirement that
the matter energy density $\mu$ be non-negative, i.e.,
\be
\l{5.14}
\Omega (\vec{y}) = \frac{\mu}{3H^{2}} \geq 0 \ ,
\ee
where the {\bf density parameter} $\Omega$ (see (\r{eq:def_par})) is a
dimensionless measure of $\mu$. \\

The {\bf vacuum boundary}, defined by $\Omega(\vec{y}) = 0$,
describes the evolution of vacuum Bianchi models, and is an
{\bf invariant set} which plays an important role in the qualitative
analysis because vacuum models can be asymptotic states for perfect
fluid models near the big-bang or at late times. There are other
invariant sets which are also specified by simple restrictions on
$\vec{y}$ which play a special role: the subsets representing each
Bianchi type (Figure 2), and the subsets representing higher
symmetry models, specifically the FLRW models and the LRS Bianchi
models (according to Figure 3).\\

It is desirable that the dimensionless state space $D$ in
$\mathbb{R}^{n}$ is a {\bf compact set}. In this case each orbit
will have non-empty future and past limit sets, and hence there
will exist a {\bf past attractor} and a {\bf future attractor}
in state space.  When using Hubble-normalised variables,
compactness of the state space has a direct physical meaning for
ever-expanding models: if the state space is compact, then at the
big-bang no physical or geometrical quantity diverges more rapidly
than the appropriate power of $H$, and at late times no such
quantity tends to zero less rapidly than the appropriate power
of $H$. This will happen for many models; however, the state space
for Bianchi Type VII$_{0}$ and Type VIII models is
{\em non-compact\/}. This lack of compactness manifests itself in
the behaviour of the Weyl curvature at late times.

\subsubsection{Equilibrium points and self-similar cosmologies}
Each ordinary orbit in the dimensionless state space corresponds to
a one-parameter family of physical models, which are conformally
related by a constant rescaling of the metric. On the other hand,
for an {\bf equilibrium point} $\vec{y}^{*}$ of the DE (\r{5.12})
(which satisfies $\vec{f}(\vec{y}^{*}) = \vec{0}$), the deceleration
parameter $q$ is a constant, i.e., $q(\vec{y}^{*}) = q^{*}$, and we
find
$$
H(\tau) = H_{0}\,e^{(1+q^{*})\tau} \ .
$$
In this case, however, the parameter $H_{0}$ is no longer
essential, since it can be set to unity by a translation of $\tau$,
$\tau \rightarrow \tau + \mbox{const}$; then (\r{5.10}) implies
that
\be
\l{5.32}
H\,t = \frac{1}{1+q^{*}} \ ,
\ee
so that by (\r{5.3}) and (\r{5.5}) the commutation functions are of
the form $(\mbox{const})\times t^{-1}$. It follows that the
resulting cosmological model is {\bf self-similar}. It then turns
out that {\it to each equilibrium point of the DE (\r{5.12}) there
corresponds a unique self-similar cosmological model}. In such a
model the physical states at different times differ only by an
overall change in the length scale. Such models are expanding, but
in such a way that their dimensionless state does not change. They
include the spatially flat FLRW model ($\Omega = 1$) and the
Milne model ($\Omega = 0$). All vacuum and orthogonal perfect fluid
self-similar Bianchi solutions have been given by Hsu and
Wainwright \ct{hsu86}. The equilibrium points determine the
asymptotic behaviour of other more general models.

\subsubsection{Phase planes}
Many {\bf phase planes} can be constructed explicitly. The reader is
referred to Wainright and Ellis \ct{waiell97} for a comprehensive
presentation and survey of results attained so far. Several
interesting points emerge: \\

* {\it Relation to lower dimensional spaces}: it seems that the
 lower dimensional spaces, delineating higher symmetry models, can
 be skeletons guiding the development of the higher dimensional
 spaces (the more generic models). This is one reason why study of
 the exact higher symmetry models is of significance. The way this
 occurs is the subject of ongoing investigation (the key issue
 being how the finite dimensional dynamical systems corresponding
 to models with symmetry are imbedded in and relate to the infinite
 dimensional dynamical system describing the evolution of models
 without symmetry).\\

* {\it Identification of models in state space}: the analysis of
 the phase planes for Bianchi models shows that the procedure
 sometimes adopted of identifying all points in state space
 corresponding to the same model is not a good idea. For example
 the Kasner ring that serves as a framework for evolution of many
 other Bianchi models contains multiple realizations of the same
 Kasner model. To identify them as the same point in state space
 would make the evolution patterns very difficult to follow. It is
 better to keep them separate, but to learn to identify where
 multiple realizations of the same model occur (which is just the
 {\bf equivalence problem} for cosmological models).\\

* {\it Isotropisation} is a particular issue that can be studied by
  use of these planes \ct{waietal98,golell99}. It turns out that
  even in the classes of non-inflationary Bianchi models that
  contain FLRW models as special cases, not all models isotropise
  at some period of their evolution; and of those that do so, most
  become anisotropic again at late times. Only an inflationary
  equation of state will lead to such isotropisation for a fairly
  general class of models (but in the tilted case it is not clear
  that the tilt angle will die away \ct{golell99}); once it has
  turned off, anisotropic modes will again occur.\\

An important idea that arises out of this study is that of {\bf
intermediate isotropisation}: namely, models that become very like
a FLRW model for a period of their evolution but start off and end
up quite unlike these models. It turns out that many Bianchi types
allow intermediate isotropisation, because the FLRW models are
{\bf saddle points} in the relevant phase planes. This leads to the
following two interesting results:

\begin{quotation}
{\bf Bianchi Evolution Theorem (1)}:  Consider a family of Bianchi
models that allow intermediate isotropisation. Define an
$\epsilon$-neighbourhood of a FLRW model as a region in state space
where all geometrical and physical quantities are closer than
$\epsilon$ to their values in a FLRW model. Choose a time scale
$L$. Then no matter how small $\epsilon$ and how large $L$, there
is an open set of Bianchi models in the state space such that each
model spends longer than $L$ within the corresponding
$\epsilon$-neighbourhood of the FLRW model.
\end{quotation}

(This follows because the saddle point is a {\bf fixed point} of the
phase flow; consequently, the phase flow vector becomes arbitrarily
close to zero at all points in a small enough open region around
the FLRW point in state space.)\\

Consequently, although these models are quite unlike FLRW models at
very early and very late times, there is an open set of them that
are observationally indistinguishable from a FLRW model (choose $L$
long enough to encompass from today to last coupling or
nucleosynthesis, and $\epsilon$ to correspond to current
observational bounds). Thus, there exist many such models that are
viable as models of the real Universe in terms of compatibility
with astronomical observations.

\begin{quotation}
{\bf Bianchi Evolution Theorem (2)}: In each set of Bianchi models
of a type admitting intermediate isotropisation, there will be
spatially homogeneous models that are linearisations of these
Bianchi models about FLRW models. These perturbation modes will
occur in any almost-FLRW model that is generic rather than
fine-tuned; however, the exact models approximated by these
linearisations will be quite unlike FLRW models at very early and
very late times.
\end{quotation}

(Proof is by linearising the equations above (see the following
section) to obtain the Bianchi equations linearised about the FLRW
models that occur at the saddle point leading to the intermediate
isotropisation. These modes will be the solutions in a small
neighbourhood about the saddle point permitted by the linearised
equations (given existence of solutions to the non-linear
equations, linearisation will not prevent corresponding linearised
solutions existing).)\\

The point is that these modes can exist as linearisations of the
FLRW model; if they do not occur, then initial data has been chosen
to set these modes precisely to zero (rather than being made very
small), which requires very special initial conditions. Thus, these
modes will occur in almost all almost-FLRW models. Hence, if one
believes in generality arguments, they will occur in the real
Universe. When they occur, they will at early and late times grow
until the model is very far from a FLRW geometry (while being
arbitrarily close to an FLRW model for a very long time, as per the
previous theorem).\\

{\it Exercise}: Most studies of CBR anisotropies and
nucleosynthesis are carried out for the Bianchi types that allow
FLRW models as special cases (see Figure 3). Show that Bianchi
models can approximate FLRW models for extended periods even if
they do not belong to those types. What kinds of CBR anisotropies
can occur in these models? (See, e.g., \ct{waiell97}.)

\section{Almost-FLRW models}
The real Universe is {\em not\/} FLRW because of all the structure
it contains, and (because of the non-linearity of Einstein's field
equations) the other exact solutions we can attain have higher
symmetry than the real Universe. Thus, in order to obtain realistic
models we can compare with detailed observations, {\em we need to
approximate\/}, aiming to obtain `almost-FLRW' models representing
a universe that is FLRW-like on a large scale but allowing for
generic inhomogeneities on a small scale.

\subsection{Gauge problem}
The major problem in studying perturbed models is the {\bf gauge
problem}, due to the fact that there is {\em no identifiable fixed
background model\/} in General Relativity. One can start with a
unique FLRW Universe model with metric $\overline{g}_{ab}$ in some
local coordinate system, and perturb it to obtain a more realistic
model: $\overline{g}_{ab}\rightarrow g_{ab} = \overline{g}_{ab}
+ \delta g_{ab}$, but then the process has no unique inverse: the
background model $\overline{g}_{ab}$ is not uniquely determined
by the lumpy Universe model $g_{ab}$ (no unique tensorial averaging
process has been defined that will recover $\overline{g}_{ab}$ from
$g_{ab}$). Many choices can be made. However, the usual variables
describing perturbations depend on the way the (fictitious)
background model $\overline{g}_{ab}$ is fitted to the metric of the
real Universe, $g_{ab}$; these variables can be given any values 
one wants by changing this correspondence.\footnote{This is often
represented implicitly rather than explicitly, by assuming that
points with the same coordinate values in the background space and
more realistic model map to each other; then the gauge freedom is
contained in the coordinate freedom available in the realistic
universe model.} For example, the dimensionless {\bf energy density
contrast} $\delta$ representing a density perturbation is usually
defined by
\be
\l{eq:def}
\delta(x) = {\mu(x)
- \overline{\mu}(x) \over \overline{\mu}(x)} \ ,
\ee
where $\mu(x)$ is the actual value of the energy density at the
point $x$, while $\overline{\mu}(x)$ is the (fictitious) background
value there, determined by the chosen mapping of the background model
into the realistic lumpy model (Ellis and Bruni \ct{eb89}). This
quantity can be given any value we desire by altering that map; we
can, e.g., set it to zero by choosing the real surfaces of
constant energy density (provided these are spacelike) to be the
background surfaces of constant time and, hence, of constant energy
density. Consequently, perturbation equations written in terms of this
variable have as solution both {\bf physical modes} and {\bf gauge
modes}, the latter corresponding to variation of gauge choice rather
than to physical variation. \\

One way to solve this is by very carefully keeping track of the
gauge choice used and the resulting gauge freedom; see Ma and
Bertschinger \ct{mabe96} and Prof. Bertschinger's lectures here.
The alternative is to use gauge-invariant variables. A widely used and
fundamentally important set of such variables are those introduced
by Bardeen \ct{bardeen}, and used for example by Bardeen,
Steinhardt and Turner \ct{bst}. Another possibility is use of {\bf
gauge-invariant and $1+3$ covariant} (`GIC') {\bf variables},
i.e., variables that are gauge-invariant and also $1+3$ covariantly
defined so that they have a clear geometrical meaning, and can be
examined in any desired (spatial) coordinate system. That is what
will be pursued here.\\

In more detail: our aim is to examine perturbed models by using
$1+3$ covariant variables defined in the {\em real\/} space-time
(not the background), deriving {\em exact\/} equations for these
variables in that space-time, and then {\em approximating\/} by
{\bf linearising} about a RW geometry to get the linearised equations
describing the evolution of energy density inhomogeneities in
almost-FLRW universes. How do we handle {\bf gauge invariance} in
this approach? We rely on the
\begin{quotation}
{\bf Gauge Invariance Lemma} (Stewart and Walker \ct{sw74}): If a
quantity $T^{\dots}{}_{\dots}$ vanishes in the background
space-time, then it is gauge-invariant (to \underline{all} orders).
\end{quotation}
[The proof is straightforward :
If $\overline{T}^{\dots}{}_{\dots} = 0$, then
$\delta T^{\dots}{}_{\dots} = T^{\dots}{}_{\dots} -
\overline{T}^{\dots}{}_{\dots} = T^{\dots}{}_{\dots}$,
which is manifestly independent of the mapping $\Phi$ from
$\overline{S}$ to $S$ (it does not matter how we map
$\overline{T}^{\dots}{}_{\dots}$ from $\overline{S}$ to $S$ when
$\overline{T}^{\dots}{}_{\dots}$ vanishes).] The application to
almost-FLRW models follows (see Ellis and Bruni \ct{eb89}), where
we use an order-of-magnitude notation as follows: Given a
{\bf smallness parameter} $\epsilon$, ${\cal O}[n]$ denotes ${\cal
O}(\epsilon^{n})$, and $A \approx B$ means $A - B = {\cal O}[2]$
(i.e., these variables are equivalent to ${\cal O}[1]$). When $A
\approx 0$ we shall regard $A$ as vanishing when we linearise (for
it is zero to the accuracy of relevant first-order
calculations). Then,
\begin{itemize}

\item 
{\bf Zeroth-order variables} are $\mu$, $p$, $\Th$, and their
covariant time derivatives, $\dot{\mu}$, $\dot{p}$, $\dot{\Th}$,

\item
{\bf First-order variables} are $\udot^{a}$, $\sig_{ab}$, $\om^{a}$,
$q^{a}$, $\pi_{ab}$, $E_{ab}$, $H_{ab}$, $X_{a}$, $Z_{a}$, and
their covariant time and space derivatives.
\end{itemize}
As these first-order variables all vanish in exact FLRW models,
provided $u^{a}$ is {\em uniquely\/} defined in the realistic
(lumpy) almost-FLRW Universe model, they are all uniquely defined GIC
variables. Thus, this set of variables provides what we wanted:
$1+3$ covariant variables characterising departures from a FLRW
geometry (and, in particular, the spatial inhomogeneity of a
universe) that are gauge-invariant when the universe is
almost-FLRW. Because they are tensors defined in the {\em real\/}
space-time, we can evaluate them in any local coordinate system
we like in that space-time.

\subsubsection{Key variables}
Two simple gauge-invariant quantities give us the information we
need to discuss the time evolution of energy density fluctuations.
The basic quantities we start with are the orthogonal projections
of the energy density gradient, i.e., the vector $ X_{a} \equiv
\3nab_{a}\mu$, and of the expansion gradient, i.e., the vector
$Z_{a} \equiv \3nab_{a}\Th$. The first can be determined (a) from
virial theorem estimates and large-scale structure observations
(as, e.g., in the POTENT programme), (b) by observing gradients in
the numbers of observed sources and estimating the mass-to-light
ratio (\,Kristian and Sachs \ct{krisac66}, Eq. (39)\,), and (c) by
gravitational lensing observations. However, these do not directly
correspond to the quantities usually calculated; but two closely
related quantities do. The first is the {\bf matter-comoving
fractional energy density gradient\/}:
\be
\l{eq:def3}
{\cal D}_{a} \equiv S\,\frac{X_{a}}{\mu} \ ,
\ee
which is gauge-invariant and dimensionless, and represents the
spatial energy density variation over a fixed comoving scale. Note
that $S$, and so ${\cal D}_{a}$, is defined only up to a constant by
equation (\r{eq:ell}); this allows it to represent the energy
density variation between any neighbouring worldlines. The vector
${\cal D}_{a}$ can be separated into a magnitude ${\cal D}$ and
direction $e_{a}$
\be
{\cal D}_{a} = {\cal D}\,e_{a} \ , ~e_{a}e^{a} = 1 \ ,
~e_{a}u^{a} = 0 \hsp5 \Rightarrow \hsp5
{\cal D} = ({\cal D}_{a}{\cal D}^{a})^{1/2} \ .
\ee
The magnitude ${\cal D}$ is the gauge-invariant
variable\footnote{Or, equivalently in the linear case, the spatial
divergence of ${\cal D}_{a}$.} that most closely corresponds to the
intention of the usual $\delta = \delta\mu/\overline{\mu}$ given in
(\r{eq:def}). The crucial difference from the usual definition is
that ${\cal D}$ represents a (real) spatial fluctuation, rather
than a (fictitious) time fluctuation, and does so in a GIC
manner. An important auxiliary variable in what follows is the {\bf
matter-comoving spatial expansion gradient}:
\be
\l{eq:def4}
{\cal Z}_{a} \equiv S\,Z_{a} \ .
\ee

The issue now is, can we find a set of equations determining how
these variables evolve? Yes we can; they follow from the exact
$1+3$ covariant equations of subsection \r{subsec:13ceqs}.

\subsection{Dynamical equations}
We can determine {\em exact\/} propagation equations along the
fluid flow lines for the quantities defined in the previous
section, and then linearise these to the almost-FLRW case. The
basic linearised equations are given by Hawking \ct{haw66} (see
his equations (13) to (19)); we add to them the linearised
propagation equations for the gauge-invariant spatial gradients
defined above \ct{eb89}.

\subsubsection{Growth of inhomogeneity}
Taking the spatial gradient of the equation of energy conservation
(\r{eq:en}) (for the case of a perfect fluid), we find \ct{eb89}
$$
\3nab_{a}(\dot{\mu}) + \Th\,\3nab_{a}(\mu+p)
+ (\mu+p)\,\3nab_{a}\Th = 0 \ ,
$$
i.e.,
$$
h_{a}{}^{b}\nabla_{b}(u^{c}\nabla_{c}\mu)
+ \Th\,(X_{a}+\3nab_{a}p) + (\mu+p)\,Z_{a} = 0 \ .
$$
Using Leibniz' Rule and changing the order of differentiation in
the second-derivative term (and noting that the pressure-gradient
term cancels on using the momentum conservation equation
(\r{eq:en1})), we obtain the {\bf fundamental equation for the
growth of inhomogeneity}:
\be
\l{eq:grow1}
\dot{X}{}^{\la a\ra} + {\sfrac43}\,\Th\,X^{a}
= -\,(\mu+p)\,Z^{a} - \sig^{a}{}_{b}\,X^{b}
+ \eta^{abc}\,\om_{b}\,X_{c} \ ,
\ee
with source term $Z^{a}$. On taking the spatial gradient of the
Raychaudhuri equation (\r{eq:ray}), we find the companion equation
for that source term:
\be
\l{eq:grow2}
\dot{Z}{}^{\la a\ra} + \Th\,Z^{a}
= -\,\sfrac{1}{2}\,X^{a} - \sig^{a}{}_{b}\,Z^{b}
+ \eta^{abc}\,\om_{b}\,Z_{c} + \udot^{a}\,{\cal R}
- 2\,\3nab^{a}(\sig^{2} - \om^{2})
+ \3nab^{a}(\3nab_{b}\udot^{b} + \udot_{b}\udot^{b}) \ ,
\ee
where
\be
\l{eq:def1}
{\cal R} = - \,\sfrac{1}{3}\,\Th^{2} + \3nab_{a}\udot^{a}
+ (\udot_{a}\udot^{a}) - 2\,\sig^{2} + 2\,\om^{2} + \mu
+ \Lambda \ .
\ee
These {\bf exact equations} contain no information not implied by the
others already given; nevertheless, they are useful in that they are
exact equations directly giving the rate of growth of inhomogeneity in
the generic (perfect fluid) case, the second, together with the
evolution equations above, giving the rate of change of all the
source terms in the first. \\

The procedure now is to {\em systematically approximate\/} all the
dynamical equations, and, in particular, the structure growth
equations given above, by dropping all terms of second order or
higher in the implicit\footnote{One could make this variable
explicit, but there does not seem to be much gain in doing so.}
expansion variable $\epsilon$. Thus, we obtain the {\bf linearised
equations} as approximations to the exact equations above by noting
that in the almost-FLRW case,
\bea
\l{eq:prop1}
\dot{X}{}^{\la a\ra} + {\sfrac43}\,\Th\,X^{a}
& = & -\,(\mu+p)\,Z^{a} + {\cal O}[2] \\
\l{eq:prop2}
\dot{Z}{}^{\la a\ra} + \Th\,Z^{a}
& = & -\,\sfrac{1}{2}\,X^{a} + \udot^{a}\,{\cal R}
+ \3nab^{a}(\3nab_{b}\udot^{b}) + {\cal O}[2] \ ,
\eea
where
\be
\l{eq:def2}
{\cal R} = - \,\sfrac{1}{3}\,\Th^{2} + \3nab_{a}\udot^{a}
+ \mu + \Lambda + {\cal O}[2] \ .
\ee
Then we linearise the equations by dropping the terms ${\cal
O}[2]$, {\em so from now on in this section\/} `$=$' {\em means
equal up to terms of order\/} $\epsilon^2$.

\subsection{Dust}
In the case of dust, $p = 0 \Rightarrow \dot{u}^a = 0$, and the
equations (\r{eq:prop1}) and (\r{eq:prop2}) for growth of
inhomogeneity become
\bea
S^{-4}\,h^{a}{}_{b}(S^{4}X^{b})\,\dot{} & = & -\,\mu\,Z^{a} \\
S^{-3}\,h^{a}{}_{b}(S^{3}Z^{b})\,\dot{} & = &
-\,\sfrac{1}{2}\,X^{a} \ ,
\eea
This closes up to give a second-order equation (take the covariant
time derivative of the first and substitute from the second and the
energy conservation equation (\r{eq:en})). To compare with the
usual equations, change to the variables ${\cal D}_{a}$ and ${\cal
Z}_{a}$ (\,see (\r{eq:def3}) and (\r{eq:def4})\,). Then the equations
become
\bea
\dot{{\cal D}}{}^{\la a\ra} & = & -\,{\cal Z}^{a} \\
\dot{{\cal Z}}{}^{\la a\ra} & = & -\,\sfrac{2}{3}\,\Th\,{\cal
Z}^{a} - \sfrac{1}{2}\,\mu\,{\cal D}^{a} \ .
\eea
These directly imply the second-order equation (take the covariant
time derivative of the first equation!)
\be
\l{eq:dustp}
0 = \ddot{{\cal D}}_{\la a\ra} + \sfrac{2}{3}\,\Th\,
\dot{{\cal D}}_{\la a\ra} - \sfrac{1}{2}\,\mu\,{\cal D}_{a} \ ,
\ee
which is the usual equation for growth of energy density
inhomogeneity in dust universes, and has the usual solutions:
when $k = 0$, then\footnote{This is the background rather than
the real value of this quantity, which is what should really be
used here; it is determined in the real space-time by
$\dot{S}/S = \sfrac{1}{3}\,\Th$. However, as we are linearising,
the difference makes only a second-order change in the
coefficients in the equations, which we can neglect.} $S(t)
\propto t^{2/3}$, and we obtain
\be
\l{eq:dust}
{\cal D}_{a}(t,x^\alpha) = d_+{}_{a}(x^\alpha)\ t^{2/3}
+ d_-{}_{a}(x^\alpha)\ t^{-1} \ , \hsp5
\dot{d}_i{}_{a} = 0 \ ,
\ee
(where $t$ is proper time along the flow lines).\footnote{We can
also see that if $\Th = 0$, there will be an exponential rather
than power-law growth.} This shows the {\bf growing mode} that
leads to structure formation and the {\bf decaying mode} that
dissipates previously existing inhomogeneities. It has been
obtained in a GIC way: all the first-order variables, including
in particular those in this equation, are gauge-invariant, and
there are no gauge modes. Furthermore, we have available the
fully non-linear equations, Eqs. (\r{eq:grow1}) and (\r{eq:grow2}),
and so can estimate the errors in the neglected terms, and set up
a systematic higher-order approximation scheme for solutions of these
equations. Solutions for other background models (with $k = \pm\,1$
or $\Lambda \neq0$) can be obtained by substituting the appropriate
values in (\r{eq:dustp}) for the background variables $\Th$ and
$\mu$, possibly changing to conformal time to simplify the
calculation.\\

{\it Exercise}: What is the growth rate at late times in a
low-density universe, when the expansion is curvature dominated and
so is linear: $S(t) = c_0\,t$? What if there is a cosmological
constant, so the late time expansion is exponential: $S(t) =
c_0\,\exp(H_0\,t)$, where $c_0$ and $H_0$ are constants?

\subsubsection{Other quantities}
We have concentrated here on the growth of inhomogeneities in the
energy density and the expansion rate.
However, all the $1+3$ covariant equations of the previous sections
apply and can be linearised in a straightforward way to the
almost-FLRW case (and one can find suitable coordinates and tetrads
in order to employ the complete set of tetrad equations in this
context, too).  Doing so, one can in particular study {\bf vorticity
perturbations} and {\bf gravitational wave perturbations}; see the
pioneering paper by Hawking \ct{haw66}. We will not consider these
further here; however, a series of interesting issues arise. The
GIC approach to gravitational waves is examined in Hogan and Ellis
\ct{ellhog96} and Dunsby, Bassett and Ellis \ct{dunbasell}.
\\

A further important issue is the effect of perturbations on
observations (apart from the CBR anisotropies, discussed below). It
has been known for a long time that anisotropies can affect area
distances as well as redshifts (see Bertotti \ct{ber66}, Kantowski
\ct{kan69a}); the {\bf Dyer--Roeder formula} (\ct{bi:dyer73}, see
also \ct{sef}) can be used at any redshift for those many rays that
propagate in the lower-density regions between inhomogeneities;
however, this formula is not accurate for those ray bundles that
pass very close to matter, where shearing becomes important. This
is closely related to the {\bf averaging problem} (see, e.g.,
Ellis \ct{ell84} or Boersma \ct{boe98}): how can dynamics and
observations of a Universe model which is basically empty almost
everywhere average out correctly to give the same dynamics and
observations as a Universe model which is exactly spatially
homogeneous? What differences will there be from the FLRW case?
We will not pursue this further here except to state that it is
believed this does in fact work out OK: in the fully inhomogeneous
case it is the Weyl curvature that causes distortions in the empty
spaces between astrophysical objects (as in gravitational lensing),
and, hence, causes convergence of both timelike and null geodesic
congruences; these, however, average out to give zero average
distortion and the same convergence effect as a FLRW space-time
with zero Weyl curvature, but with the Ricci curvature causing
focusing of these curves (see \ct{wei76} for a discussion of the
null case; however, some subtleties arise here in terms of the
way areas are defined when strong lensing takes place
\ct{lens1,lens2,lens3}).

\subsection{Perfect fluids}
A GIC analysis similar to the dust case has been given by Ellis,
Hwang and Bruni determining FLRW perturbations for the perfect
fluid case \ct{ehb89,ebh90}. This gives the single-fluid equation
for growth of structure in an almost-FLRW Universe model (again,
derived in a GIC manner), and includes as special cases a fully
$1+3$ covariant derivation of the {\bf Jeans length} and of the
{\bf speed of sound} for barotropic perfect fluids. To evaluate
the last two terms in (\r{eq:prop2}) when $p\neq 0$, using
(\r{eq:en1}) we see that, to first order,
\be
\3nab_{a}(\3nab_{b}\udot^{b})
= -\,\frac{\3nab_{a}(\3nab_{b}\3nab^{b} p)}{(\mu+p)} \ .
\ee
But, for simplicity considering only the case of vanishing
vorticity, $\om^{a}=0$,\footnote{When it is non-zero,
(\r{eq:tilt}) must be taken into account when commuting
derivatives; see \ct{ebh90}.} we have
\be
\3nab_b\3nab^b\3nab_a p
= \3nab_b\3nab_a\3nab^b p \ ,
\ee
and, on using the Ricci identities for the $\3nab$-derivatives and
the zeroth-order relation ${}^{3}\!R_{ab} =
\sfrac{1}{3}\,{}^{3}\!R\, h_{ab}$ for the 3-dimensional Ricci
tensor, we obtain
\be
\3nab_{a}\3nab_{b}\3nab^{b}p = \3nab_{b}\3nab_{a}\3nab^{b}p
- \sfrac{1}{3}\,{}^{3}\!R\ \3nab_{a}p \ .
\ee
Thus, on using ${}^{3}\!R = {\cal K} = 6k/S^2$, we find
\be
\l{eq:grad}
\sfrac{1}{2}\,{\cal K}\,\udot_{a} = -\,\frac{1}{(\mu+p)}\,
\frac{3k}{S^2}\,\3nab_{a}p \ , \hsp5
\3nab_{a}(\3nab_{b}\udot^{b}) = \frac{1}{(\mu+p)}\,
\left(\frac{2k}{S^2}\,\3nab_{a}p - \3nab^{2}\3nab_{a}p\right) \ , 
\ee
introducing the notation $\3nab^{2}\3nab_{a}p \equiv
\3nab_{b}\3nab^{b}\3nab_{a}p$. 
In performing this calculation, note that there will {\em not} be
3-spaces orthogonal to the fluid flow if $\om^{a} \neq 0$, but
still we can calculate the 3-dimensional orthogonal derivatives as
usual (by using the projection tensor $h^{a}{}_{b}$); the
difference from when $\om^{a} = 0$ will be that the quantity we
calculate as a curvature tensor, using the usual definition from
commutation of second derivatives, will not have all the usual
curvature tensor symmetries. Nevertheless, the zeroth-order
equations, representing the curvature of the 3-spaces orthogonal to
the fluid flow in the background model, will agree with the
linearised equations up to the required accuracy.\\

Now if $p = p(\mu,s)$, where $s$ is the {\bf entropy per particle},
we find
\be
\3nab_{a}p = \left(\frac{\p p}{\p\mu}\right)_{s = {\rm const}}
\3nab_{a}\mu
+ \left(\frac{\p p}{\p s}\right)_{\mu = {\rm const}}\3nab_{a}s \ .
\ee
We assume we can ignore the second term (pressure variations caused
by spatial entropy variations) relative to the first (pressure
variations caused by energy density variations) and spatial
variations in the scale function $S$ (which would at most cause
second-order variations in the propagation equations). Then
(ignoring terms due to the spatial variation of $\p p/\p\mu$, which
will again cause second-order variations) we find in the
zero-vorticity case,
\be
\l{eq:mom2}
S\,[\ \sfrac{1}{2}\,{\cal K}\,\udot_{a} + \3nab_{a}(\3nab_{b}
\udot^{b})\ ] = -\,\frac{1}{(1+p/\mu)} \,
\left(\frac{\p p}{\p\mu}\right)\left(\frac{k}{S^2}\,{\cal D}_{a}
+ \3nab^{2}{\cal D}_{a}\right) \ .
\ee
This is the result that we need in proceeding with (\r{eq:prop2}).

\subsubsection{Second-order equations}
The equations for propagation can now be used to obtain
second-order equations for ${\cal D}_{a}$.\footnote{And the
variable $\Phi_{a} = \mu\,S^{2}\,{\cal D}_{a}$ that corresponds
more closely to Bardeen's variable; see \ct{ebh90}.} For easy
comparison, we follow Bardeen \ct{bardeen} by defining
\be
\l{eq:def6}
w = \frac{p}{\mu} \ , \hsp5 \cs = \frac{\p p}{\p\mu}
\hsp5 \Rightarrow \hsp5
\left(\frac{p}{\mu}\right)\dot{} ~\equiv ~\dot{w}
= -\,\Th\,(1+w)\,(\cs-w) \ .
\ee

Now covariant differentiation of (\r{eq:prop1}), projection
orthogonal to $u^{a}$, and linearisation gives a second-order
equation for ${\cal D}_{a}$ (we use Eqs. (\r{eq:ray}),
(\r{eq:prop2}), (\r{eq:def6}) and (\r{eq:mom2}) in the
process). We find
\bea
\l{eq:second}
0 & = & \ddot{{\cal D}}_{\la a\ra} + (\sfrac{2}{3} - 2w
+ \cs)\,\Th\,\dot{{\cal D}}_{\la a\ra}
- [\ (\sfrac{1}{2}+4w-\sfrac{3}{2}w^2 -3\cs)\,\mu
+ (c_s^2-w)\,\frac{12k}{S^2}\ ]\ {\cal D}_{a} \\
& & \hspace{10mm} + \ \cs\,[\ \frac{2k}{S^2}\,{\cal
D}_{a} - \3nab^{2}{\cal D}_{a}\ ] \ . \nonumber
\eea
This equation is the {\em basic result of this subsection\/}; the
rest of the discussion examines its properties and special
cases. It is a second-order equation determining the evolution of
the GIC energy density variation variable ${\cal D}_{a}$ along
the fluid flow lines. It has the form of a {\bf wave equation} with
extra terms due to the expansion, gravitation and the
spatial curvature of the Universe model.\footnote{We have dropped
$\Lambda$ in these equations; it can be represented by setting
$w = -\,1$.} We bracket the last two terms together, because when
we make a harmonic decomposition these terms together give the
harmonic eigenvalues $n^2$.\\

This form of the equations allows for a variation of $w = p/\mu$
with time. However, if $w = \mbox{const}$, then from (\r{eq:def6})
$\cs = w$, and the equation simplifies to
\be
\l{eq:second1}
0 = \ddot{{\cal D}}_{\la a\ra} + (\sfrac{2}{3} - w)\,\Th\,
\dot{{\cal D}}_{\la a\ra} - \sfrac{1}{2}\,(1-w)\,(1+3w)\,\mu\,
{\cal D}_{a} + w\,[\ \frac{2k}{S^2}\,{\cal D}_{a} - \3nab^{2}
{\cal D}_{a}\ ] \ .
\ee
The matter source term vanishes if $w = 1$ (the case of `stiff
matter' $\Leftrightarrow p = \mu$) or $w = -\,\sfrac{1}{3}$ (the
case $p = -\,\sfrac{1}{3}\,\mu$, corresponding to matter with no
active gravitational mass). Between these two limits (`ordinary
matter'), the matter term is positive and tends to cause the
energy density gradient to increase (`gravitational aggregation');
outside these limits, the term is negative and tends to cause
the energy density gradient to decrease (`gravitational
smoothing'). Also, the sign of the damping term (giving the
adiabatic decay of inhomogeneities) is positive if
$\sfrac{2}{3} > w$ (that is, $ 2\mu > 3p $) but negative
otherwise (they adiabatically grow rather than decay in
this case). The equation reduces correctly to the corresponding
dust equation in the case $w = 0$. Two other cases of importance
are:\\

* {\it Speed of sound}: when $\Theta$, $\mu$, and $k/S^2$ can be
 neglected, we see directly from ({\r{eq:second}) that $c_{s}$
 introduced above {\em is} the speed of sound (and that imaginary
 values of $c_{s}$, i.e., negative values of $\p p/\p\mu$,
 then lead to exponential growth or decay rather than
 oscillations).\\

* {\it Radiation}: In the case of pure radiation, $\gamma =
 \sfrac{4}{3}$ and $w = \sfrac{1}{3} = \cs$. Then we find
 from (\r{eq:second1})
 \be
 \l{eq:radiation}
 0 = \ddot{{\cal D}}_{\la a\ra} + \sfrac{1}{3}\,\Th\,
 \dot{{\cal D}}_{\la a\ra} - \sfrac{2}{3}\,(\sfrac{1}{3}\,\Th^{2}
 + \frac{3k}{S^2})\,{\cal D}_{a} + \sfrac{1}{3}\,
 [\ \frac{2k}{S^2}\,{\cal D}_{a} - \3nab^{2}{\cal D}_{a}\ ] \ .
 \ee

\subsubsection{Harmonic decomposition}
It is standard (see, e.g., \ct{haw66} and \ct{bardeen}) to
decompose the variables harmonically, thus effectively {\em
separating out the time and space variations\/} by turning the
differential equations for time variation of the perturbations
as a whole into separate time variation equations for each
component of spatial variation, characterised by a
matter-comoving wavenumber. This conveniently represents
the idea of a matter-comoving wavelength for the matter
inhomogeneities. In our case we do so by writing ${\cal D}_{a}$ in
terms of harmonic vectors $Q^{(n)}_a$, from which the background
expansion has been factored out. \\

We start with the defining equations of the $1+3$ covariant {\bf
scalar harmonics} $Q^{(n)}$,
\be
\l{eq:defhar}
\dot{Q}{}^{(n)} = 0 \ , \hsp5
\3nab^{2}Q^{(n)} = -\,\frac{n^2}{S^2}\,Q^{(n)} \ ,
\ee
corresponding to Bardeen's {\bf scalar Helmholtz equation} (2.7)
\ct{bardeen}, but expressed $1+3$ covariantly following Hawking
\ct{haw66}. From these quantities we define the $1+3$ covariant
{\bf vector harmonics} (cf. \ct{bardeen}, equations (2.8) and (2.10);
we do not divide by the wavenumber, however, so our equations are
valid even if $n = 0$)
\be
\l{eq:defhar1}
Q^{(n)}_a \equiv S\,\3nab_{a}Q^{(n)} \hsp5 \Rightarrow \hsp5
Q^{(n)}_{a}\,u^{a} = 0 \ , \hsp5
\dot{Q}{}^{(n)}_{\la a\ra} \approx 0 \ , \hsp5
\3nab^{2}Q^{(n)}_{a} = -\,\frac{(n^2-2k)}{S^2}\,Q^{(n)}_{a} \ ,
\ee
(the factor $S$ ensuring these vector harmonics are approximately
covariantly constant along the fluid flow lines in the almost-FLRW
case). We write ${\cal D}_{a}$ in terms of these harmonics:
\be
{\cal D}_{a} = \sum_n\,{\cal D}^{(n)}\,Q^{(n)}_{a} \ , \hsp5
\3nab_{a}{\cal D}^{(n)} \approx 0 \ ,
\ee
where ${\cal D}^{(n)}$ is the harmonic component of ${\cal D}_{a}$
corresponding to the {\bf matter-comoving wavenumber} $n$,
containing the time variation of that component; to first order, ${\cal
D}^{(n)} \equiv \delta\mu^{(n)}/\overline{\mu}$. Putting this
decomposition in the linearised equations (\r{eq:second}) and
(\r{eq:second1}), the harmonics decouple. Thus, e.g., we obtain
from (\r{eq:second1}) the $n$-th harmonic equation
\be
\l{eq:harmonic}
0 = \ddot{{\cal D}}^{(n)} + (\sfrac{2}{3} - w)\,\Th\,\dot{{\cal 
D}}^{(n)} - [\ \sfrac{1}{2}\,(1-w)\,(1+3w)\,\mu - w\,\frac{n^2}{S^2} 
\ ]\ {\cal D}^{(n)} \ ,
\ee
(valid for each $n \geq 0$), where one can, if one wishes,
substitute for $\mu$ from the zeroth-order Friedmann equation
in terms of $\Th$ and $k/S^2$. This equation shows how the growth
of the inhomogeneity depends on the matter-comoving wavelength.
For the case of radiation, $w = \sfrac{1}{3}$, this is
\be
\l{eq:harmonics}
0 = \ddot{{\cal D}}^{(n)} + \sfrac{1}{3}\,\Th\,\dot{{\cal D}}^{(n)}
- [\ \sfrac{2}{3}\,\mu - \sfrac{1}{3}\,\frac{n^2}{S^2} 
\ ]\ {\cal D}^{(n)} \ .
\ee
%

\subsection{Implications}
To determine the solutions explicitly, we have to substitute for
$\mu$, $\Th$ and $S$ from the zeroth-order equations. The most
important issue is which terms dominate.

\subsubsection{Jeans instability}
Jeans' criterion is that {\bf gravitational collapse} will tend
to occur if the combination of the matter term and the term
containing the Laplace operator in (\r{eq:second}) or
(\r{eq:second1}) is {\em positive\/} \ct{jea02}; i.e., if
\be
\l{eq:jeans}
\sfrac{1}{2}\,(1-w)\,(1+3w)\,\mu\,{\cal D}_{a} > w\,
[\ \frac{2k}{S^2}\,{\cal D}_{a} - \3nab^{2}{\cal D}_{a}\ ] \ ,
\ee
when $\cs = w$. Using the harmonic decomposition, this can
be expressed in terms of an equivalent scale: from
(\r{eq:harmonic}) gravitational collapse tends to occur for a mode
${\cal D}^{(n)}$ if
\be
\l{eq:jeans1}
\sfrac{1}{2}\,(1-w)\,(1+3w)\,\mu > w\,\frac{n^2}{S^2} \ ,
\ee
i.e., if
\be
\l{eq:jeans2}
n_J \equiv \left[\ (1-w)\left(\frac{1}{w} + 3\right)
\frac{\mu(t)}{2}\ \right]^{1/2}S(t) > n \ .
\ee
In terms of wavelengths the {\bf Jeans length} is defined by
\be
\l{eq:jeans3}
\lambda_J \equiv \frac{2\pi\,S(t)}{n_J}
= c_{s}\,c\,\sqrt{\,\frac{\pi}{G\mu(t)}\,
\frac{1}{(1-w)(1+3w)}\,} \ , 
\ee
where we have re-established the fundamental constants $c$ and $G$
($w = (c_{s}/c)^{2}$). Thus, {\bf gravitational collapse}
will occur for small $n$ (wavelengths longer than $\lambda_J$), but
not for large $n$ (wavelengths less than $\lambda_J$), for the
pressure gradients are then large enough to resist the collapse and
lead to {\bf acoustic oscillations} instead. \\

For {\em non-relativistic matter\/}, $|w| \ll 1$ and $\mu =
\rho_{m}c^{2}$, where $\rho_m$ is the mass density of the matter,
so
\be
\lambda_J = c_{s}\,\sqrt{\,\frac{\pi}{G\rho_{m}(t)}\,} \ .
\ee
Then the {\bf Jeans mass} will be
\be
M_J = \frac{4\pi}{3}\,\rho_m\,\lambda_J^3
= \frac{4\pi}{3}\,c_{s}^{3}\left(\frac{\pi}{G}\right)^{3/2}
\rho_m^{-1/2} \ .
\ee

For {\em radiation\/}, where $w = \sfrac{1}{3}$ and $\mu =
\rho_{r}c^{2}$, collapse will occur if
\be
(2\mu)^{1/2} < {n_J\over S} \hsp5 \Leftrightarrow \hsp5
\lambda > \lambda_J = c_{s}\,
\sqrt{\,\frac{3\pi}{4G\rho_{r}(t)}\,} \ .
\ee
The corresponding {\bf Jeans mass} for matter coupled to the
radiation will be
\be
M_J = \frac{4\pi}{3}\,\rho_m\,\lambda_J^3
= \frac{4\pi}{3}\,c_{s}^{3}\left(\frac{3\pi}{4G\rho_r}
\right)^{3/2}\rho_m \ .
\ee
%

\subsubsection{Short-wavelength solutions}
For wavelengths much shorter than the Jeans length, equation
(\r{eq:harmonic}) becomes the damped harmonic equation
\be
\l{eq:harmonic1}
0 = \ddot{{\cal D}}^{(n)} + (\sfrac{2}{3}-w)\,\Th\,\dot{{\cal 
D}}^{(n)} + w\,\frac{n^2}{S^2}\,{\cal D}^{(n)} \ ,
\ee
giving oscillations. In the early Universe, during
radiation dominated expansion before decoupling, the tight coupling
of the dominant radiation and matter leads to a fluid with $w =
\sfrac{1}{3}$; then the short-wavelength equation becomes
\be
\l{eq:harmonics1}
0 = \ddot{{\cal D}}^{(n)} + \sfrac{1}{3}\,\Th\,\dot{{\cal 
D}}^{(n)} + \sfrac{1}{3}\,\frac{n^2}{S^2}\,{\cal D}^{(n)} \ ,
\ee
giving the {\bf acoustic oscillations} during that era for modes
such that $\lambda < \lambda_J = c_{s}c\,(3\pi/4G\mu)^{1/2}$.

\subsubsection{Long-wavelength solutions}
For wavelengths much longer than the Jeans length, we can drop the
Laplace operator terms in (\r{eq:harmonic}) to obtain
\be
\l{eq:harmonic2}
0 = \ddot{{\cal D}}^{(n)} + (\sfrac{2}{3}-w)\,\Th\,\dot{{\cal 
D}}^{(n)} - \sfrac{1}{2}\,(1-w)\,(1+3w)\,\mu\,{\cal D}^{(n)} \ .
\ee
Thus, the second-order propagation equations become ordinary
differential equations along the fluid flow lines, easily solved
for particular equations of state. In the case of radiation ($w =
\sfrac{1}{3}$) we find
\be
\l{eq:scalar1}
0 = \ddot{{\cal D}}^{(n)} + \sfrac{1}{3}\,\Th\,\dot{{\cal D}}^{(n)}
- \sfrac{2}{3}\left(\sfrac{1}{3}\,\Th^2 + \frac{3k}{S^2}\right)
{\cal D}^{(n)} \ ,
\ee
when $\lambda > \lambda_J = c_{s}c\,(3\pi/4G\mu)^{1/2}$. When
$k = 0$, then $S(t) \propto t^{1/2}$, and we obtain in this
long-wavelength limit
\be
\l{eq:rad3}
{\cal D}_{a} = d_+{}_{a}(x^\alpha)\ t
+ d_-{}_{a}(x^\alpha)\ t^{-1/2} \ ,
\hsp5 \dot{d}_i{}_{a} = 0 \ ,
\ee
(where $t$ is proper time along the flow lines). The corresponding
standard results in the synchronous and matter-comoving proper time
gauges differ, being modes proportional to $t$ and to $t^{1/2}$; we
obtain the same growth law as derived in the matter-comoving time
orthogonal gauge and equivalent gauges. As our variables are GIC,
we believe they show the latter gauges represent the physics more
accurately than any other. Note that, moreover, we obtain no
fictitious modes (proportional to $t^{-1}$) because we are using
GIC variables.

\subsubsection{Change of behaviour with time}
Any particular inhomogeneity will have a constant matter-comoving
size and, hence, constant matter-comoving wavelength $\lambda$ and
constant matter-comoving wavenumber $n$ as defined above. However,
the Jeans length will {\em vary with time\/}. \\

During the {\bf radiation era},\footnote{This is early enough
that we can ignore the curvature term in the Friedmann equation.}
$S \propto t^{1/2}$ and $\mu = \sfrac{3}{4}\,t^{-2}$
(see (\r{eq:early})), so (dropping the dimensional constants)
$\rho_m \propto t^{-3/2}$, and the matter-comoving Jeans length
\be
\lambda_J = \sqrt{\,\sfrac{1}{3}\,\pi\,t^2\,} \propto t
\ee
will steadily grow to a value $\lambda_J^{\rm max}$ at
{\bf matter--radiation equality}, while the Jeans mass of
coupled matter will grow as
\be
M_J = \frac{4\pi}{3}\,\rho_m\,\lambda_J^3 \propto
t^{-3/2}\,t^3 = t^{3/2} \ . 
\ee
Thereafter, until {\bf recombination}, the Jeans mass stays
constant: the matter and radiation are still tightly coupled
but now the Universe model is {\bf matter-dominated} and the speed
of sound of the coupled fluid depends on the matter density: $c_{s}  
(c/\sqrt{3})\,(\,1+(3\rho_m/4\rho_r)\,)$ (see Rees \ct{ree71}).
After recombination the Jeans length and mass will rapidly die away
because the matter and radiation decouple, leading to $c_{s}
\rightarrow 0$ and so $\lambda_J \rightarrow 0$. Each wavelength
$\lambda$ longer than $\lambda_J^{\rm max}$ will have a growing
mode as in (\r{eq:rad3}) until the Jeans wavelength becomes greater
than $\lambda$; it will then stop growing and undergo acoustic
oscillations which will last until decoupling when the Jeans length
drops towards zero and matter-dominated growth starts according to
(\r{eq:dust}). Growth of small perturbations eventually slows down
when the Universe model becomes {\bf curvature-dominated} at late
times (when this happens depends on $\Omega_0$; in a critical
density universe, it never occurs). \\

Thus, the {\em key times\/} for any wavelength after the initial
perturbations have been seeded\footnote{The usual assumption is
that perturbations are essentially unaffected by all the strong
interactions in the early universe after the end of inflation,
including the ending of pair production (when matter ceases to be
relativistic), decoupling of neutrinos, and the irreversible
interactions during baryosynthesis and nucleosynthesis, and
decoupling with Hot Dark Matter (`HDM') or CDM.} are (i) $t_{J}$,
when they become smaller than the Jeans length (if they are small
enough that this occurs), (ii) $t_{\rm equ}$, when the
matter-dominated era starts (which will be before decoupling of
matter and radiation, because $\Omega_0 \geq 0.1$),
(iii) $t_{\rm dec}$, when decoupling takes place. The acoustic
oscillations have constant amplitude in the radiation-dominated
era from $t_{J}$ until $t_{\rm equ}$, and then die away as
$t^{-1/6}$ in the matter-dominated era until they end at
$t_{\rm dec}$ \ct{ree71}. Baryonic inhomogeneities `freeze out'
at that time; they then start growing by damped gravitational
attraction. If they grow large enough, changing to non-linear
collapse and ultimately star formation, then local energy
generation starts. \\

{\it Exercise}: Establish these behaviours from the equations
given above.\\

By contrast CDM freezes out at $t_{\rm equ}$ and starts growth at
that time (Rees \ct{ree95}). Thus, in a {\bf CDM-dominated}
Universe model, as is often supposed, the CDM fluctuations that
govern {\bf structure formation} start gravitational growth
{\em earlier\/} than the baryons. They then govern the growth of
inhomogeneities, attracting the baryons into their potential
wells; a 2-fluid description representing the separate average
velocities and their relative motion (see below) is needed to
examine this.\\

This picture has to be modified, however, by allowing for diffusion
effects. Kinetic theory is the best way to tackle this. The result
is damping of perturbations below diffusion scales which depend on
whether or not Hot Dark Matter (`HDM') is present; baryonic
fluctuations on small scales are attenuated by photon viscosity and
free-streaming of neutrinos \ct{ree71,ree95}.

\subsection{Other matter}
Many other cases have been examined in this GIC formalism. We list
them with major references.

\subsubsection{Scalar fields}
The case of scalar fields is dealt with in a GIC way by Bruni,
Ellis and Dunsby \ct{bed92}. This analysis leads to the usual
conserved quantities and theory of growth of inhomogeneities in an
inflationary era. A {\em key element\/} here is choice of
4-velocity; for small perturbations there is a unique obvious
choice, namely, choosing $u^a$ orthogonal to the surfaces on which
the scalar field $\phi$ is constant (assuming these are spacelike).
The energy-momentum tensor then has the form of a perfect fluid,
but with energy density and pressure depending on both kinetic and
potential energy terms for $\phi$.

\subsubsection{Multi-fluids and imperfect fluids}
The physically important case of multi-fluids is dealt with by
Dunsby, Bruni and Ellis \ct{dbe92}; e.g., enabling modelling
of perturbations that include a matter--radiation interaction. The
{\em key element\/} again is choice of 4-velocity. Each component
has a separate 4-velocity $u_{(i)}^a$, and there are various
options now for the reference 4-velocity $u^a$. When linear changes
are made in this choice of 4-velocity, the essential effect is to
alter the measured momentum in the $u^a$-frame (Maartens
{\em et al\/} \ct{mageel98}). The equations are simplified most
by choosing $u^a$ as the centre of mass 4-velocity for the sum of all
components, and the 4-velocity $u_{(i)}^a$ of the $i$-th component
as its centre of mass 4-velocity.  One must then carefully check
the separate momentum and energy equations for each component, as
well as for the matter as a whole. These determine the evolution of
the $1+3$ covariantly defined {\bf relative velocities}:
$V_{ij}^a= u_{(i)}^a- u_{(j)}^a$, and the separate matter densities
$\mu_{(i)}$.\\

{\it Exercise}: Establish the equations for the relative velocity
and the energy density inhomogeneities in the 2-fluid case.\\

The case of imperfect fluids is closely related, and the same issue
of choice of 4-velocity arises. As pointed out earlier, it is
essential to use realistic equations of state in studying
perturbations of an imperfect fluid, such as described by the
M\"{u}ller--Israel--Stewart theory \ct{mue67,isrste79} (see
Maartens and Triginier \ct{matr98} for a detailed GIC analysis of
such imperfect fluids).

\subsubsection{Magnetic fields}
These have been examined in a GIC way by Barrow and Tsagas
\ct{bats97}, using the $1+3$ covariant splitting of the
electromagnetic field and Maxwell's field equations \ct{ell73}.

\subsubsection{Newtonian version}
A Newtonian version of the analysis can be developed fully in
parallel to the relativistic version \ct{ell90}, based on the
Newtonian analogue of the approach to cosmology presented in 
\ct{ell71}, and including derivations of the Newtonian Jeans length
and Newtonian formulae for the growth of inhomogeneities.\\

{\it Exercise}: Establish these equations, and hence determine the
main differences between the Newtonian and relativistic versions of
structure formation.

\subsubsection{Alternative gravity}
The same GIC approach can be used to analyze higher-derivative
gravitational theories; details are in \ct{him94}.

\subsection{Relation to other formalisms}
The relation between the GIC approach to perturbations and the very
influential gauge-invariant formalism by Bardeen \ct{bardeen} has
been examined in depth \ct{bde92}. The essential points are that \\

* as might be expected, the implications of both approaches for
 structure formation are the same,

* the implications of the GIC formalism can be worked out in any
 desired local coordinate system, including Bardeen's coordinates
 (which are incorporated into that approach in an essential way
 from the start),

* Bardeen's approach is essentially based on the linearised
 equations, while the GIC starts with the full non-linear equations
 and linearises them, as explained above. This enables an estimate
 of the errors involved, and a systematic $n$-th order
 approximation scheme,

* the GIC formalism {\em does not use a non-local splitting\/}
  \ct{stew86} into {\bf scalar}, {\bf vector} and {\bf tensor
  modes}, and only uses a harmonic splitting (into wavelengths)
  at a late stage of the analysis; these are both built into
  Bardeen's approach {\it ab initio}.\\

Both approaches have the advantage over gauge approaches that they
do not involve gauge modes, and the differential equations are of
minimal order needed to characterise the physics of the problem.
Many papers on the use of various gauges are rather confused;
however, the major paper by Ma and Bertschinger \ct{mabe96}
clarifies the relations between important gauges in a clear way. As
shown there, the answers obtained for large-scale growth of
inhomogeneities is indeed gauge-dependent, and this becomes
significant particularly on very large scales. The GIC formalism
obviates this problem. However, whatever formalism is used, the
issue is how the results relate to observations. The perturbation
theory predicts structure growth, gravitational wave emission,
gravitational lensing, and background radiation anisotropies. The
latter are one of the most important tests of the geometry and
physics of perturbed models, and are the topic of the final
section.

\section{CBR anisotropies}
Central to present day cosmology is the study of the information
obtainable from measurements of {\bf CBR anisotropies}. A GIC
version of the pioneering Sachs--Wolfe paper \ct{sw67}, based on
photon path integration and calculation of the redshift along these
paths (cf. the integration in terms of Bardeen's variables by
Panek \ct{pan86}), is given by Dunsby \ct{dun97} and Challinor and
Lasenby \ct{chalas98a}. However, a {\bf kinetic theory approach}
enables a more in-depth study of the photons' evolution and
interactions with the matter inhomogeneities, and so is the
dominant way of analyzing CBR anisotropies.

\subsection{Covariant relativistic kinetic theory}
{\bf Relativistic kinetic theory} (see, e.g., \ct{ehl71} and
\ct{li66}--\ct{ste71}) provides a self-consistent microscopically
based treatment where there is a natural unifying framework in
which to deal with a gas of particles in circumstances ranging from
hydrodynamical to free-streaming behaviour. The photon gas
undergoes a transition from hydrodynamical tight coupling with
matter, through the process of decoupling from matter, to
non-hydrodynamical free-streaming. This transition is
characterised by the evolution of the {\bf photon mean free path
length} from effectively zero to effectively infinity. This range
of behaviour can appropriately be described by kinetic theory with
non-relativistic classical {\bf Thomson scattering} (see,
e.g., Jackson \ct{jac62} or Feynman I \ct{fey1}), and the baryonic
matter with which radiation interacts can reasonably be described
hydrodynamically during these times. \\

In this approach, the single-particle {\bf photon distribution
function} $f(x^i,p^a)$ over a 7-dimensional {\bf phase space}
\ct{ehl71} represents the number of photons measured in the
3-volume element $dV$ at the event $x^i$ that have momenta in the
momentum space volume element $\pi$ about the momentum $p^a$
through the equation
\be
dN = f(x^i,p^a)\,(-p_au^a)\,\pi\,dV \ ,
\ee
where $u^a$ is the observer's 4-velocity and the redshift factor
$(-p_au^a)$ makes $f$ into a (observer-independent) scalar. The
rate of change of $f$ in photon phase space is determined by the
relativistic generalisation of {\bf Boltzmann's equation}
\be
\l{boltzmann}
L(f) = C[f] \ ,
\ee
where the {\bf Liouville operator}
\be
L(f) = \frac{\p f}{\p x^i}\,\frac{dx^i}{dv}
+ \frac{\p f}{\p p^a}\,\frac{dp^a}{dv} 
\ee
gives the change of $f$ in parameter distance $dv$ along the
geodesics that characterise the particle motions. The {\bf
collision term} $C[f]$ determines the rate of change of $f$ due
to emission, absorption and scattering processes; it can represent
Thomson scattering, binary collisions, etc. Over the period of
importance for CBR anisotropies, i.e., considerably after
electron--positron annihilation, the average photon energy is much
less than the electron rest mass and the electron thermal energy
may be neglected so that the quantised Compton interaction between
photons and electrons (the dominant interaction between radiation
and matter) may reasonably be described in the non-quantised
Thomson limit.\footnote{The $1+3$ covariant treatment described in
this section also neglects polarisation effects.} After decoupling,
there is very little interaction between matter and the CBR so we
can use {\bf Liouville's equation}:
\be
C[f] = 0 ~~\Rightarrow ~~L(f)=0 \ .
\l{eq:12}
\ee

The energy-momentum tensor of the photons is
\be
\l{eq:stressr}
T_{\!_R}^{ab} = \int_{T_x} p^a\,p^b\,f\,\pi \ ,
\ee
where $\pi$ is the volume element in the momentum tangent space
$T_x$. This satisfies the conservation equations (\r{eq:cons1}) and
(\r{eq:cons2}), and is part of the total energy-momentum tensor
$T^{ab}$ determining the space-time curvature by Einstein's
field equations (\r{eq:efe}).\\

{\em Exercise\/}: The same theory applies to particles with
non-zero rest-mass. (i) What is the form of Boltzmann's
collision term for binary particle collisions? (ii) Show that
energy-momentum conservation in individual collisions will lead to
conservation of the particle energy-momentum tensor $T^{ab}$. (iii)
Using the appropriate integral definition of entropy, determine an
H-theorem for this form of collision. (iv) Under what conditions
can equilibrium exist for such a gas of particles, and what is the
equilibrium form of the particle distribution function? (See
\ct{ehl61}.)

\subsection{Angular harmonic decomposition}
In the $1+3$ covariant approach of \ct{emt83,etm83},\footnote{Based
on the Ph.D. Thesis of R Treciokas, Cambridge University, 1972,
combined with the covariant formalism of F A E Pirani \ct{pir64}
(see also K S Thorne \ct{tho80}) by G F R Ellis when in Hamburg
(1st Institute of Theoretical Physics) in 1972; see also
\ct{egs68,treell71}. A similar formalism has been developed by M L
Wilson \ct{wil83}.} the photon 4-momentum $p^{a}$ (where
$p_{a}p^{a}=0$) is split relative to an observer moving with
4-velocity $u^{a}$ as
\be
\l{E}
p^{a} = E\,(u^{a}+e^{a}) \ , \hsp5 e_{a}e^{a} = 1 \ , \hsp5
e_{a}u^{a} = 0 \ ,
\ee
where $E = (-p_{a}u^{a})$ is the {\bf photon energy} and $e^{a}
= E^{-1}\,h^{a}{}_{b}\,p^{b}$ is the photon's {\bf spatial
propagation direction}, as measured by a matter-comoving
(fundamental) observer. Then the {\em photon distribution
function is decomposed into $1+3$ covariant\/} {\bf harmonics}
via the expansion \ct{emt83,ge98}
\be
\l{r3}
f(x,p) = f(x,E,e) = \sum_{\ell\geq
0}F_{A_\ell}(x,E)\,{e}^{A_\ell}
= F + F_ae^a + F_{ab}e^ae^b + \cdots\,
= \sum_{\ell\geq0}F_{A_\ell}e^{\la A_\ell\ra} \ ,
\ee
where ${e}^{A_\ell}\equiv e^{a_1}e^{a_2}\cdots e^{a_\ell}$, and
$e^{\la A_\ell\ra}$ is the symmetric trace-free part of
${e}^{A_\ell}$. The $1+3$ covariant {\bf distribution function
anisotropy multipoles} $F_{A_\ell}$ are irreducible since they
are Projected, Symmetric, and Trace-Free (`PSTF'), i.e.,
\[
F_{a\cdots b} = F_{\la a\cdots b\ra} \hsp5 \Leftrightarrow \hsp5
F_{a\cdots b} = F_{(a\cdots b)} \ , ~F_{a\cdots b}\,u^b = 0
= F_{a\cdots bc}\,h^{bc} \hsp5 \Rightarrow \hsp5
F_{A_{\ell}} = F_{\la A_{\ell}\ra} \ . 
\]                                                            
They encode the anisotropy structure of the photon distribution
function in the same way as the usual {\bf spherical harmonic
expansion}
\[
f = \sum_{\ell\geq0}\,
\sum_{m=-\ell}^{+\ell}f_\ell^m(x,E)\ Y_\ell^m(e) \ ,
\]
but with two major advantages: (a) the $F_{A_\ell}$ are $1+3$
covariant, and thus independent of any choice of coordinates in
momentum space, unlike the $f_\ell^m$; (b) $F_{A_\ell}$ is a
rank-$\ell$ tensor field on space-time for each {\it fixed} $E$,
and directly determines the $\ell$-multipole of radiation
anisotropy after integration over $E$. The multipoles can be
recovered from the photon distribution function via
\be
\l{r6}
F_{A_\ell} = \Delta_\ell^{-1}\int f(x,e)\,e_{\la A_\ell\ra}\,
d\Omega \ , \hsp5 \mbox{with} \hsp5 \Delta_{\ell}
= 4\pi\,{2^{\ell}\,(\ell!)^2 \over (2 \ell+1)!} \ , 
\ee
where $d\Omega = d^2{e}$ is a solid angle in {\bf momentum
space}. A further useful identity is \ct{emt83}
\be
\l{r7}
\int {e}^{A_{\ell}}\,d\Omega
= {4\pi\over \ell+1}\left\{
\begin{array}{ll}
0 & \ell ~\mbox{odd} \ , \\
{}&{} \\
h^{(a_1a_2}\,
h^{a_3a_4}\cdots h^{a_{\ell-1}a_\ell)} & \ell ~\mbox{even} \ .
\end{array}
\right.
\ee

The first three multipoles determine the radiation energy-momentum
tensor, which is, from (\r{eq:stressr}) and (\r{eq:stress}),
\be
\l{eq:star}
T_{\!_R}^{ab}(x) = \int p^a\,p^b\,f(x,p)\,d^3p
= \mu_{\!_R}\,(u^a\,u^b + {\sfrac13}\,h^{ab})
+ 2\,q_{\!_R}^{(a}\,u^{b)} + \pi_{\!_R}^{ab} \ ,
\ee
where $d^3p = E\,dE\,d\Omega$ is the covariant momentum space
volume element on the future null cone at the event $x$. It follows
from (\r{r3}) and (\r{eq:star}) that the dynamical quantities of
the radiation (in the $u^{a}$-frame) are:
\bea
\l{em3}
\mu_{\!_R} = 4\pi\int_0^\infty E^3\,F\,dE \ , \hsp5
q_{\!_R}^{a} = {4\pi\over 3}\int_0^\infty E^3\,F^a\,dE \ , \hsp5
\pi_{\!_R}^{ab} = {8\pi\over 15}\int_0^\infty E^3\,F^{ab}\,dE \ .
\eea
We extend these dynamical quantities to all multipole orders by
defining the $1+3$ covariant {\bf brightness anisotropy
multipoles}\footnote{Because photons are massless, we do not need
the complexity of the moment definitions used in \ct{emt83}. From
now on, all energy integrals will be understood to be over the
range $0\leq E\leq\infty$.} \ct{ge98}
\be
\l{r10}
\Pi_{A_{\ell}} = \int_0^\infty E^3\,F_{A_{\ell}}\,dE
= \Pi_{\la A_{\ell}\ra}\ ,
\ee
so that $\Pi = \mu_{\!_R}/4\pi$, $\Pi^a = 3\,q_{\!_R}^a/4\pi$ and
$\Pi^{ab} = 15\,\pi_{\!_R}^{ab}/8\pi$.

Writing Boltzmann's equation (\r{boltzmann}) in the form
\be
\l{boltz}
{df\over dv} \equiv p^{a}\,\vec{e}_{a}(f) - \Gamma^{a}{}_{bc}\,
p^{b}\,p^{c}\,{\p f\over \p p^{a}} = C[f] \ ,
\ee
the {\em collision term is also decomposed into $1+3$ covariant
harmonics\/}:
\be
\l{scatt}
C[f] = \sum_{\ell\geq 0} b_{A_\ell}(x,E)\,e^{A_\ell}
= b + b_ae^a + b_{ab}e^ae^b + \cdots \ ,
\ee
where the $1+3$ covariant {\bf scattering multipoles}
$b_{A_{\ell}} = b_{\la A_{\ell}\ra}$ encode irreducible properties
of the particle interactions. Then Boltzmann's equation is
equivalent to an {\em infinite hierarchy of $1+3$ covariant\/}
{\bf multipole equations}
\[
L_{A_\ell}(x,E) = b_{A_\ell}[\,F_{A_m}(x,E)\,] \ ,
\]
where $L_{A_{\ell}} = L_{\la A_{\ell}\ra}$ are the anisotropy
multipoles of $df/dv$, and will be given in the next
subsection. These multipole equations are tensor field equations on
space-time for each value of the photon energy $E$ (but note that
energy changes along each photon path). Given the solutions
$F_{A_\ell}(x,E)$ of the equations, the relation (\r{r3}) then
determines the full photon distribution function $f(x,E,e)$ as a
scalar field over phase space.

\subsection{Non-linear $1+3$ covariant multipole equations}          
The full Boltzmann equation in photon phase space contains {\em
more\/} information than necessary to analyze radiation
anisotropies in an inhomogeneous Universe model. For that purpose,
when the radiation is close to {\bf black body}, we do not require
the full spectral behaviour of the photon distribution function
multipoles, but only the {\bf energy-integrated multipoles}. The
monopole leads to the average (all-sky) temperature, while the
higher-order multipoles determine the temperature anisotropies.
The GIC definition of the {\bf average temperature} $T$ is given
according to the {\bf Stefan--Boltzmann law} by
\be
\l{r27}
\mu_{\!_R}(x) = 4\pi\int E^3\,F(x,E)\,dE = r\,T^{4}(x) \ ,
\ee
where $r$ is the {\bf radiation constant}. If $f$ is close to a
Planckian distribution, then $T$ is the thermal black body average
temperature. But note that {\em no\/} notion of background
temperature is involved in this definition. There is an all-sky
average implied in (\r{r27}). {\bf Fluctuations} across the sky
are measured by energy-integrating the higher-order multipoles (a
precise definition is given below), i.e., the fluctuations are
determined by the $\Pi_{A_{\ell}}$ ($\ell\geq1$) defined in
(\r{r10}).\\

The form of $C[f]$ in (\r{scatt}) shows that $1+3$ covariant
equations for the temperature fluctuations arise from decomposing
the {\bf energy-integrated Boltzmann equation}
\be
\l{ibe}
\int E^2\,{df \over dv}\,dE = \int E^2\,C[f]\,dE
\ee
into $1+3$ covariant multipoles. We begin with the right-hand side,
which requires the $1+3$ covariant form of the Thomson scattering
term. Defining the $1+3$ covariant {\bf energy-integrated
scattering multipoles}
\be
K_{A_\ell} = \int E^2\,b_{A_\ell}\,dE
= K_{\la A_{\ell}\ra} \ ,
\ee
we find that \ct{mageel98}
\bea
\l{rr1}
K & = & \ne\st\left[\ {\ts{4\over3}}\,\Pi\vb^2
- {\ts{1\over3}}\,\Pi^av_{\!_Ba}\ \right] + {\cal O}[3] \ , \\
\l{rr2}
K^a & = & -\,\ne\st\left[\ \Pi^a-4\,\Pi\vb^a - {\ts{2\over5}}\,
\Pi^{ab}v_{\!_Bb}\ \right] + {\cal O}[3] \ , \\
\l{rr3}
K^{ab} & = & -\,\ne\st\left[\ {\ts{9\over10}}\,\Pi^{ab}
- {\ts{1\over2}}\,\Pi^{\la a}\vb^{b\ra} - {\ts{3\over7}}\,
\Pi^{abc}v_{\!_Bc}\ - 3\Pi v_{_B}^{\la a}v_{_B}^{b\ra}
\ \right] + {\cal O}[3] \ ,\\
\l{rr3a}
K^{abc} & = & -\,\ne\st\left[\ \,\Pi^{abc}
- {\ts{3\over2}}\,\Pi^{\la ab}\vb^{c\ra} - {\ts{4\over9}}\,
\Pi^{abcd}v_{\!_Bd}\ \right] + {\cal O}[3] \ ,
\eea
and, for $\ell > 3$,                                     
\be
\l{r21}
K^{A_\ell}
= -\,\ne\st\left[\ \Pi^{A_\ell} - \Pi^{\la A_{\ell-1}}
\vb^{a_\ell\ra} - \left({\ell+1\over 2\ell+3}\right)
\Pi^{A_\ell a}v_{\!_Ba}\ \right] + {\cal O}[3] \ ,
\ee
where the expansion is in terms of the {\bf peculiar velocity}
$v_B^a$ of the baryons relative to the reference frame $u^a$.
Parameters are the {\bf free electron number density} $\ne$
and the {\bf Thomson scattering cross section} $\st$, the
latter being proportional to the square of the classical
electron radius \ct{jac62,fey1}. The first three multipoles
are affected by Thomson scattering differently than the
higher-order multipoles.
                                                              
Equations (\r{rr1})--(\r{r21}), derived in \ct{mageel98}, are a
non-linear generalisation of the results given by Challinor and
Lasenby \ct{cl98}. They show the {\em coupling of baryonic bulk
velocity to the radiation multipoles, arising from $1+3$ covariant
non-linear effects in Thomson scattering.} If we linearise fully,
i.e., neglect all terms containing $\vb$ except the
$\mu_{\!_R}\vb^a$ term in the dipole $K^a$, which is first-order,
then our equations reduce to those in \ct{cl98}. The generalised
non-linear equations apply to the analysis of $1+3$ covariant
second-order effects on an FLRW background, to first-order effects
on a spatially homogeneous but anisotropic background, and more
generally, to any situation where the baryonic frame is in
non-relativistic motion relative to the fundamental $u^a$-frame. \\

Next we require the anisotropy multipoles $L_{A_{\ell}}$ of
$df/dv$. These can be read off for photons directly from the
general expressions in \ct{emt83}, which are exact, $1+3$
covariant, and also include the case of massive particles. For
clarity and completeness, we outline an alternative, $1+3$
covariant derivation (the derivation in \ct{emt83} uses
tetrads). We require the rate of change of the photon energy along
null geodesics, given by \ct{egs68,treell71}
\be
\l{dEdv}
{dE \over dv} = -\,\left[\ \sfrac{1}{3}\,\Th
+ (\udot_{a}e^{a}) + (\sig_{ab}e^{a}e^{b})\ \right] E^{2} \ ,
\ee
which follows directly from $p^{b}\nabla_{b}p^{a} = 0$ with (\r{E}) 
and (\r{eq:kin}). Then
\begin{eqnarray*}
{d\over dv}\left[\ F_{a_1\cdots a_\ell}(x,E)\,
e^{a_1}\cdots e^{a_{\ell}}\ \right]
& = &
{d\over dv}\left[\ E^{-\ell}\,F_{a_1\cdots a_\ell}(x,E)\,
p^{a_1}\cdots p^{a_{\ell}}\ \right] \\
{} & = & E\left\{\ [\ {\ts{1\over3}}\,\Th + \udot_be^b
+ \sig_{bc}e^be^c\ ]\left(\ell\,F_{a_1\cdots a_\ell}
- E\,F'_{a_1\cdots a_\ell}\right)e^{a_1}\cdots{e}^{a_\ell}
\right. \\
&&{}\left.
+ \ \left(u^{a_1}+e^{a_1}\right)\cdots\left(u^{a_\ell}+
e^{a_\ell}\right)\,[\ \dot{F}_{a_1\cdots a_\ell}
+e^b\nabla_bF_{a_1\cdots a_\ell}\ ]\ \right\} \ ,
\end{eqnarray*}
where a prime denotes $\p/\p E$. The first term is readily put into
irreducible PSTF form using identities in \ct{emt83}, p~470.  In
the second term, when the round brackets are expanded, only those
terms with at most one $u^{a_r}$ survive, and
\[
u^{a}\dot{F}_{a\cdots} = -\,\udot^{a}F_{a\cdots} \ , \hsp5
u^{b}\nabla^{a}F_{b\cdots} = -\,(\,{\ts{1\over3}}\,\Th\,h^{ab}
+ \sig^{ab} + \eta^{abc}\,\omega_c\,)\,F_{b\cdots} \ .
\]
Thus, the $1+3$ covariant {\bf anisotropy multipoles}
$L_{A_\ell}$ of $df/dv$ are
\bea
\l{r25}
E^{-1}\,L_{A_\ell}
& = &
\dot{F}_{\la A_\ell \ra} - {\ts{1\over3}}\,\Th\,E\,F'_{A_\ell}
+ \3nab_{\la a_\ell}F_{A_{\ell-1}\ra} + {(\ell+1)\over(2\ell+3)}\,
\3nab^bF_{A_\ell b} \nonumber\\
&&{}
-\ {(\ell+1)\over(2\ell+3)}\,E^{-(\ell+1)}\left[E^{\ell+2}\,
F_{A_\ell b}\right]'\udot^b - E^\ell\left[E^{1-\ell}\,
F_{\la A_{\ell-1}}\right]'\udot_{a_\ell\ra} \nonumber\\
&&{}
+ \ \ell\,\omega^b\,\eta_{bc\la a_\ell}\,F_{A_{\ell-1}\ra}{}^c
- {(\ell+1)(\ell+2)\over(2\ell+3)(2\ell+5)}\,E^{-(\ell+2)}
\left[E^{\ell+3}\,F_{A_\ell bc}\right]'\sig^{bc} \nonumber\\
&&{}
- \ {2\ell\over (2\ell+3)}\,E^{-1/2}\left[E^{3/2}\,F_{b\la A_{\ell-1}}
\right]'\sigma_{a_\ell\ra}{}^b - E^{\ell-1}\left[E^{2-\ell}\,
F_{\la A_{\ell-2}}\right]'\sigma_{a_{\ell-1}a_\ell\ra} \\
& = & E^{-1}\,b_{A_\ell} \ . \nonumber
\eea

This regains the result of \ct{emt83} (\,equation (4.12)\,) in the
massless case. The form given here benefits from the streamlined
version of the $1+3$ covariant formalism. We re-iterate that this
result is {\em exact\/} and holds for any photon or (massless)
neutrino distribution in {\em any\/} space-time.

We now multiply (\r{r25}) by $E^3$ and integrate over all photon
energies, using integration by parts and the fact that
$E^n\,F_{a\cdots}\rightarrow 0$ as $E\rightarrow\infty$ for any
positive $n$. We obtain the evolution equations that determine the
brightness anisotropies multipoles $\Pi_{A_{\ell}}$:
\bea
\l{r26}
\dot{\Pi}_{\la A_\ell\ra}
+ {\ts{4\over3}}\,\Th\,\Pi_{A_\ell}
+ \3nab_{\la a_\ell}\Pi_{A_{\ell-1}\ra}
+ {(\ell+1)\over(2\ell+3)}\,\3nab^b\Pi_{A_\ell b}
&&{} \\
- \ {(\ell+1)(\ell-2)\over(2\ell+3)}\,\udot^b\,\Pi_{A_\ell b}
+ (\ell+3)\,\udot_{\la a_\ell}\,\Pi_{A_{\ell-1}\ra}
+ \ell\,\omega^b\,\eta_{bc\la a_\ell}\,\Pi_{A_{\ell-1}\ra}{}^c
&&{} \nonumber\\
- \ {(\ell-1)(\ell+1)(\ell+2)\over(2\ell+3)(2\ell+5)}\,
\sig^{bc}\,\Pi_{A_\ell bc} + {5\ell\over(2\ell+3)}\,
\sig^b{}_{\la a_\ell}\,\Pi_{A_{\ell-1}\ra b}
- (\ell+2)\,\sigma_{\la a_{\ell}a_{\ell-1}}\,\Pi_{A_{\ell-2}\ra}
& = & K_{A_\ell} \ . \nonumber
\eea

Once again, this is an exact result, and it holds also for any
collision term, i.e., any $K_{A_\ell}$. For decoupled neutrinos, we
have $K_{\!_N}^{A_\ell}=0$ in this equation. For photons undergoing
Thomson scattering, the right-hand side of (\r{r26}) is given by
(\r{r21}), which is exact in the kinematical and dynamical
quantities, but first-order in the relative baryonic velocity. The
equations (\r{r21}) and (\r{r26}) thus constitute a non-linear
generalisation of the FLRW-linearised case given by Challinor and
Lasenby \ct{cl98}.\\

The monopole and dipole of equation (\r{r26}) give the evolution
equations for the energy and momentum densities:
\bea
\l{r26a}
\dot{\Pi} + {\ts{4\over3}}\,\Th\,\Pi
+ {\ts{1\over3}}\,\3nab_a\Pi^a
+ {\ts{2\over3}}\,\udot_a\Pi^a
+ {\ts{2\over15}}\,\sigma_{ab}\Pi^{ab}
& = & K \ , \\
\l{r26b}
\dot{\Pi}^{\la a\ra} + {\ts{4\over3}}\,\Th\,\Pi^a
+ \3nab^a\Pi + {\ts{2\over5}}\,\3nab_{b}\Pi^{ab}
{}&& \nonumber \\
+ \ {\ts{2\over5}}\,\udot_b\,\Pi^{ab}
+ 4\,\Pi\,\udot^a + \eta^{abc}\,\om_{b}\,\Pi_{c}
+ \sigma^{ab}\,\Pi_b
& = & K^{a} \ ,
\eea
(these are thus the equations giving the divergence of $T^{ab}$
in (\r{eq:stressr})). For {\em photons\/}, the equations are:
\bea
\l{nl5}
\dot{\mu}_{\!_R} + {\ts{4\over3}}\,\Th\,\mu_{\!_R}
+ \3nab_{a}q_{\!_R}^a + 2\,\udot_aq_{\!_R}^a
+ \sigma_{ab}\pi_{\!_R}^{ab}
& = & \ne\st\,(\,{\ts{4\over3}}\,\mu_{\!_R}\vb^2
- q_{\!_R}^av_{\!_Ba}\,) + {\cal O}[3] \ , \\
\l{nl6}
\dot{q}_{\!_R}^{\la a\ra} + {\ts{4\over3}}\,\Th\,q_{\!_R}^a
+ {\ts{4\over3}}\,\mu_{\!_R}\,\udot^a
+ {\ts{1\over3}}\,\3nab^a\mu_{\!_R}
+ \3nab_{b}\pi_{\!_R}^{ab}
&&\nonumber\\
+ \ \sigma^a{}_b\,q_{\!_R}^b + \eta^{abc}\,\om_{b}\,q_{\!_Rc}
+ \udot_b\,\pi_{\!_R}^{ab}
& = & \ne\st\,(\,{\ts{4\over3}}\,\mu_{\!_R}\,\vb^a
- q_{\!_R}^a + \pi_{\!_R}^{ab}\,v_{\!_Bb}\,) + {\cal O}[3] \ ,
\eea
(the present versions of the fluid energy and momentum
conservation equations (\r{eq:cons1}) and (\r{eq:cons2})).

The non-linear dynamical equations are completed by the
energy-integrated Boltzmann multipole equations. For {\em
photons\/}, the quadrupole evolution equation is
\bea
\l{nl8}
&&\dot{\pi}_{\!_R}^{\la ab\ra} + {\ts{4\over3}}\,\Th\,
\pi_{\!_R}^{ab} + {\ts{8\over15}}\,\mu_{\!_R}\,\sig^{ab}
+ {\ts{2\over5}}\,\3nab^{\la a}q_{\!_R}^{b\ra}
+ {8\pi\over35}\,\3nab_{c}\Pi^{abc} \nonumber\\
&&{} + \ 2\,\udot^{\la a}\,q_{\!_R}^{b\ra}
+ 2\,\om^c\,\eta_{cd}{}^{\la a}\,\pi_{\!_R}^{b\ra d}
+ {\ts{2\over7}}\,\sig_c{}^{\la a}\,\pi_{\!_R}^{b\ra c}
- {32\pi\over315}\,\sig_{cd}\,\Pi^{abcd} 
\nonumber\\
{}&&{} = -\,\ne\st\,(\,{\ts{9\over10}}\,\pi_{\!_R}^{ab}
- {\ts{1\over5}}\,q_{\!_R}^{\la a}\,\vb^{b\ra}
- {8\pi\over35}\,\Pi^{abc}\,v_{\!_Bc}
- {\textstyle{2\over5}}\,\rho_{_R}\,v_{_B}^{\langle a}\,
v_{_B}^{b\rangle}) + {\cal O}[3] \ ,
\eea
(a fluid description gives no analogue of this and the following
equations). The higher-order multipoles ($\ell> 3$) evolve
according to
\bea
\l{nl9}
&&\dot{\Pi}^{\la A_\ell\ra} + {\ts{4\over3}}\,\Th\,\Pi^{A_\ell}
+ \3nab^{\la a_\ell}\Pi^{A_{\ell-1}\ra}
+ {(\ell+1)\over(2\ell+3)}\,\3nab_b\Pi^{A_\ell b}
\nonumber\\
&&{} - \ {(\ell+1)(\ell-2)\over(2\ell+3)}\,\udot_b\,\Pi^{A_\ell b}
+ (\ell+3)\,\udot^{\la a_\ell}\,\Pi^{A_{\ell-1}\ra}
+ \ell\,\omega^b\,\eta_{bc}{}^{\la a_\ell}\,\Pi^{A_{\ell-1}\ra c}
\nonumber\\
&&{} - \ {(\ell-1)(\ell+1)(\ell+2)\over(2\ell+3)(2\ell+5)}\,
\sig_{bc}\,\Pi^{A_\ell bc} + {5\ell\over(2\ell+3)}\,
\sigma_b{}^{\la a_\ell}\,\Pi^{A_{\ell-1}\ra b} 
- (\ell+2)\,\sig^{\la a_{\ell}a_{\ell-1}}\,\Pi^{A_{\ell-2}\ra}
\nonumber\\
&&{} = -\,\ne\st\left[\ \Pi^{A_\ell} - \Pi^{\la A_{\ell-1}}\,
\vb^{a_\ell\ra} - \left({\ell+1\over 2\ell+3}\right)
\Pi^{A_\ell a}\,v_{\!_Ba}\ \right] + {\cal O}[3] \ .
\eea
For $\ell=3$, the $\Pi^{\la A_{\ell-1}}v_{_B}^{{a_\ell}\ra}$ term
on the right-hand side of equation (\r{nl9}) must be multiplied by
${3\over2}$.  For neutrinos, the equations are the same except
without the Thomson scattering terms. Note that these equations
link angular multipoles of order $\ell-2$, $\ell-1$, $\ell$,
$\ell+1$, $\ell+2$, i.e., they link {\em five\/} successive harmonic
terms. This is the source of the {\bf harmonic mixing} that
occurs as the radiation propagates.

These equations for the radiation (and neutrino) multipoles
generalise the equations given by Challinor and Lasenby \ct{cl98},
to which they reduce when we remove all terms ${\cal O}(\epsilon
v_{\!_I})$ and ${\cal O}(\epsilon^2)$. In this case, on introducing
a {\bf FLRW-linearisation}, there is major simplification of the
equations:
\bea
\l{nl5a}
\dot{\mu}_{\!_R} + {\ts{4\over3}}\,\Th\,\mu_{\!_R}
+ \3nab_{a}q_{\!_R}^{a}
& \approx & 0 \ , \\
\l{nl6a}
\dot{q}_{\!_R}^{\la a\ra} + \sfrac{4}{3}\,\Th\,q_{\!_R}^a
+ {\ts{4\over3}}\,\mu_{\!_R}\,\udot^a
+ {\ts{1\over3}}\,\3nab^a\mu_{\!_R} + \3nab_{b}\pi_{\!_R}^{ab} 
& \approx & \ne\st\,(\,{\ts{4\over3}}\,\mu_{\!_R}\,\vb^a
- q_{\!_R}^a\,) \ , \\
\l{nl8a}
\dot{\pi}_{\!_R}^{\la ab\ra} + \sfrac{4}{3}\,\Th\,\pi_{\!_R}^{ab}
+ {\ts{8\over15}}\,\mu_{\!_R}\,\sigma^{ab}
+ {\ts{2\over5}}\,\3nab^{\la a}q_{\!_R}^{b\ra}
+ {8\pi\over35}\,\3nab_{c}\Pi^{abc}
& \approx & - \,{\ts{9\over10}}\,\ne\st\,\pi_{\!_R}^{ab} \ ,
\eea
and, for $\ell\geq3$,
\be
\l{nl9a}
\dot{\Pi}^{\la A_\ell\ra} + \sfrac{4}{3}\,\Th\,\Pi^{A_\ell}
+ \3nab^{\la a_\ell}\Pi^{A_{\ell-1}\ra}
+ {(\ell+1)\over(2\ell+3)}\,\3nab_b\Pi^{A_\ell b}
\approx -\,\ne\st\,\Pi^{A_\ell} \ .
\ee
Note that these equations now link only angular multipoles of order
$\ell-1$, $\ell$, $\ell+1$, i.e., they link {\em three\/} successive
terms. This is a major qualitative difference from the full
non-linear equations.

\subsection{Temperature anisotropy multipoles}
Finally, we return to the definition of temperature anisotropies.
As noted above, these are determined by the $\Pi_{A_\ell}$.  We
define the {\bf temperature fluctuation} $\tau(x,e)$ via the
directional temperature which is determined by the directional
bolometric brightness:
\be
\l{r28}
T(x,e) = T(x)\,[\ 1 + \tau(x,e)\ ]
= \left[\ {4\pi\over r}\int E^3\,f(x,E,e)\,dE\ \right]^{1/4} \ .
\ee
This is a GIC definition which is also exact. We can rewrite it
explicitly in terms of the $\Pi_{A_\ell}$:
\be
\l{r29}
\tau(x,e) = \left[\ 1+{4 \pi \over\mu_{\!_R}}\sum_{\ell\geq 1}
\Pi_{A_\ell}{e}^{A_\ell}\ \right]^{1/4} - 1
= \tau_ae^a + \tau_{ab}e^ae^b + \cdots
= \sum_{\ell\geq 1} \tau_{A_{\ell}}(x)\,e^{A_{\ell}} \ .
\ee
In principle, we can extract the $1+3$ covariant irreducible PSTF
{\bf temperature anisotropy multipoles} $\tau_{A_{\ell}} =
\tau_{\la A_{\ell}\ra}$ by using the inversion (\r{r6}):
\be
\l{r30}
\tau_{A_\ell}(x) = {\Delta_{\ell}^{-1}} \int \tau(x,e)\,
e_{\la A_\ell\ra}\,d\Omega \ .                                 
\ee

In the almost-FLRW case, when $\tau$ is ${\cal O}[1]$, we regain
from (\r{r29}) the linearised definition given in \ct{mes95a}:
\be
\l{r31}
\tau_{A_\ell} \approx \left({\pi \over \mu_{\!_R}}\right) 
\Pi_{A_\ell} \ ,
\ee
where $\ell\geq 1$. In particular, the dipole and quadrupole are
\be
\l{dip-quad}
\tau^a \approx {3q_{\!_R}^a\over 4\mu_{\!_R}} \hsp5
\mbox{and} \hsp5
\tau^{ab} \approx {15\pi_{\!_R}^{ab}\over 2\mu_{\!_R}} \ . 
\ee

We can normalise the dynamical brightness anisotropy multipoles
$\Pi_{A_\ell}$ of the radiation to define the dimensionless $1+3$
covariant {\bf brightness temperature anisotropy multipoles}
($\ell\geq1$)
\be
{\cal T}_{A_\ell} = \left({\pi\over r\,T^4}\right)\Pi_{A_\ell}
\approx \tau_{A_\ell} \ .
\ee
Thus, the ${\cal T}_{A_{\ell}} = {\cal T}_{\la A_{\ell}\ra}$ are
equal to the temperature anisotropy multipoles plus non-linear
corrections. In terms of these quantities, the {\bf hierarchy of
radiation multipoles} becomes}:
\bea
\l{nl10}
{\dot{T}\over T} & = & -\,{\ts{1\over3}}\,\Th - {\ts{1\over3}}\,
\3nab_a{\cal T}^a 
- {\ts{4\over3}}\,{\cal T}^a\,{\3nab_a T\over T}
- {\ts{2\over3}}\,\udot_a\,{\cal T}^a
- {\ts{2\over15}}\,\sigma_{ab}\,{\cal T}^{ab} \nonumber\\
&&{} + {\ts{1\over3}}\,\ne\st\,v_{\!_Ba}\,(\,\vb^a - {\cal T}^a\,)
+ {\cal O}[3] \ , \\
\l{nl11}
\dot{{\cal T}}^{\la a\ra} & = & -\,4\left({\dot T\over T}
+{\ts{1\over3}}\,\Th\right){\cal T}^a
- {\3nab^aT\over T} - \udot^a - {\ts{2\over5}}\,\3nab_b{\cal T}^{ab}
+ \ne\st\,(\,\vb^a-{\cal T}^a\,) \nonumber\\
&&{} + \ {\ts{2\over5}}\,\ne\st\,{\cal T}^{ab}\,v_{\!_Bb}
- \sigma^a{}_b\,{\cal T}^b - {\ts{2\over5}}\,\udot_b\,{\cal T}^{ab}
- \eta^{abc}\,\om_{b}\,{\cal T}_{c}
- {\ts{8\over5}}\,{\cal T}^{ab}\,{\3nab_b T\over T}
+ {\cal O}[3] \ , \\
\l{r38}
\dot{{\cal T}}^{\la ab\ra}  & = &
- \,4\left({\dot{T}\over T} + {\ts{1 \over 3}}\,\Th\right)
{\cal T}^{ab} - \sigma^{ab} - \3nab^{\la a}\,{\cal T}^{b\ra} 
- {\ts{3 \over 7}}\,\3nab_c{\cal T}^{abc}
- {\ts{9\over10}}\,\ne\st\,{\cal T}^{ab} \nonumber \\
&&{} + \ \ne\st\,(\,{\ts{1\over2}}\,{\cal T}^{\la a}\,\vb^{b\ra}
+ {\ts{3\over7}}\,{\cal T}^{abc}\,v_{\!_Bc}\,
+{\textstyle{3\over4}}v_{_B}^{\langle a}v_{_B}^{b\rangle}
)
- 5\,\udot^{\la a}\,{\cal T}^{b\ra}
- {\ts{4 \over 21}}\,\sigma_{cd}\,{\cal T}^{abcd} \nonumber\\
&&{} - \ 2\,\omega^c\,\eta_{cd}{}^{\la a}\,{\cal T}^{b\ra d} 
- {\ts{10 \over 7}}\,\sigma_c{}^{\la a}\,{\cal T}^{b\ra c}
- {\ts{12\over7}}\,{\cal T}^{abc}\,{\3nab_c T\over T} 
+ {\cal O}[3] \ ,
\eea
and, for $\ell>3$: 
\bea
\l{r36}
\dot{\cal T}^{\la A_{\ell}\ra} & = &
-\,4 \left({\dot{T}\over T} + {\ts{1\over 3}}\,\Th\right) 
{\cal T}^{A_\ell}
- \3nab^{\la a_\ell}{\cal T}^{A_{\ell-1}\ra}
- {(\ell+1)\over (2 \ell+3)}\,\3nab_b{\cal T}^{A_{\ell}b} 
- \,\ne\st\,{\cal T}^{A_\ell} \nonumber\\
&&{} + \ \ne\st\left[\ {\cal T}^{\la A_{\ell-1}}\,\vb^{a_\ell\ra}
+ \left({\ell+1\over 2\ell+3}\right){\cal T}^{A_\ell b}\,
v_{\!_Bb}\ \right] + {(\ell+1)\,(\ell-2) \over (2\ell+3)}\,
\udot_b\,{\cal T}^{A_{\ell}b} \nonumber\\
&&{} -\ (\ell+3)\,\udot^{\la a_\ell}\,{\cal T}^{A_{\ell-1} \ra} 
- \ell\,\omega^b\,\eta_{bc}{}^{\la a_{\ell}}\,
{\cal T}^{A_{\ell-1}\ra c}
+ (\ell+2)\,\sigma^{\la a_{\ell} a_{\ell-1}}\,
{\cal T}^{A_{\ell-2}\ra} \nonumber \\
&&{} + \ {(\ell-1)\,(\ell+1)\,(\ell+2) \over (2\ell+3)\,(2\ell+5)} 
\,\sig_{bc}\,{\cal T}^{A_\ell bc}
- { 5\ell \over (2\ell+3)}\,\sigma_b{}^{\la a_{\ell}}\,
{\cal T}^{A_{\ell-1}\ra b} \nonumber \\
&&{} - \ 4\,{(\ell+1) \over (2\ell+3)}\,
{\cal T}^{A_\ell b}\,{\3nab_b T\over T} + {\cal O}[3] \ .
\eea
For $\ell=3$, the Thomson scattering term 
${\cal T}^{\langle A_{\ell-1}}v_{_B}^{{a_\ell}\rangle}$ must be 
multiplied by a factor ${3\over2}$.

The $1+3$ covariant non-linear multipole equations given in this
form show more clearly the evolution of temperature anisotropies
(including the monopole, i.e., the average temperature $T$), in
general linking five successive harmonics. Although the ${\cal
T}_{A_\ell}$ only determine the actual temperature fluctuations
$\tau_{A_\ell}$ to linear order, they are a useful dimensionless
measure of anisotropy. Furthermore, equations (\r{nl10})--(\r{r36})
apply as the evolution equations for the temperature anisotropy
multipoles when the radiation anisotropy is small (i.e., ${\cal
T}_{A_\ell} \approx \tau_{A_\ell}$), but the space-time
inhomogeneity and anisotropy are not restricted.  This includes the
particular case of small CBR anisotropies in general Bianchi
models, or in perturbed Bianchi models.

{\bf FLRW-linearisation}, i.e., the case when only first-order
effects relative to the FLRW limit are considered, reduces the
above equations to the linearised form, generically linking three
successive harmonics:
\bea
\l{nl10a}
{\dot{T}\over T} & \approx & -\,{\ts{1\over3}}\,\Th
- {\ts{1\over3}}\,\3nab_a{\tau}^a \ , \\
\l{nl11a}
\dot{{\tau}}^{\la a\ra} & \approx & -\,{\3nab^aT\over T}
- \udot^a - {\ts{2\over5}}\,\3nab_b{\tau}^{ab}
+ \ne\st\,(\,\vb^a-{\tau}^a\,) \ , \\
\l{r38a}
\dot{{\tau}}^{\la ab\ra}  & \approx & -\,\sigma^{ab}
- \3nab^{\la a}{\tau}^{b \ra} 
- {\ts{3 \over 7}}\,\3nab_c{\tau}^{abc}
- {\ts{9\over10}}\,\ne\st\,{\tau}^{ab} \ ,
\eea
and, for $\ell\geq3$: 
\bea
\dot{\tau}^{\la A_{\ell}\ra} & \approx &
-\,\3nab^{\la a_\ell}{\tau}^{A_{\ell-1}\ra}
- {(\ell+1)\over (2 \ell+3)}\,\3nab_b{\tau}^{A_{\ell}b} 
- \ne\st\,{\tau}^{A_\ell} \ .
\l{r36a}
\eea
These are GIC multipole equations leading to the {\bf Fourier mode
formulation} of the energy-integrated Boltzmann equations used in
the standard literature (see, e.g., \ct{HS95a} and the references
therein) when they are decomposed into spatial harmonics and
associated wavelengths.\footnote{They contain the free-streaming
subcase for $\ne = 0$.} This then allows examination of diffusion
effects, which are wavelength-dependent.

These linearised equations, together with the linearised equations
governing the kinematical and gravitational quantities, may be
$1+3$ covariantly split into {\bf scalar}, {\bf vector} and
{\bf tensor modes}, as described in \ct{bde92,cl98}. The modes can
then be expanded in $1+3$ covariant eigentensors of the
matter-comoving Laplacian, and the Fourier coefficients obey
ordinary differential equations.\\

{\em Exercise\/}: Determine the resulting hierarchy of mode
equations.  Show from these equations that there is a wavelength
$\lambda_S$ such that for shorter wavelengths the perturbations are
heavily damped (the physical reason is photon diffusion).\\

Numerical integrations of these equations are performed for scalar
modes by Challinor and Lasenby \ct{cl98}, with further analytic
results given in \ct{ge2}. These are the Sachs--Wolfe family of
integrations, of fundamental importance in determining CBR
anisotropies, as discussed in many places; see, e.g.,  Hu and
Sugiyama \ct{HS95a}, Hu and White \ct{hw96} (the GIC version will
be given in the series of papers by Gebbie, Maartens, Dunsby and
Ellis). However, they assume an almost-FLRW geometry. We turn now
to justifying that assumption.

\subsection{Almost-EGS-Theorem and its applications}
One of our most important understandings of the nature of the
Universe is that at recent times it is well-represented by the
standard spatially homogeneous and isotropic FLRW models. The basic
reason for this belief is the observed high degree of isotropy of
the CBR, together with a fundamental result of Ehlers, Geren and
Sachs \ct{egs68} (hereafter `EGS'), taken nowadays (see,
e.g., \ct{he73}) to establish that the Universe is almost-FLRW at
recent times (i.e., since decoupling of matter and radiation). \\
        
The {\bf EGS programme} can be summarized as follows: Using (a) the
measured high isotropy of the CBR at our space-time position, and
(b) the {\bf Copernican assumption} that we are not at a privileged
position in the Universe, the aim is to deduce that the Universe is
accurately FLRW. EGS gave an exact theorem of this kind: if a
family of freely-falling observers measure self-gravitating
background radiation to be everywhere {\it exactly} isotropic in
the case of non-interacting matter and radiation, then the Universe
is {\it exactly} FLRW. This is taken to establish the desired
conclusion, in view of the measured near-isotropy of the radiation
at our space-time location. However, of course the CBR is not
exactly isotropic. Generally, we want to show stability of
arguments we use \ct{tavell88}; in this case, we wish to do so by
showing the EGS result remains {\it nearly} true if the radiation
is {\it nearly} isotropic, thus providing the foundation on which
further analyses, such as that in the famous Sachs--Wolfe paper
\ct{sw67}, are based. More precisely, we aim to prove the following
theorem \ct{sme95}.

\begin{quotation}
{\bf Almost-EGS-Theorem}: {\it If} the Einstein--Liouville
equations are satisfied in an expanding Universe model, where
there is present pressure-free matter with 4-velocity vector
field $u^a$ ($u_au^a = -\,1$) such that (freely-propagating)
background radiation is everywhere almost-isotropic relative
to $u^a$ in some domain $U$, {\it then} the space-time is
almost-FLRW in $U$.
\end{quotation}
        
This description is intended to represent the situation in the
Universe since decoupling to the present day. The pressure-free
matter represents the galaxies on which fundamental observers
live, who measure the radiation to be almost isotropic.

\subsubsection{Assumptions}
In detail, we consider matter and radiation in a space-time region
$U$. In our application, we consider $U$ to be the region within
and near our past light cone from decoupling to the present day
(this is the observable space-time region where we would like to
prove the Universe is almost-FLRW; before decoupling a different
analysis is needed, for collisions dominate there, and we do not
have sufficient data to comment on the situation far from our past
light cone).  Our assumptions are that, in the region considered,\\
        
(1) Einstein's field equations are satisfied, with the total
energy-momentum tensor $T_{ab}$ composed of
{\em non-interacting\/} matter and radiation components:
\be
\l{eq:7}
T^{ab} = T_{\!_M}^{ab} + T_{\!_R}^{ab} \ , \hsp5
\nabla_{b}T_{\!_M}^{ab} = 0 \ , \hsp5
\nabla_{b}T_{\!_R}^{ab} = 0 \ .
\ee
The independent conservation equations express the decoupling of
matter from radiation; the matter energy-momentum tensor is
$T_{\!_M}^{ab} = \rho\,u^a\,u^b$, ($\mu_{\!_M} = \rho$ is the
matter energy density) and, at each point, the radiation
energy-momentum tensor is $T_{\!_R}^{ab} = \int f\,p^a\,p^b\,
\pi$, where $\pi$ is the momentum space volume element. The
only non-zero energy-momentum tensor contribution from the matter,
relative to the 4-velocity $u^a$, is the energy density; so the
total energy density is
\be
\l{eq:10}
\mu = \mu_{\!_R} + \rho = {\cal O}[0] \ ,
\ee
while all other energy-momentum tensor components are simply equal
to the radiation contributions. Without confusion we will {\it
omit} a subscript `$R$' on these terms.\\
        
(2) The matter 4-velocity field $u^{a}$ is geodesic and expanding:
\be
\l{eq:11}
\udot^{a} = 0 \ , \hsp5 \Th = {\cal O}[0] > 0 \ ,
\ee
(the first requirement in fact follows from momentum conservation
for the pressure-free matter; see (\r{eq:en1})). The assumption
of an expanding Universe model is essential to what follows, for
otherwise there are counter-examples to the result \ct{emn78}.
\\
        
(3) The radiation obeys Liouville's equation (\r{eq:12}): $L(f) =
0$, where $L$ is the Liouville operator. This means that there is
{\em no\/} entropy production, so that $q^a$ and $\pi_{ab}$ are
{\em not\/} dissipative quantities, but measure the extent to which
$f$ deviates from isotropy.\\
        
(4) Relative to $u^a$ --- i.e., for all matter-comoving observers
--- the photon distribution function $f$ is almost isotropic
everywhere in the region $U$. Formally, in this region, $F =
F(x,E)$ and its time derivatives are zeroth-order, while
$F_{A_\ell} = F_{A_\ell}(x,E)$ plus their time and space
derivatives are at most first-order for $\ell > 0$. In brief:
\be
\l{eq:13}
F, \,\dot{F} = {\cal O}[0] \ , \hsp5
F_{A_\ell}, \,\dot{F}_{\la A_\ell\ra}, \,
\3nab_{a}F_{A_\ell} = {\cal O}[1] \ ,
\ee
(and we assume all higher derivatives of $F_{A_\ell}$ are also
${\cal O}[1]$ for $\ell > 0$).\\
        
It immediately follows, under reasonable assumptions about the
phase space integrals of the harmonic components, that the scalar
moments are zeroth-order but the tensor moments and their derivatives
are first-order. We assume the conditions required to ensure this
are true, i.e., we additionally suppose
\be
\l{eq:14}
\mu_{\!_R}, \,p_{\!_R} = {\cal O}[0] \ , \hsp5 q_{a},
\,\dot{q}_{\la a\ra},\,\3nab_{a}q_{b} =  {\cal O}[1] \ , \hsp5
\pi_{ab}, \,\dot{\pi}_{\la ab\ra}, \,\3nab_{a}\pi_{bc} = {\cal O}[1]
\ee
are satisfied as a consequence of (\r{eq:13}), and that the same
holds for the higher time and space derivatives of $q^{a}$,
$\pi_{ab}$ and for all higher-order moments (specifically, those
defined in (\r{eq:22}) below). The {\em first aim\/} now is to
show that
      
  (a) the kinematical quantities and the Weyl curvature are
  almost-FLRW, i.e., $\sig_{ab}$ and $\om^{a}$ are first-order,
  which then implies that $E_{ab}$ and $H_{ab}$ are also
  first-order. The {\em second aim\/} is to show that then,
        
  (b) there are coordinates such that the metric tensor takes a
  perturbed RW form. 

\subsubsection{Proving almost-FLRW kinematics}
Through an appropriate integration over momentum space, the zeroth
harmonic of Liouville's equation (\r{eq:12}) gives the energy
conservation equation for the radiation (\,cf. (\r{nl5})\,),
\be
\l{eq:15}
\dot{\mu}_{\!_R} + \sfrac{4}{3}\,\Th\,\mu_{\!_R}
+ \3nab_{a}q^{a} + \sig_{ab}\pi^{ab} = 0 \ ,
\ee
while the first harmonic gives the momentum conservation equation
(\,cf. (\r{nl6})\,),
\be
\l{eq:16}
\dot{q}^{\la a\ra} + \sfrac{4}{3}\,\Th\,q^a + \sfrac{1}{3}\,
\3nab^{a}\mu_{\!_R} + \3nab_{b}\pi^{ab} + \sig^{a}{}_{b}\,q^{b}
+ \eta^{abc}\,\om_{b}\,q_{c} = 0 \ ,
\ee
which implies by (\r{eq:14}),
\be
\l{eq:17}
\3nab_{a}\mu_{\!_R} = {\cal O}[1] \ .
\ee
Taking spatial derivatives of (\r{eq:16}), the same result
follows for the higher spatial derivatives of $\mu_R$; in
particular, its second derivatives are at most first-order. \\
        
Now the definition of the $\3nab$-derivative leads to the identity
(\r{eq:tilt}), giving
\be
\l{eq:18}
(\3nab_a\3nab_b - \3nab_b\3nab_a)\mu_{\!_R}
= 2\,\eta_{abc}\,\om^{c}\,\dot{\mu}_{\!_R} \ .
\ee
In an expanding Universe model, by (\r{eq:14}), the energy
conservation equation (\r{eq:15}) for the radiation shows
\be
\l{eq:19}
\dot{\mu}_{\!_R} + \sfrac{4}{3}\,\Th\,\mu_{\!_R} = {\cal O}[1] \ .
\ee
Thus, because the model is expanding, $\dot{\mu}_R$ is
zeroth-order. However, the left-hand side of (\r{eq:18}) is
first-order; consequently, by the higher-derivative version of
(\r{eq:17}),
\be
\l{eq:20}
\om^{a} = {\cal O}[1] \ .
\ee
The second harmonic of Liouville's equation (\r{eq:12}) leads,
after an appropriate integration over momentum space, to an
evolution equation for the anisotropic stress tensor $\pi_{ab}$
(which is the present version of (\r{nl8})):
\be
\l{eq:21}
\dot{\pi}^{\la ab\ra} + \sfrac{4}{3}\,\Th\,\pi^{ab}
+ \sfrac{8}{15}\,\mu_{\!_R}\,\sig^{ab}
+ \sfrac{2}{5}\,\3nab^{\la a}q^{b\ra}
+ J^{ab}
+ 2\,\om^{c}\,\eta_{cd}{}^{\la a}\,\pi^{b\ra d}
+ \sfrac{2}{7}\,\sig_{c}{}^{\la a}{}\,\pi^{b\ra c}
- \frac{32\pi}{315}\,\sig_{cd}\,\Pi^{abcd} = 0 \ ,
\ee
where (cf. (\r{r10}))
\be
\l{eq:22}
J^{ab} = \frac{8\pi}{35}\int_0^\infty E^3\,\3nab_{c}F^{abc}\,dE \ ,
\hsp5 \Pi^{abcd} = \int_0^\infty E^3\,F^{abcd}\,dE \ .
\ee
Consequently, by (\r{eq:20}), (\r{eq:13}) and (\r{eq:14}),
\be
\l{eq:23}
\sig_{ab} = {\cal O}[1] \ ,
\ee
and (on taking derivatives of the above equation) the same is true
for its time and space derivatives. Then, to first order, the
evolution equations become
\bea
\dot{\mu}_{\!_R} + \sfrac{4}{3}\,\Th\,\mu_{\!_R} + \3nab_{a}q^{a}
& \approx & 0 \ , \\
\dot{q}^{\la a\ra} + \sfrac{4}{3}\,\Th\,q^{a} + \sfrac{1}{3}\,
\3nab^{a}\mu_{\!_R} + \3nab_{b}\pi^{ab}
& \approx & 0 \ , \\
\dot{\pi}^{\la ab\ra} + \sfrac{4}{3}\,\Th\,\pi^{ab}
+ \sfrac{8}{15}\,\mu_{\!_R}\,\sig^{ab}
+ \sfrac{2}{5}\,\3nab^{\la a}q^{b\ra} + J^{ab}
& \approx & 0 \ ,
\eea
showing the equations can {\em only\/} close at first order {\em
if\/} we can argue that $J^{ab} = {\cal O}[2]$. {\em Any\/} such
approximation must be done with utmost care because of the {\bf
radiation multipole truncation theorem} (see below). \\
        
It now follows from the shear propagation equation (\r{eq:sigdot})
and the $H_{ab}$-equation (\r{hconstr}) that all the Weyl curvature
components are also at most first-order:
\be
\l{eq:24}
E_{ab} = {\cal O}[1]\ , \hsp5 H_{ab} = {\cal O}[1] \ .
\ee
Consequently, the $(\div\,E)$-equation (\r{eq:divE}) shows that
\be
\l{eq:25}
X_a \equiv \3nab_a\mu = {\cal O}[1] \hsp5 \Rightarrow \hsp5
\3nab_a\rho = {\cal O}[1] \ ,
\ee
by (\r{eq:10}). Finally, the $(0\alpha)$-equation (\r{eq:onu}), or
spatial derivatives of (\r{eq:19}), show that
\be
\l{eq:26}
Z_a \equiv \3nab_a\Th = {\cal O}[1] \ .
\ee
This establishes that the kinematical quantities for the matter
flow and the Weyl curvature are almost-FLRW: all the quantities that
vanish in the FLRW case are at most first-order here. Thus, the
only zeroth-order $1+3$ covariantly defined quantities are those that
are non-vanishing in FLRW models.

\subsubsection{Proving almost-FLRW dynamics}
It follows that the zeroth-order equations governing the dynamics are
just those of a FLRW model. Because the kinematical quantities
and the Weyl curvature are precisely those we expect in a perturbed
FLRW model, we can linearise the $1+3$ covariant equations about
the FLRW values in the usual way, in which the background model
will --- as we have just seen --- obey the usual FLRW
equations. Hence, the usual linearised $1+3$ covariant FLRW
perturbation analyses can be applied \ct{eb89,bde92}, leading to
the usual results for growth of inhomogeneities in almost-FLRW
Universe models.

\subsubsection{Finding an almost-RW metric}
Given that the kinematical quantities and the Weyl curvature take an
almost-FLRW form, the {\em key issue\/} in proving existence of an
almost-RW metric is choice of a time function to use as a cosmic
time (in the realistic, inhomogeneous Universe model). The problem
is that although we have shown the vorticity will be small, it will
in general not be zero; hence, there will be no time surfaces
orthogonal to the matter flow lines \ct{ehl61,ell71}. The problem
can equivalently be viewed as the need to find a vorticity-free
(i.e., hypersurface-orthogonal) congruence of curves to use as a
kinematical reference frame, which, if possible, one would like to
also be geodesic. As we cannot assume the matter 4-velocity
fulfills this condition, we need to introduce another congruence of
curves, say $\hat{u}^a$, that {\it is} hypersurface-orthogonal and
that does not differ too greatly from $u^a$, so the matter is
moving slowly (non-relativistically) relative to that frame. Then
we have to change to that frame, using a `hat' to denote its
kinematical quantities. (See \ct{sme95}.)

\subsubsection{Result}
This gives the result we want; we have shown that freely
propagating almost-isotropic background radiation everywhere in a
region $U$ implies the Universe model is almost-FLRW in that
region. Thus, we have proved the stability of the Ehlers, Geren and
Sachs \ct{egs68} result. This result is the foundation for the
important analysis of Sachs and Wolfe \ct{sw67} and all related
analyses, determining the effect of inhomogeneities on the CBR by
integration from last scattering till today in an almost-FLRW
Universe model, for these papers {\em start off\/} with the
assumption that the Universe {\em is\/} almost-FLRW since
decoupling, and build on that basis. \\
        
It should also be noted that the assumption that radiation is
isotropic about distant observers can be partially checked by
testing how close the CBR spectrum is to black body in those
directions where we detect the {\bf Sunyaev--Zel'dovich effect}
\ct{sunzel70} (see also Goodman \ct{goo95}). Since that effect
mixes and scatters incoming radiation, substantial radiation
anisotropies relative to those clusters of galaxies inducing the
effect would result in a significant distortion of the outgoing
spectrum.\\

{\em Exercise\/}: How good are the limits we might obtain on the
CBR anisotropy at distance points on our past light cone by this
method?  Are there any other ways one can obtain such limits?

\subsection{Other CBR calculations}
One can extend the above analysis to obtain model-independent
limits on the CBR anisotropy (Maartens {\em et al\/} \ct{mes95a}).
However, there are also a whole series of model-dependent analyses
available.

\subsubsection{Sachs--Wolfe and related effects}
We will not pursue here the issue of {\bf Sachs--Wolfe effect}
\ct{sw67} type calculations, integrating up the redshift from the
surface of last scattering to the observer and so determining the
CBR anisotropy, and, hence, cosmological parameters \ct{jung96},
because they are so extensively covered in the literature. The GIC
version is covered by Challinor and Lasenby \ct{cl98} and
developed in depth in the series of papers by Gebbie, Dunsby,
Maartens and Ellis \ct{mageel98,ge98,ge2}. However, a few comments
are in order. \\

One can approach this topic $1+3$ covariantly by a direct
generalisation of the photon type of calculation (see
\ct{dun97,chalas98a}), or by use of the relativistic
kinetic theory formalism developed above, as in the
papers just cited. Two things need to be very carefully
considered. These are,\\

  (i) The issue of putting {\em correct limits\/} on the
  Sachs--Wolfe integral. This integral needs to be taken to the
  {\em real\/} surface of last scattering, not the background surface
  of last scattering (which is fictitious). If this is not done,
  the results may be gauge-dependent. One should place the limits
  on this integral properly, which can be done quite easily by
  calculating when the optical depth due to Thomson scattering is
  unity \ct{pan86,xu94}.\\

  (ii) In order to obtain {\em solutions\/}, one has somehow to cut
  off the harmonic series, for otherwise there is an infinite
  regress whereby higher-order harmonics determine the evolution of
  lower-order ones, as is clear from the multipole equations
  above. However, truncation of harmonics in a kinetic theory
  description needs to be approached with great caution, because of
  the following theorem for collision-free radiation \ct{etm83}:

\begin{quotation}
{\bf Radiation Multipole Truncation Theorem}: In the exact
Einstein--Liouville theory, truncation of the angular harmonics
seen by a family of geodesic observers {\it at any order whatever}
leads to the vanishing of the shear of the family of observers;
hence, this can only occur in highly restricted spaces.
\end{quotation}
The proof is given in \ct{etm83}. This is an exact result of the
full theory. However, {\it it remains true in the linear theory\/}
if the linearisation is done carefully, {\it even though the term
responsible for the conclusion is second-order} and so is omitted
in most linearised calculations. \\

The point is that in that equation, at the relevant order, there
are {\em only\/} second-order terms in the equation, so one cannot
drop the terms responsible for this conclusion (one can only drop
them in an equation where there are also linear terms, so the
second-order terms are negligible compared with the linear terms;
that is not the case here). Thus, a proper justification for the
acceptability of some effective truncation procedure relies on
showing how a specific approximation procedure gives an acceptable
approximation to the results of the exact theory, despite this
disjuncture.\\

{\em Exercise\/}: Give such a justification.

\subsubsection{Other models}
As mentioned previously in these lectures, there is an extensive
literature on the CBR anisotropy:
\begin{itemize}
\item in Bianchi (exact spatially homogeneous) models; see,
e.g., \ct{bar83}--\ct{bunn96}, 
\item in Swiss-Cheese (exact inhomogeneous) models; see,
e.g., \ct{dye76} and \ct{pan92}, 
\item in small universes (FLRW models with compact spatial topology
that closes up on a small scale, so that we have seen round the
universe already since decoupling \ct{ellsch86}); see,
e.g., \ct{ste93} and \ct{cor97}.
\end{itemize}
These provide important parametrized sets of models that one can
use to test and exploit the restrictions the CBR observations place
on alternative Universe models that are not necessarily close to
the standard models at early or late times, but are still
potentially compatible with observations, and deserve full
exploration. \\

In each case, whether carrying out Sachs--Wolfe type or more
specific model calculations, we get tighter limits than in the
almost-EGS case, but they are also more model-dependent. In each
case they can be carried out using the $1+3$ covariantly defined
harmonic variables and can be used to put limits on the $1+3$
covariantly defined quantities that are the theme of this
article. \\

The overall conclusion is that --- given the Copernican assumption
underlying use of the almost-FLRW models --- the high degree of
observed CBR isotropy confirms use of these models for the
observable Universe and puts limits on the amplitude of anisotropic
and inhomogeneous modes in this region (i.e., within the horizon
and since decoupling). Other models remain viable as
representations of the Universe at earlier and later times and
outside the particle horizon.

\section{Conclusion and open issues}
The sections above have presented the $1+3$ covariant variables and
equations, together with the tetrad equations, a series of
interesting exact cosmological models, and a systematic procedure
for obtaining approximate almost-FLRW models and examining
observations in these models. It has been shown how a number of
anisotropic and inhomogeneous cosmological models are useful in
studying observational limits on the geometry of the real Universe,
and how interesting dynamical and observational issues arise in
considering these models; indeed issues such as whether inflation
in fact succeeds in making the Universe isotropic or not cannot be
tackled without examining such models \ct{anetal91}.

\subsection{Conclusion}
The {\bf standard model of cosmology} is vindicated in the
following sense (cf. \ct{pee88,colell97}):

\begin{itemize}
\item The Universe is expanding and evolving, as evidenced by the 
({\em magnitude, redshift\/})--relation for galaxies and other
sources, together with number count observations and a broad
compatibility of age estimates;

\item It started from a hot big bang early stage which evolved to
the presently observed state, this early era being evidenced by the
CBR spectrum and concordance with nucleosynthesis observations;

\item Given a\footnote{Necessarily philosophically based \ct{ell75}.} 
Copernican assumption, the high degree of isotropy of the CBR and
other observations support an almost-FLRW (nearly homogeneous and
isotropic) model within the observable region (inside our past
light cone) since decoupling and probably back to nucleosynthesis
times;

\item There are globally inhomogeneous spherically symmetric models
that are also compatible with the observations if we do not
introduce the Copernican assumption \ct{muheel98}, and there are
inhomogeneous and anisotropic modes that suggest the Universe is
not almost-FLRW at very early and very late times
\ct{waiell97,waietal98}, and on very large scales (outside the
particle horizon) \ct{lin90}.
\end{itemize}
The issue of cosmological parameters may be resolved in the next
decade due to the flood of new data coming in --- from deep space
number counts and redshift measurements, observations of supernovae
in distant galaxies, measurement of strong and weak gravitational
lensing, and CBR anisotropy measurements (see Coles and Ellis
\ct{colell97} for a discussion). \\

The further extensions of this model proposed from the {\bf particle
physics} side are at present mainly compatible but not yet as
compelling from an observational viewpoint, namely:
\begin{itemize}
\item An early inflationary era helps resolve some philosophical
puzzles about the structure of the Universe related specifically to
its homogeneity and isotropy \ct{gut80}--\ct{lin90}, but the link
to particle physics will not be compelling until a specific inflaton
candidate is identified;
\item Structure formation initiated by quantum fluctuations at very
early stages in an inflationary model are a very attractive idea,
with the proposal of a CDM-dominated late Universe and associated
predictions for the CBR anisotropy giving a strong link to
observations that may be confirmed in the coming decade; however,
theoretical details of this scheme, and particularly the associated
issues of biasing and the normalisation of matter to radiation,
are still to be resolved;
\item The density of matter in the Universe is probably below the
critical value predicted by the majority of inflationary models
\ct{colell97}; compatibility with inflation may be preserved by
either moving to low density inflationary models\footnote{For an
early proposal, see \ct{ellmij} and \ct{ell91}.} or by confirmation
of a cosmological constant that is dynamically dominant at the
present time;\footnote{But way below that predicted by present
field theory \ct{weinberg}.} the latter proposal needs testing
relative to number counts, lensing observations, and structure
formation scenarios;
\item The vibrant variety of pre-inflationary proposals involve a
huge extrapolation of presently known physics to way beyond the
testable domain and, given their variety, do not as yet give a
compelling unique view of that era.
\end{itemize}
Hence, it is suggested \ct{colell97} that these proposals should
not be regarded as part of the standard model but rather as
interesting avenues under investigation.

\subsection{Open issues}
Considered from a broader viewpoint, substantial issues remain
unresolved. Among them are, \\

(1) The Newtonian theory of cosmology is not yet adequately
resolved.  Newtonian theory is only a good theory of gravitation
when it is a good approximation to General Relativity; obtaining
this limit in non-linear cosmological situations raises a series of
questions and issues that still need clarifying \ct{vanell98a},
particularly relating to boundary conditions in realistic Newtonian
cosmological models.\\

(2) We have some understanding of how the evolution of families of
inhomogeneous models relates to that of families of higher symmetry
models.  It has been indicated that a skeleton of higher symmetry
models seems to guide the evolution of lower symmetry models in the
state space (the space of cosmological space-times)
\ct{waiell97}. This relation needs further elucidation. Also,
anisotropic and inhomogeneous inflationary models are relatively
little explored and problems remain \ct{raymod88,pen89b};\\

(3) We need to find a suitable measure of probability in the full
space of cosmological space-times, and in its involutive
subspaces. The requirement is a natural measure that is
plausible. Some progress has been made in the FLRW subcase
\ct{gihast87}--\ct{cou95}, but even here it is not
definitive. Closely related to this is the issue of the stability
of the results we derive from cosmological modelling
\ct{tavell88};\\

(4) We need to be able to relate descriptions of the same
space-time on different scales of description. This leads to the
issue of averaging and the resulting effective (polarization)
contributions to the energy-momentum tensor, arising because
averaging does not commute with calculating the field equations for
a given metric \ct{ell84}, and we do not have a good procedure for
fitting a FLRW model to a lumpy realistic model of the Universe
\ct{ellsto87}. It includes the issue (discussed briefly above) of
how the almost-everywhere empty real Universe can have dynamical
and observational relations that average out to high precision to
the FLRW relations on a large scale. \\

(5) Related to this is the question of definition of entropy for
gravitating systems in general,\footnote{In the case of black
holes, there is a highly developed theory; but there is no
definition for a general gravitational field; see,
e.g., \cite{pen99}.} and cosmological models in particular. This may
be expected to imply a coarse-graining in general, and so is
strongly related to the averaging question. It is an important
issue in terms of its relation to the spontaneous formation of
structure in the early Universe, and also relates to the still
unresolved arrow of time problem \ct{pen99}, which in turn relates
to a series of further issues concerned with the effect on local
physics of boundary conditions at the beginning of the Universe
(see, e.g., \ct{ellsch72}).\\

(6) One can approach relating a model cosmology to astronomical
observations by a strictly observational approach
\ct{krisac66,ensmw85}, as opposed to the more usual model-based
approach as envisaged in the main sections above and indeed in most
texts on cosmology. Intermediate between them is a best-fitting
approach \ct{ellsto87}. Use of such a fitting approach is probably
the best way to tackle modelling the real Universe \ct{maelst95},
but it is not yet well-developed.\\

(7) Finally, underlying all these issues is the series of problems
arising because of the uniqueness of the Universe, which is what
particularly gives cosmology its unique character and underlies the
special problems in cosmological modelling and application of
probability theory to cosmology \ct{ell91a}. Proposals to deal with
this by considering an ensemble of Universe models realized in one
or other of a number of possible ways are in fact untestable and,
hence, of a metaphysical rather than physical nature; but this
needs further exploration.\\

There is interesting work still to be done in all these areas. It
will be important in tackling these issues to, as far as possible,
use gauge-invariant and $1+3$ covariant methods, because coordinate
methods can be misleading.

\section*{Acknowledgements}
GFRE and HvE thank Roy Maartens, John Wainwright, Peter Dunsby, Tim
Gebbie, and Claes Uggla for work done together, some of which is
presented in these lectures, Charles Hellaby for useful comments
and references, Tim Gebbie for helpful comments, and Anthony
Challinor and Roy Maartens for corrections to some equations. We
are grateful to the Foundation for Research and Development (South
Africa) and the University of Cape Town for financial support. HvE
acknowledges the support by a grant from the Deutsche
Forschungsgemeinschaft (Germany).

This research has made use of NASA's Astrophysics Data System.



\end{document}